\IfFileExists{/dev/null}{\immediate\write18{sh makeLinks.sh}}{\immediate\write18{makeLinks.bat}}

\documentclass[a4paper,journal,twocolumn,10pt,final]{IEEEtran}

\makeatletter
\def\subsubsection{%
	\@startsection
	{subsubsection}                 
	{3}                             
	{\z@}                           
	{2.5ex plus 1.5ex minus 1.5ex}  
	{1ex plus .5ex minus 0ex}     
	{\normalfont\normalsize\itshape}
}
\makeatother

\usepackage[none]{hyphenat}
\usepackage[utf8]{inputenc}
\usepackage[T1]{fontenc}
\usepackage{textcomp}
\usepackage{scalefnt}
\usepackage{amsmath,amsfonts,amssymb,amsbsy,amstext}

\usepackage{mathtools}          
\usepackage{cmap}               
\usepackage[dvipsnames,svgnames,x11names]{xcolor}   
\usepackage{varwidth}

\let\originalleft\left
\let\originalright\right
\renewcommand{\left}{\mathopen{}\mathclose\bgroup\originalleft}
\renewcommand{\right}{\aftergroup\egroup\originalright}

\usepackage{graphicx}
\graphicspath{{.}{.\figs}}
\usepackage{placeins} 

\usepackage{multirow}
\usepackage{tabularx}
\usepackage{booktabs}
\newcolumntype{C}{>{\centering\arraybackslash}X}
\newcolumntype{R}{>{\flushright\arraybackslash}X}
\newcolumntype{L}{>{\flushleft\arraybackslash}X}
\newcolumntype{P}{>{\centering\arraybackslash} p{0.5\linewidth}}

\usepackage[%
style=ieee,
backend=biber, 
]{biblatex}

\defbibheading{bibliography}[\refname]{\section*{#1}}

\makeatletter
\g@addto@macro{\UrlBreaks}{\UrlOrds}
\makeatother

\usepackage{acro}

\NewDocumentCommand{\acro}{m o m o}
{%
	\IfValueTF{#2}{%
		\IfValueTF{#4}{%
			\DeclareAcronym{#1}{short={#2},long={#3},#4}
		}{%
			\DeclareAcronym{#1}{short={#2},long={#3}}
		}
	}{%
		\IfValueTF{#4}{%
			\DeclareAcronym{#1}{short={#1},long={#3},#4}
		}{%
			\DeclareAcronym{#1}{short={#1},long={#3}}
		}
	}
}

\acro{2D}{two-dimensional}
\acro{3D}{three-dimensional}
\acro{4D}{four-dimensional}
\acro{6D}{six-dimensional}
\acro{1G}{first generation}
\acro{2G}{second generation}
\acro{3G}{third generation}
\acro{4G}{fourth generation}
\acro{5G}{fifth generation}
\acro{5GC}{5G core network}
\acro{5G-StoRM}{5G stochastic radio channel for dual mobility}
\acro{6G}{sixth generation}
\acro{3GPP}{3rd generation partnership project}
\acro{3GPP2}{3rd generation partnership project 2}
\acro{A2A}{air-to-air}
\acro{A2G}{air-to-ground}
\acro{AA}{antenna array}
\acro{AC}{admission control}
\acro{AD}{attack-decay}
\acro{AE}{antenna element}
\acro{AF}{amplify and forward}
\acro{ABS}{almost blank subframe}
\acro{ACF}{autocorrelation function}
\acro{ADSL}{asymmetric digital subscriber line}
\acro{AHW}{alternate hop-and-wait}
\acro{AI}{artificial intelligence}
\acro{AMC}{adaptive modulation and coding}
\acro{AP}{access point}
\acro{APA}{adaptive power allocation}
\acro{API}{application protocol interface}
\acro{ARQ}{automatic repeat request}
\acro{AT}{averaged time}
\acro{ARMA}{autoregressive moving average}
\acro{ASE}{average squared error}
\acro{ASC}{adaptive satisfaction control}
\acro{ATES}{adaptive throughput-based efficiency-satisfaction trade-off}
\acro{AWGN}{additive white gaussian noise}
\acro{BAP}{backhaul adaptation protocol}
\acro{BB}{branch and bound}
\acro{BC}{branch and cut}
\acro{BD}{block diagonalization}
\acro{BER}{bit error rate}
\acro{BF}{best fit}
\acro{BL}{bit loading}
\acro{BLER}{block error rate}
\acro{BLPC-1}{bit loading with power constraint in hop 1}
\acro{BLPC-2}{bit loading with power constraint in hop 2}
\acro{BPC}{binary power control}
\acro{BPSK}{binary phase-shift keying}
\acro{BRA}{balanced random allocation}
\acro{BS}{base station}
\acro{BSP}{\acs*{BS} placement}
\acro{BSR}{buffer status report}
\acro{CAP}{combinatorial allocation problem}
\acro{CAPEX}{capital expenditure}
\acro{CB}{contextual bandit}
\acro{CBF}{coordinated beamforming}
\acro{CBR}{constant bit rate}
\acro{CBS}{class based scheduling}
\acro{CC}{congestion control}
\acro{CCL}{common cell list}
\acro{CDF}{cumulative distribution function}
\acro{CDL}{clustered delay line}
\acro{CDMA}{code-division multiple access}
\acro{CIR}{channel impulse response}
\acro{CH}{channel hardening}
\acro{CHO}{conditional handover}
\acro{C-RAN}{cloud-based radio access network}
\acro{CL}{closed loop}
\acro{CLT}{central limit theorem}
\acro{CLPC}{closed loop power control} 
\acro{CN}{core network}       
\acro{CNR}{channel-to-noise ratio}
\acro{CPA}{cellular protection algorithm}
\acro{CPICH}{common pilot channel}
\acro{CoMP}{coordinated multi-point}
\acro{CQI}{channel quality indicator}
\acro{CRE}{cell range expansion}
\acro{CRM}{constrained rate maximization}
\acro{CRN}{cognitive radio network}
\acro{C-RNTI}{cell radio network temporary identifier}
\acro{CRRM}{centralized/common radio resource management}
\acro{CRS}{cell-specific reference signal}
\acro{CS}{coordinated scheduling}
\acro{CSI}{channel state information}
\acro{CTS}{clear to send}
\acro{CU}{centralized unit}
\acro{CUE}{cellular user equipment}
\acro{CWND}{congestion window size}
\acro{D2D}{device-to-device}
\acro{D2N}{drone-to-network}
\acro{DAA}{drone antenna array}
\acro{DAG}{directed acyclic graph}
\acro{DBS}{drone mounted base station}
\acro{DC}{dual connectivity}
\acro{DCA}{dynamic channel allocation}
\acro{DE}{differential evolution}
\acro{DF}{decode and forward}
\acro{DFT}{discrete fourier transform}
\acro{DIST}{distance}
\acro{DL}{downlink}
\acro{DMA}{double moving average}
\acro{DMRS}{demodulation reference signal}
\acro{D2DM}{\acs*{D2D} mode}
\acro{DMS}{\acs*{D2D} mode selection}
\acro{DPC}{dirty paper coding}
\acro{DRA}{dynamic resource assignment}
\acro{DSA}{dynamic spectrum access}
\acro{DSM}{delay-based satisfaction maximization}
\acro{DU}{distributed unit}
\acro{E2E}{end-to-end}
\acro{ECC}{electronic communications committee}
\acro{EDF}{earliest deadline first}
\acro{EE}{energy efficiency}
\acro{EFLC}{error feedback based load control}
\acro{EI}{efficiency indicator}
\acro{e-ICIC}{enhanced inter-cell interference coordination}
\acro{eMBB}{enhanced mobile broadband}
\acro{eNB}{evolved node B}
\acro{EXP}{exponential}
\acro{EPA}{equal power allocation}
\acro{EPC}{evolved packet core}
\acro{EPS}{evolved packet system}
\acro{E-UTRAN}{evolved universal terrestrial radio access network}
\acro{ES}{exhaustive search}
\acro{FCP}{fundamental counting principle}
\acro{FCA}{flow control algorithm}
\acro{FD}{full duplex}
\acro{FDD}{frequency division duplex}
\acro{FDM}{frequency division multiplexing}
\acro{FDMA}{frequency division multiple access}
\acro{FER}{frame erasure rate}
\acro{FIFO}{first in first out}
\acro{FF}{fast fading}
\acro{FRS}[FS]{fast-rat scheduling}
\acro{FS}{fast switching}
\acro{FSB}{fixed switched beamforming}
\acro{FST}{fixed \acs*{SNR} target}
\acro{FT}{fourier transform}
\acro{FTP}{file transfer protocol}
\acro{GA}{genetic algorithm}
\acro{GAP}{generalized assignment problem}
\acro{GAP-MQ}{generalized assignment problem with minimum quantities}
\acro{GATES}{generalized adaptive throughput-based efficiency-satisfaction trade-off}
\acro{GBR}{guaranteed bit rate}
\acro{GLR}{gain to leakage ratio}
\acro{gNB}{gNodeB}
\acro{GOS}{generated orthogonal sequence}
\acro{GP}{gaussian process}
\acro{GPL}{GNU general public license}
\acro{GPS}{global positioning system}
\acro{GRP}{grouping}
\acro{GSM}{global system for mobile communications}
\acro{GTEL}{wireless telecommunications research group}
\acro{HARQ}{hybrid automatic repeat request}
\acro{HBF}{hybrid beamforming}
\acro{HCPP}{hardcore point process}
\acro{HetNet}{heterogeneous network}
\acro{HD}{half duplex}
\acro{HH}{hughes-hartogs}
\acro{HardH}[HH]{hard handover}
\acro{HMS}{harmonic mode selection}
\acro{HO}{handover}
\acro{HOL}{head of line}
\acro{HPBW}{half power beamwidth}
\acro{HSDPA}{high speed downlink packet access}
\acro{HSR}{high-speed railway}
\acro{HSPA}{high speed packet access}
\acro{HTTP}{hypertext transfer protocol}
\acro{IAB}{integrated access and backhaul}
\acro{ICMP}{internet control message protocol} 
\acro{ICI}{intercell interference}
\acro{ICIC}{inter-cell interference coordination}
\acro{ID}{identification}
\acro{i.i.d.}{independent and identically distributed}
\acro{IETF}{internet engineering task force}
\acro{IPC}{individual power constraint}
\acro{UID}{unique identification}
\acro{IID}{independent and identically distributed}
\acro{IIR}{infinite impulse response}
\acro{ILP}{integer linear problem}
\acro{IMT}{international mobile telecommunications}
\acro{INV}{inverted norm-based grouping} 
\acro{IoT}{internet of things}
\acro{IP}{internet protocol}
\acro{IPv6}{internet protocol version 6}
\acro{IRA}{integrated resource allocation}
\acro{IRS}{intelligent reflecting surface}
\acro{ISD}{inter-site distance}
\acro{ISI}{inter symbol interference}
\acro{ISM}{industrial, scientific and medical}
\acro{ITU}{international telecommunication union}
\acro{JOAS}{joint opportunistic assignment and scheduling}
\acro{JOS}{joint opportunistic scheduling}
\acro{JP}{joint processing}
\acro{JRAPAP}{joint rb assignment and power allocation problem}
\acro{JS}{jump-stay}
\acro{JSM}{joint satisfaction maximization}
\acro{KKT}{karush-kuhn-tucker}
\acro{KPI}{key performance indicator}
\acro{LAC}{link admission control}
\acro{LA}{link adaptation}
\acro{LBS}{location based service}
\acro{LC}{load control}
\acro{LOS}{line of sight}
\acro{LP}{linear programming}
\acro{LTE}{long term evolution}
\acro{LTE-A}{\acs*{LTE}-advanced}
\acro{LTE-Advanced}{\ac{LTE-A}}
\acro{LTE-R}{\ac{LTE} railway}
\acro{LSP}{large scale parameter}
\acro{MADRDPG}{multi-agent deep recurrent deterministic policy gradient}
\acro{MeNB}{master \acs*{ENB}}
\acro{M2M}{machine-to-machine}
\acro{MAC}{medium access control}
\acro{MANET}{mobile ad hoc network}
\acro{MEDS}{method of exact doppler spread}
\acro{MC}{modular clock}
\acro{MCP}{minimal cost power}
\acro{MCS}{modulation and coding scheme}
\acro{MDB}{measured delay based}
\acro{MDI}{minimum \acs*{D2D} interference}
\acro{MDSM}{modified delay-based satisfaction maximization}
\acro{MDU}{max-delay-utility}
\acro{METIS}{mobile and wireless communications enablers for the twenty-twenty information society \acs*{5G}}
\acro{MF}{matched filter}
\acro{MFAP}{mobile femtocell access point}
\acro{MG}{maximum gain}
\acro{MH}{multi-hop}
\acro{mIAB}{mobile \ac{IAB}}
\acro{MILP}{mixed integer linear programming}
\acro{MIMO}{multiple input multiple output}
\acro{MINLP}{mixed integer nonlinear programming}
\acro{MIP}{mixed integer programming}
\acro{MISO}{multiple input single output}
\acro{MIT}{mobility interruption time}
\acro{ML}{machine learning}
\acro{MLWDF}{modified largest weighted delay first}
\acro{MME}{mobility management entity}
\acro{MMF}{max-min fairness}
\acro{mmMAGIC}{millimetre-wave based mobile radio access network for fifth generation integrated communications}
\acro{MMRP}{maximizing the minimal residual power}
\acro{MMRP-LB}{maximizing the minimal residual power with lower bound}
\acro{MMSE}{minimum mean square error}
\acro{mMTC}{massive machine-type communications}
\acro{mmWave}{millimeter wave}
\acro{MN}{master node}
\acro{MOS}{mean opinion score}
\acro{MPF}{multicarrier proportional fair}
\acro{MPRP}{maximization of the product of the residual powers}
\acro{MRA}{maximum rate allocation}
\acro{MR}{maximum rate}
\acro{MRC}{maximum ratio combining}
\acro{MRN}{mobile relay node}
\acro{MRT}{maximum ratio transmission}
\acro{MRUS}{maximum rate with user satisfaction}
\acro{MS}{mode selection}
\acro{MSE}{mean squared error}
\acro{MCG}{master cell group} 
\acro{MSI}{multi-stream interference}
\acro{MT}{mobile termination}
\acro{MTC}{machine-type communication}
\acro{IMS}{\acs*{IP} multimedia subsystem}
\acro{MTSI}{multimedia telephony services over \acs*{IMS}}
\acro{MTSM}{modified throughput-based satisfaction maximization}
\acro{MU-MIMO}{multi-user multiple input multiple output}
\acro{MU}{multi-user}
\acro{Multi-CUT}{multi-cell and multi-user and multi-tier}
\acro{NAS}{non-access stratum}
\acro{NB}{node B}
\acro{nBS}{neighbor base station}
\acro{NCL}{neighbor cell list}
\acro{NG}{next generation}
\acro{NGC}{next generation core network}
\acro{NLP}{nonlinear programming}
\acro{NLOS}{non-line of sight}
\acro{NMSE}{normalized mean square error}
\acro{NN}{neural network}
\acro{NOMA}{non-orthogonal multiple access}
\acro{NORM}{normalized projection-based grouping}
\acro{NP}{non-polynomial time}
\acro{NR}{new radio}
\acro{NRT}{non-real time}
\acro{NSA}{non-stand-alone}
\acro{NSPS}{national security and public safety services}
\acro{OBF}{opportunistic beamforming}
\acro{OFDMA}{orthogonal frequency division multiple access}
\acro{OFDM}{orthogonal frequency division multiplexing}
\acro{OFPC}{open loop with fractional path loss compensation}
\acro{O2I}{outdoor-to-indoor}
\acro{OL}{open loop}
\acro{OLPC}{open-loop power control}
\acro{OL-PC}{open-loop power control}
\acro{OM}[O\&M]{operational \& maintenance}
\acro{OMA}{orthogonal multiple access}
\acro{OPEX}{operational expenditure}
\acro{ORA}{orthogonal resource allocation}
\acro{ORB}{orthogonal random beamforming}
\acro{JO-PF}{joint opportunistic proportional fair}
\acro{OSI}{open systems interconnection}
\acro{PA}{power allocation}
\acro{PAIR}{\acs*{D2D} pair gain-based grouping}
\acro{PAPR}{peak-to-average power ratio}
\acro{P2P}{peer-to-peer}
\acro{PBCH}{physical broadcast channel}
\acro{PBS}{pico base station}        
\acro{PC}{power control}
\acro{PCI}{physical cell \acs*{ID}}
\acro{PDCP}{packet data convergence protocol}
\acro{PDF}{probability density function}
\acro{PDU}{protocol data unit}
\acro{PER}{packet error rate}
\acro{PF}{proportional fair}
\acro{PGRA}{probabilistic graph based resource allocation}
\acro{P-GW}{packet data network gateway}
\acro{PHY}{physical}
\acro{PL}{pathloss}
\acro{PLR}{packet loss ratio}
\acro{PRABE}{power and resource allocation based on quality of experience}
\acro{PRB}{physical resource block}
\acro{PROJ}{projection-based grouping}
\acro{PSD}{power spectral density}
\acro{PSDe}{positive semi-definite}
\acro{ProSe}{proximity services}
\acro{PS}{packet scheduling}
\acro{PSO}{particle swarm optimization}
\acro{PSS}{primary synchronization signal}
\acro{PTAS}{polynomial-time approximation scheme}
\acro{PTP}{point-to-point}
\acro{PZF}{projected zero-forcing}
\acro{QAM}{quadrature amplitude modulation}
\acro{QHMLWDF}{queue-HOL-MLWDF}
\acro{QoE}{quality of experience}
\acro{QoS}{quality of service}
\acro{QPSK}{quadri-phase shift keying}
\acro{QSM}{queue-based satisfaction maximization}
\acro{QuaDRiGa}{quasi deterministic radio channel generator}
\acro{RACH}{random access}
\acro{RAISES}{reallocation-based assignment for improved spectral efficiency and satisfaction}
\acro{RAN}{radio access network}
\acro{RA}{resource allocation}
\acro{RAP}{rb assignment problem}
\acro{RAT}{radio access technology}[long-plural-form={radio access technologies}]
\acro{RATE}{rate-based}
\acro{RAU}{remote antenna unit}
\acro{RB}{resource block}
\acro{RBG}{resource block group}
\acro{REF}{reference grouping}
\acro{RET}{remote electrical tilt}
\acro{RF}{radio frequency}
\acro{RL}{reinforcement learning}
\acro{RLC}{radio link control}
\acro{RLF}{radio link failure}
\acro{RM}{rate maximization}
\acro{RMa}{rural macro}
\acro{RMJ-SNR}{region of the minimum joint snr}
\acro{RMEC}{rate maximization under experience constraints}
\acro{RMSE}{root-mean-square error}
\acro{RNC}{radio network controller}
\acro{RND}{random grouping}
\acro{RNN}{recurrent neural network}
\acro{RRA}{radio resource allocation}
\acro{RRM}{radio resource management}
\acro{RSCP}{received signal code power}
\acro{RSRP}{reference signal received power}
\acro{RSRQ}{reference signal received quality}
\acro{RR}{round robin}
\acro{RRC}{radio resource control}
\acro{RSSI}{received signal strength indicator}
\acro{RT}{real time}
\acro{RTS}{request to send}
\acro{RU}{resource unit}
\acro{RUNE}{rudimentary network emulator}
\acro{RV}{random variable}
\acro{Rx}{receiver}
\acro{RZF}{regularized zero-forcing}
\acro{SA}{subcarrier assignment}
\acro{SAC}{session admission control}
\acro{sBS}{serving base station}
\acro{SC}{small cell}
\acro{SCon}[SC]{single connectivity}
\acro{SCG}{secondary cell group}
\acro{SCM}{spatial channel model}
\acro{SCS}{subcarrier spacing}
\acro{SC-FDMA}{single carrier - frequency division multiple access}
\acro{SCRV}{spatially consistent random variable}
\acro{SD}{soft dropping}
\acro{SDAP}{service	data adaptation protocol}
\acro{S-D}{source-destination}
\acro{SDPC}{soft dropping power control}
\acro{SDM}{space-division multiplexing}
\acro{SDMA}{space-division multiple access}
\acro{SE}{squared error}
\acro{SF}{shadow fading}
\acro{SMDP}{semi-markov	decision problem}
\acro{SeNB}{secondary \acs*{ENB}}
\acro{SER}{symbol error rate}
\acro{SES}{simple exponential smoothing}
\acro{S-GW}{serving gateway}
\acro{SIC}{successive interference cancellation}
\acro{SINR}{signal to interference-plus-noise ratio}
\acro{SLNR}{signal to leakage-plus-noise ratio}
\acro{SNN}{strictly non-negative}
\acro{SI}{satisfaction indicator}
\acro{SIP}{session initiation protocol}
\acro{SISO}{single input single output}
\acro{SIMO}{single input multiple output}
\acro{SIR}{signal to interference ratio}
\acro{SM}{subcarrier matching}
\acro{SMA}{simple moving average}
\acro{SMSE}{spatial-mean-square error}
\acro{SN}{secondary node}
\acro{SNR}{signal to noise ratio}
\acro{SON}{self organizing networks}
\acro{SoA}{state-of-the-art}
\acro{SoS}{sum-of-sinusoids}
\acro{SORA}{satisfaction oriented resource allocation}
\acro{SORA-NRT}{satisfaction-oriented resource allocation for non-real time services}
\acro{SORA-RT}{satisfaction-oriented resource allocation for real time services}
\acro{SP}{signal processing}
\acro{SPF}{single-carrier proportional fair}
\acro{SRA}{sequential removal algorithm}
\acro{SRB1}{signaling radio bearer~1}
\acro{SRS}{sounding reference signal}
\acro{SSB}{synchronization signal block}
\acro{SSP}{small scale parameter}
\acro{SSS}{secondary synchronization signal}
\acro{ST}{spanning tree}
\acro{STTD}{space time transmit diversity}[long-plural-form={space time transmit diversities}]
\acro{SU-MIMO}{single-user multiple input multiple output}
\acro{SU}{single-user}
\acro{tBS}{target base station}
\acro{SVD}{singular value decomposition}
\acro{TCP}{transmission control protocol}
\acro{TDD}{time division duplex}
\acro{TDM}{time division multiplexing}
\acro{TDMA}{time division multiple access}
\acro{TETRA}{terrestrial trunked radio}
\acro{TP}{transmit power}
\acro{TPC}{total power constraint}
\acro{TR}{technical report}
\acro{TTI}{transmission time interval}
\acro{TTR}{time-to-rendezvous}
\acro{TTT}{time-to-trigger}
\acro{TSM}{throughput-based satisfaction maximization}
\acro{TU}{typical urban}
\acro{TV}{television}
\acro{TVWS}{\acs*{TV} white space}
\acro{Tx}{transmitter}
\acro{UAV}{unmanned aerial vehicle}
\acro{UABSC}{user-assisted bearer split control}
\acro{UDP}{user datagram protocol}
\acro{UDN}{ultra-dense networks}
\acro{UE}{user equipment}
\acro{UBA}{\ac{UE}-\ac{BS} association}
\acro{UEPS}{urgency and efficiency-based packet scheduling}
\acro{UFC}{federal university of cear\'{a}}
\acro{ULA}{uniform linear array}
\acro{UL}{uplink}
\acro{UMa}{urban macro}
\acro{UMi}{urban micro}
\acro{UMTS}{universal mobile telecommunications system}
\acro{UPN}{user provided network}
\acro{URA}{uniform rectangular array}
\acro{URI}{uniform resource identifier}
\acro{URLLC}{ultra-reliable low-latency communications}
\acro{URM}{unconstrained rate maximization}
\acro{VBR}{variable bit rate}
\acro{VET}{variable electrical tilt}
\acro{VMR}{vehicle-mounted relay}
\acro{VR}{virtual resource}
\acro{VoIP}{voice over \acs*{IP}}
\acro{WCDMA}{wideband code division multiple access}
\acro{WF}{water-filling}
\acro{WID}{wireless infrastructure drone}
\acro{Wi-Fi}{wireless fidelity}
\acro{WiMAX}{worldwide interoperability for microwave access}
\acro{WINNER}{wireless world initiative new radio}
\acro{WLAN}{wireless local area network}
\acro{WMPF}{weighted multicarrier proportional fair}
\acro{WP}{work package}
\acro{WPF}{weighted proportional fair}
\acro{WSN}{wireless sensor network}
\acro{WWW}{world wide web}
\acro{WFQ}{weighted fair queuing}
\acro{XIXO}{(single or multiple) input (single or multiple) output}
\acro{XPR}{cross-polarization power ratio}
\acro{ZF}{zero-forcing}
\acro{ZMCSCG}{zero mean circularly symmetric complex gaussian}

\usepackage{siunitx}
\usepackage{datetime}
\usepackage{url}

\usepackage[caption=false,font=footnotesize]{subfig}

\usepackage{algorithm}
\usepackage{algorithmicx}
\usepackage{algpseudocode}

\usepackage{epstopdf}

\usepackage[hidelinks=true]{hyperref}

\usepackage[final]{changes} 
\definechangesauthor[name={Victor F. Monteiro}]{vfm}
\definechangesauthor[name={Tarcisio F. Maciel}]{tfm}
\setaddedmarkup{{\color{blue}#1}}
\setdeletedmarkup{{\color{red}\sout{#1}}}

\usepackage[referable,flushleft]{threeparttablex}
\AtBeginEnvironment{tablenotes}{\footnotesize} 

\newcommand{\SecRef}[2][]{Section#1~\ref{#2}}
\newcommand{\FigRef}[2][]{Fig.#1~\ref{#2}}

\newcommand{\TabRef}[2][]{Table#1~\ref{#2}}

\usepackage{standalone}
\usepackage{shapepar}

\usepackage{tikz}
\usetikzlibrary{arrows}
\usetikzlibrary{positioning}
\usetikzlibrary{shapes.misc}
\usetikzlibrary{shapes.geometric}
\usetikzlibrary{shapes.symbols}
\usetikzlibrary{external}

\usetikzlibrary{shadows}
\usetikzlibrary{backgrounds}
\usetikzlibrary{shapes}
\usetikzlibrary{shapes.multipart}
\usetikzlibrary{matrix}
\usetikzlibrary{intersections}
\usetikzlibrary{fit}
\usetikzlibrary{calc}
\usetikzlibrary{chains}
\usetikzlibrary{scopes}
\usetikzlibrary{decorations.pathreplacing}
\usetikzlibrary{decorations.text}
\usetikzlibrary{arrows.meta} 
\usetikzlibrary{mindmap}
\usetikzlibrary{trees}

\usepackage{pgfplots}
\pgfplotsset{%
	width=0.95\columnwidth,
	height=0.25\textheight, 
	compat=1.14,
	compat/show suggested version=false,
	filter discard warning=false,
	tick label style={font=\footnotesize},
	label style={font=\footnotesize},
	every axis label={font=\footnotesize},
	grid=major,
	grid style={dashed,gray!30},
	cycle list shift=0,
	enlargelimits=false,
	legend style={%
		font=\footnotesize,
		legend cell align=left,
		nodes={inner xsep=2pt,inner ysep=1pt,text depth=0.15em},
	},
}
\usepackage{pgfplotstable}

\newtoggle{tikzexternalize}
\togglefalse{tikzexternalize}

\iftoggle{tikzexternalize}{
	\immediate\write18{mkdir -p ./tikz-external-figs/}
	\usetikzlibrary{external}
	\tikzexternalize[prefix=./tikz-external-figs/,mode=list and make]
	\tikzset{external/system call={pdflatex \tikzexternalcheckshellescape -halt-on-error -interaction=batchmode -jobname "\image" "\texsource"}}
}{
}

\AtBeginDocument{%
	\abovedisplayskip 1.5ex plus 4pt minus 2pt
	\belowdisplayskip \abovedisplayskip
	\abovedisplayshortskip 0pt plus 4pt
	\belowdisplayshortskip 1.5ex plus 4pt minus 2pt
}

\begin{document}
\title{Paving the Way Towards Mobile IAB:\\ Problems, Solutions and Challenges}

\author{Victor F. Monteiro, Fco. Rafael M. Lima, \textit{Senior Member, IEEE}, Darlan C. Moreira, Diego A. Sousa, \\ Tarcisio F. Maciel, Behrooz Makki, \textit{Senior Member, IEEE}, and Hans Hannu
\thanks{This work was supported by Ericsson Research, Sweden, and Ericsson Innovation Center, Brazil, under UFC.49 Technical Cooperation Contract Ericsson/UFC. The work of Francisco R. M. Lima was supported by FUNCAP (edital BPI) under Grant BP4-0172-00245.01.00/20. The work of Tarcisio F. Maciel was supported by CNPq under Grant 312471/2021-1.}
\thanks{Victor F. Monteiro, Fco. Rafael M. Lima, Darlan C. Moreira, Diego A. Sousa and Tarcisio F. Maciel are with the Wireless Telecommunications Research Group (GTEL), Federal University of Cear\'{a} (UFC), Fortaleza, Cear\'{a}, Brazil. Diego A. Sousa is also with Federal Institute of Education, Science, and Technology of Cear\'{a} (IFCE), Paracuru, Brazil. Behrooz Makki and Hans Hannu are with Ericsson Research, Sweden.} \vspace*{-0.8cm} 
}

\maketitle

\begin{abstract}
Deploying access and backhaul as wireless links, a.k.a. \ac{IAB}, is envisioned \deleted{to be}\added{as} a viable approach to enable flexible and dense networks. %
Even further, \ac{mIAB} is a candidate solution to enhance the connectivity of \deleted{a cluster of }\acp{UE} \added{moving together}\deleted{with similar mobility pattern, e.g., passengers in a bus}. %
\added{In this context, different of other works from the literature, the present}\deleted{This} work overviews the basis for the deployment of \ac{mIAB} \added{by presenting: 1) the current status of \ac{IAB} standardization in the \ac{5G} \ac{NR}; 2) a new taxonomy for state-of-the-art works regarding fixed \ac{IAB} and \ac{mIAB}; 3) an extensive performance analysis of \ac{mIAB} based on simulation results; and 4) open challenges and potential future prospects of \ac{mIAB}}\deleted{and studies its potentials and challenges}. %
\added{Specifically, the proposed taxonomy classifies \ac{IAB} works according to different perspectives and categorizes \ac{mIAB} works according to the type of mobility. %
	For each type of mobility, the main studied topics are presented.} %
\deleted{Specifically, we present the current status of \ac{IAB} standardization in the \ac{5G} \ac{NR} and state-of-the-art works regarding fixed \ac{IAB} and \ac{mIAB}. %
	A taxonomy is also presented for works from the literature. %
	On the one hand, works related to fixed \ac{IAB} are classified according to, for example, problem objectives, assumptions and adopted mathematical tools. %
	On the other hand, \ac{mIAB} related works are grouped according to the type of mobility, e.g., train and bus, and for each group the main studied topics are presented, e.g., interference management. %
	Furthermore, simulation results are presented in order to evaluate the \ac{mIAB} performance.} %
\added{Regarding the performance evaluation, we}\deleted{We} consider an urban macro scenario where \ac{mIAB} nodes are deployed in buses in order to improve the passengers connection. %
\added{The results show that, compared to other network architectures, the deployment of \ac{mIAB} nodes remarkably improves the passengers throughput and latency in both downlink and uplink}. %
\deleted{Extensive performance analyses are presented, including passengers and pedestrians throughput, latency and link quality in both downlink and uplink. %
	Results related to the wireless backhaul are also presented as well as the profile of the links served by the \ac{mIAB} donors. %
	Finally, we also present the lessons learned, open issues and future directions related to \ac{mIAB}.} %
\end{abstract}

\begin{IEEEkeywords}
	5G standardization, 6G, backhaul, integrated access and backhaul (IAB), mobile IAB, mobility, moving cell.
\end{IEEEkeywords}

%
\IEEEpeerreviewmaketitle
\acresetall


\goodbreak
\section{Introduction}
\label{SEC:Survey_Intro}

\Ac{5G} networks are being designed and deployed considering a dense deployment of small cells in order to simultaneously serve more \acp{UE} with higher throughput and lower delay~\cite{Jungnickel2014}. %
However, building from scratch a completely new infrastructure is costly and takes time~\cite{Dang2020}. %
Deploying a wireless backhaul is then envisioned to be a technically viable approach to enable flexible and dense network deployments~\cite{Inoue2020}. %

Wireless backhaul has already been considered in the past. %
However, on the one hand, it has been based on non-standardized solutions, deployed on \ac{mmWave} apart of the spectrum used for access links and mainly designed for \ac{PTP} communications with good networking planning and \ac{LOS} connections, as in~\cite{Ni2012, Dhillon2014, Schulz2015, Rahman2015}. %
On the other hand, in \ac{5G}, access links and backhaul may be deployed both in \ac{mmWave} and \ac{5G} networks are expected to be dense, possibly unplanned and with low height access points, which requests the support of \ac{NLOS} backhaul. %
These \ac{5G} characteristics are the main reasons why academia and industry are now investigating a standardized solution for joint access and backhaul fitting the \ac{5G} characteristics. %
The \ac{3GPP}, the organization in charge of standardizing \ac{5G} \ac{NR}, has already started standardizing this new wireless backhaul under the term \ac{IAB}. %

\ac{IAB} is promising and desirable due to its cost effectiveness\footnote{The cost discussions and the details of the simulation parameter settings are not necessarily aligned with the Ericsson's points of interest.}, fast deployment and easy maintenance~\cite{Yin2020}. %
Regarding the cost, the authors of~\cite{Xi2018} compared the deployment cost of three typical backhaul technologies, i.e., wireless, direct fiber and passive optical. %
They showed that a wireless backhaul is the most cost-effective among the three backhaul technologies. %
Another advantage is the support of larger bandwidth in \ac{NR} without the need to proportionally densify the supporting wired transport network~\cite{Inoue2020}. %

\begin{figure}
	\centering
	\includegraphics[width=\columnwidth]{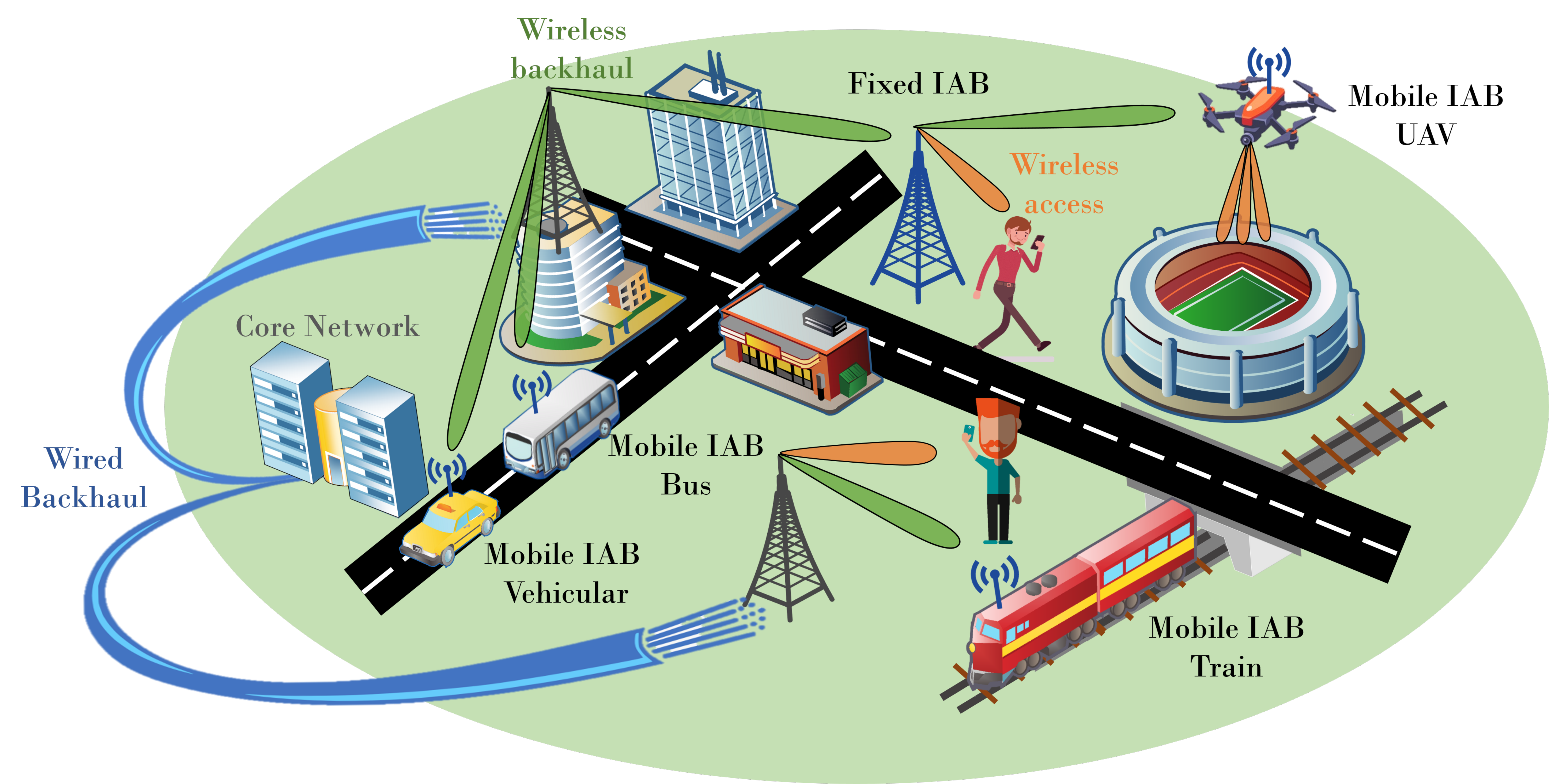}
	\caption{Overview scenario presenting examples of fixed \ac{IAB} and \ac{mIAB} use cases.}\label{FIG:IAB-Scenario}
\end{figure}

Thinking further, \ac{IAB} also allows the deployment of mobile cells, called here \ac{mIAB}, which should be available after \ac{3GPP} Release~18. %
The \ac{mIAB} is expected to enhance the service of moving \acp{UE} improving connectivity to the network and to avoid signaling storms from simultaneous \ac{HO} messages. %
Examples of \ac{mIAB} applications are illustrated in~\FigRef{FIG:IAB-Scenario}, e.g., \ac{mIAB} nodes deployed at trains, buses and \acp{UAV}\footnote{UAV-based communication will not be part of the Release~18 discussions on mIAB.}. %
In the examples presented in~\FigRef{FIG:IAB-Scenario}, the \ac{mIAB} nodes are wirelessly served either by a \ac{BS} which is connected by wire/fiber to the \ac{CN} or even by a \ac{BS} which is also wirelessly served by another wired-connected \ac{BS} to the \ac{CN}. %

In terms of mobility, one can consider \added{three}\deleted{two} distinct types of \ac{IAB} nodes: \added{fixed \ac{IAB} nodes,} nomadic \ac{IAB} nodes and \ac{mIAB} nodes. %
\added{While fixed \ac{IAB} nodes are always are the same position, allowing a system planning of long duration,}\deleted{A} nomadic \ac{IAB} \added{nodes adapt their}\deleted{refers to the cases where the \ac{IAB} node adapts its} location before the communication, \deleted{while it remains}\added{remaining} fixed during the communication. %
The main objective of nomadic \ac{IAB} is to \added{temporarily} extend the coverage and serve out-vehicle \acp{UE} in a given area, e.g., disaster area, close to stadiums, etc. %
With \ac{mIAB}, on the other hand, the \ac{mIAB} node on top of, e.g., trains, buses, trams, may communicate during the movement. %
As opposed to nomadic \ac{IAB}, the main goal for \ac{mIAB} is to serve the in-vehicle \acp{UE}. %
In this work, we concentrate on \ac{mIAB}. %

In this context, the present work aims at presenting an overview on the topic of \ac{mIAB} and on the technologies enabling its deployment at \ac{5G} networks and beyond. %
At the time of writing of this work, as far as we know, \cite{Zhang2021}~is the unique survey focused on \ac{IAB}, while \cite{Jaffry2021}~is the most complete survey covering mobile cells. %

The work in~\cite{Zhang2021} focuses on fixed \ac{IAB}. %
As opposed, we concentrate on the potentials and challenges of \ac{mIAB}. %
Particularly, 1) we review and propose a taxonomy for works on \ac{mIAB} which was not covered by~\cite{Zhang2021}. %
Also, 2) different from~\cite{Zhang2021}, we present deep performance evaluations to verify the potential benefits of \ac{mIAB}. %
Finally, while~\cite{Zhang2021} presents a single taxonomy on fixed \ac{IAB}, we present a different organization of the reviewed articles. %
More specifically, different from~\cite{Zhang2021}, we highlight that the works do not fit in a unique class of criteria as we classify the same work into different categories and over different perspectives. %

Our work is different from~\cite{Jaffry2021} from various perspectives. %
Firstly, while the authors of~\cite{Jaffry2021} only focused on moving cells, we also address fixed \ac{IAB}, which also uses a wireless backhaul. %
More specifically, we provide an overview of the architecture and protocols standardized by \ac{3GPP} to support \ac{IAB} and, we present a taxonomy for fixed \ac{IAB} works, which is the basis for \ac{mIAB} in \ac{5G} and beyond \ac{5G}. %
Secondly, \cite{Jaffry2021} only addresses land-based public transport vehicles such as trains, subways, buses, and vans, while we also address cars and \acp{UAV}. %
Finally, as~\cite{Zhang2021}, the survey on~\cite{Jaffry2021} does not present a detailed performance evaluation on \ac{mIAB} and does not compare its performance with alternative technologies. %

By presenting simulation results that take into account the \ac{3GPP} standardized architecture and protocols related to \ac{IAB}, we aim at comparing the performance of \ac{mIAB} networks with current networks and at presenting future directions of research to improve its performance. %

The present work is organized as follows. %
First, we present the basis of \ac{IAB} in Sections~\ref{SUBSEC:Survey_IAB_3GPP} and~\ref{SEC:Survey_fixed_IAB}. %
More specifically, \SecRef{SUBSEC:Survey_IAB_3GPP} presents current \ac{IAB} aspects already standardized by \ac{3GPP} for \ac{5G} \ac{NR}. %
Besides, \SecRef{SEC:Survey_fixed_IAB} presents a literature review of fixed \ac{IAB}. %
In this section, the works are classified following different criteria, such as: studied dimensions, system modeling assumptions/constraints, considered problem objectives and \acp{KPI}, solution approaches and adopted mathematical tools. %
Furthermore, \SecRef{SEC:Survey_Mobile_IAB} presents works related to \ac{mIAB}. %
The presented works are first grouped according to the type of mobility, e.g., train, bus, \ac{UAV}, etc.. %
This is due to fact that works considering similar type of mobility usually consider similar problems, e.g., \ac{UAV} positioning, and take advantage of specific characteristics of each type of mobility, e.g., previously known mobile trajectory of buses. %
For each type of mobility, we present the most recurrent topics and solutions
presented in the literature. %
Moreover, \SecRef{SEC:Perf_Eval} presents a performance evaluation of \ac{mIAB} through computational simulations results. %
Extensive performance analysis is presented, including passengers and pedestrians throughput, latency and link quality in both downlink and uplink transmission directions. %
Results related to the wireless backhaul are also presented as well as the profile of links served by the \ac{IAB} donors. %
Finally, \SecRef{SEC:Survey_Issues_Future} summarizes the lessons learned and lists some open issues and future directions. %
\SecRef{SEC:Survey_Conclusions} presents the conclusions of this work. %


\goodbreak
\section{IAB on 3GPP}
\label{SUBSEC:Survey_IAB_3GPP}

Wireless backhaul has been studied for years in the literature and the \ac{3GPP} has already taken some standardization actions on this matter in the context of \ac{LTE} Release 10~\cite{3gpp.overview.rel.10}. %
Although wireless backhaul are provided for \acp{BS} with the so called \ac{LTE} relaying~\cite{3gpp.36.806}, the commercial interest on it was not as large as expected. %
The main limitations of \ac{LTE} relaying are: 1) limited number of hops, i.e., only single hop is supported; 2) static parent to child architecture; 3) inflexible bandwidth partitioning between access and backhaul; and 4) limited available spectrum~\cite{Polese2020}. %

The standardization process on \ac{IAB} started in 2017 by means of a study item for \ac{5G} \ac{NR} within the scope of Release 15. %
The technical report~\cite{3gpp.38.874} describes \ac{IAB} architectures, radio protocols and physical layer aspects considered by that study item. %
Later, on July 2020, Release 16 was endorsed, in which protocol layers and architecture of \ac{IAB} were included in the technical specification~\cite{3gpp.38.300b}. %
Moreover, the minimum \ac{IAB} \ac{RF} characteristics and minimum \ac{NR} \ac{IAB} performance requirements were specified in~\cite{3gpp.38.174}. %

Different from \ac{LTE} relaying, \ac{NR}-\ac{IAB} provides greater flexibility by enabling multihop architecture, flexible topology, dynamic resource sharing, as well as the possibility of using \ac{mmWave} bands for both access and backhaul. %
Specifically, the wider bandwidths on \ac{mmWave} together with directionality of massive \ac{MIMO} make it possible to use aggressive spectrum reuse and reach higher transmit data rates compared to legacy 6 GHz bands. %
In the following, we provide more details on \ac{IAB} architecture and features. %

\goodbreak
\subsection{\acs{IAB} Architecture}\label{SUBSUBSEC:IAB_ARCHITECTURE}

\begin{figure}[!ht]
	\subfloat[Multiple backhaul hops illustration.]{%
		\includegraphics[width=0.85\columnwidth]{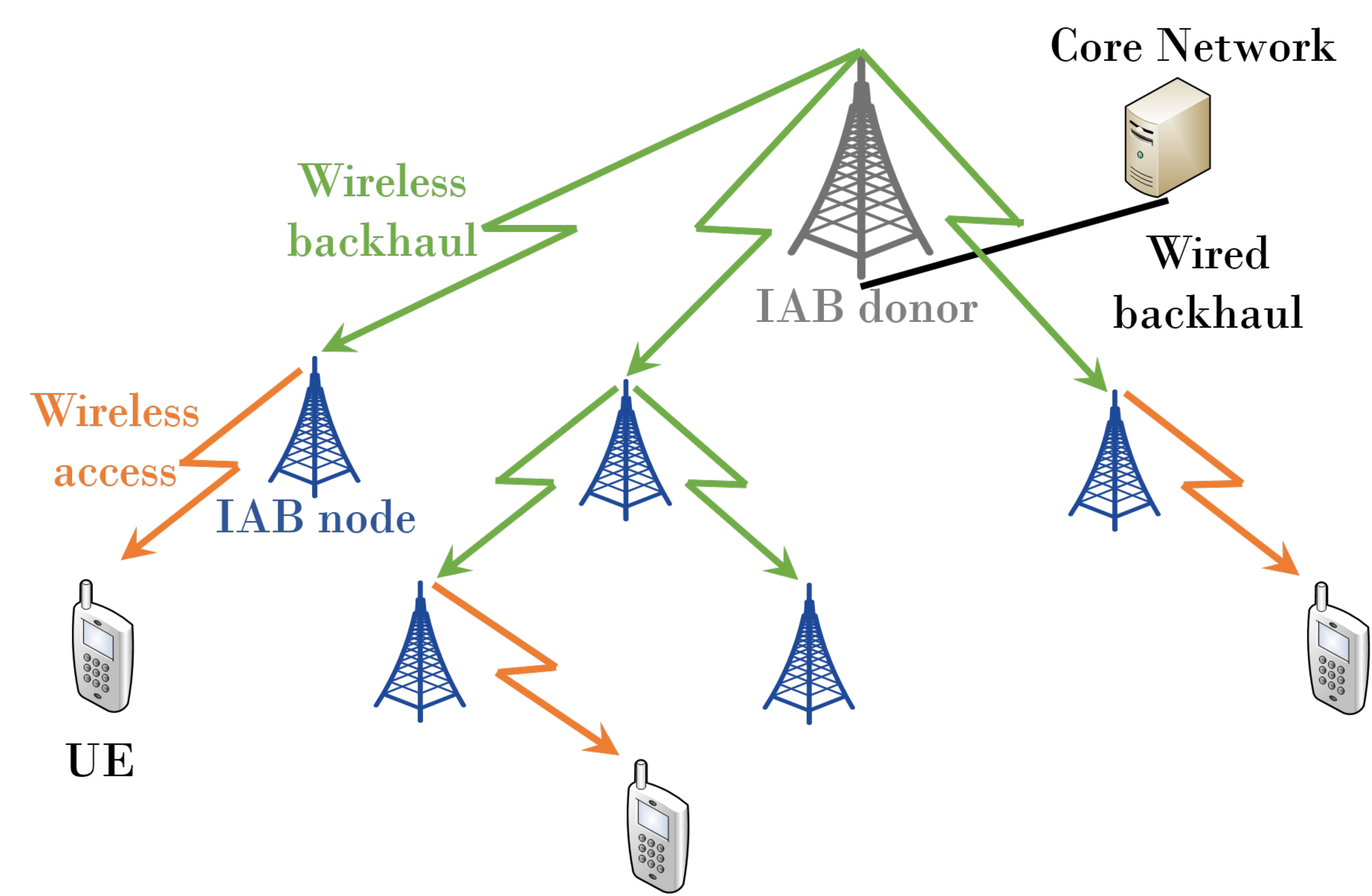}  
		\label{FIG:Architecture-system}
	}	
	
	\subfloat[\acs{IAB} donor and nodes main components.]{%
		\includegraphics[width=0.85\columnwidth]{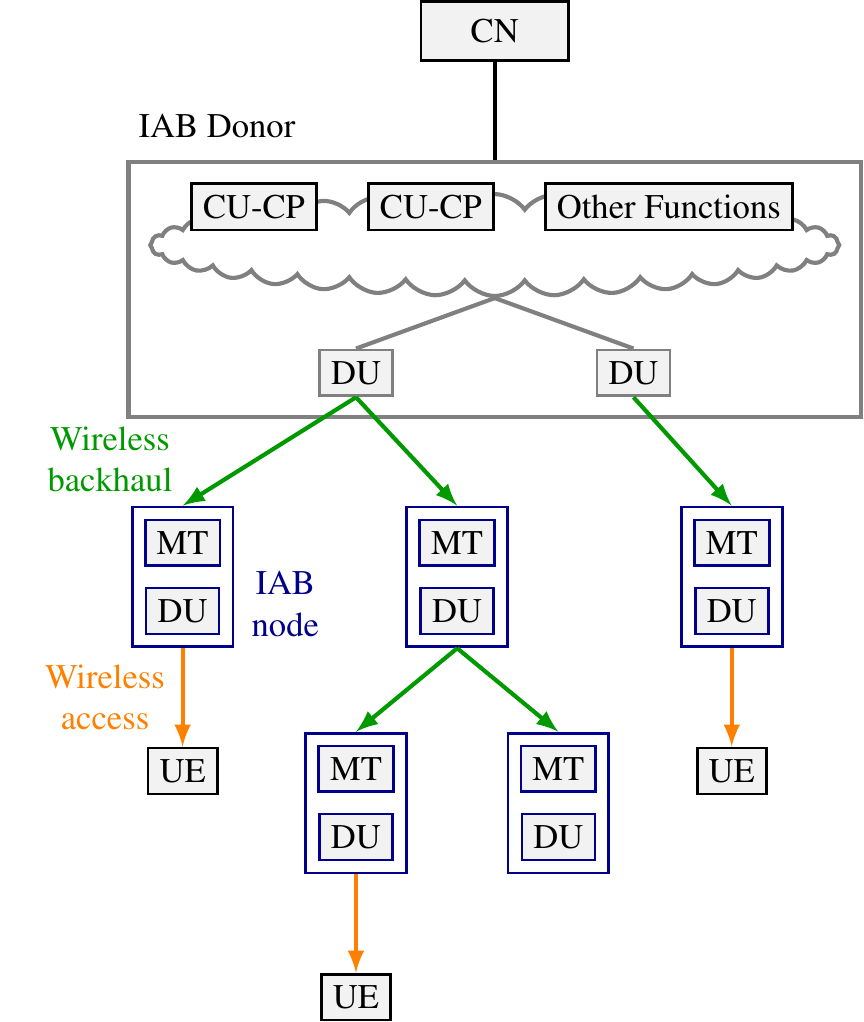} 
		\label{FIG:Architecture-functions}
	}
	\caption{\acs{IAB} \acs{NR} architecture.}\label{FIG:Architecture}	
\end{figure}

The integration of \ac{IAB} in \ac{NR} strives to reuse existing functions and interfaces. %
Figure~\ref{FIG:Architecture} illustrates the \ac{IAB} architecture adopted in \ac{NR}. %
One important aspect is the support of multiple wireless backhaul hops. %
This is illustrated in~\FigRef{FIG:Architecture-system} by the blue \acp{BS}, called \ac{IAB} nodes, which can be wireless connected one to another. %
In contrast, the terminating \ac{gNB}, i.e., the gray \ac{BS} with a wired backhaul and which provides a wireless backhaul to other \acp{BS}, is called \ac{IAB} donor. %
The \ac{IAB} \ac{NR} support of multiple hops helps to overcome two challenges of signal propagation in \acp{mmWave}: its limited range of coverage and its low capacity to contour obstacles. %
However, the main challenges to support multiple hops are related to scalability issues, e.g., increased signaling load and \ac{UL} scheduling latency. %

Regarding the \ac{IAB} nodes, they support \ac{gNB} \ac{DU} and \ac{MT} functionalities~\cite{3gpp.38.300b}, as illustrated in~\FigRef{FIG:Architecture-functions}. %
On the one hand, the \ac{MT} part of an \ac{IAB} node manages the radio interface layers of its backhaul towards an upstream \ac{IAB} donor or other \ac{IAB} node. %
On the other hand, the \ac{DU} part provides the \ac{NR} interface to \acp{UE} and to \ac{MT} parts of downstream \ac{IAB} nodes. %

For compatibility purposes with legacy networks, the \ac{MT} part of an \ac{IAB} node acts as a regular \ac{UE} from the point-of-view of its serving \ac{BS} (either an \ac{IAB} donor or another \ac{IAB} node). %
From a \ac{UE} point of view, the \ac{DU} part of an \ac{IAB} node looks like as the \ac{DU} of a regular \ac{gNB}. %

\begin{figure}[!ht]
	\subfloat[User plane.]{%
		\includegraphics[width=0.95\columnwidth]{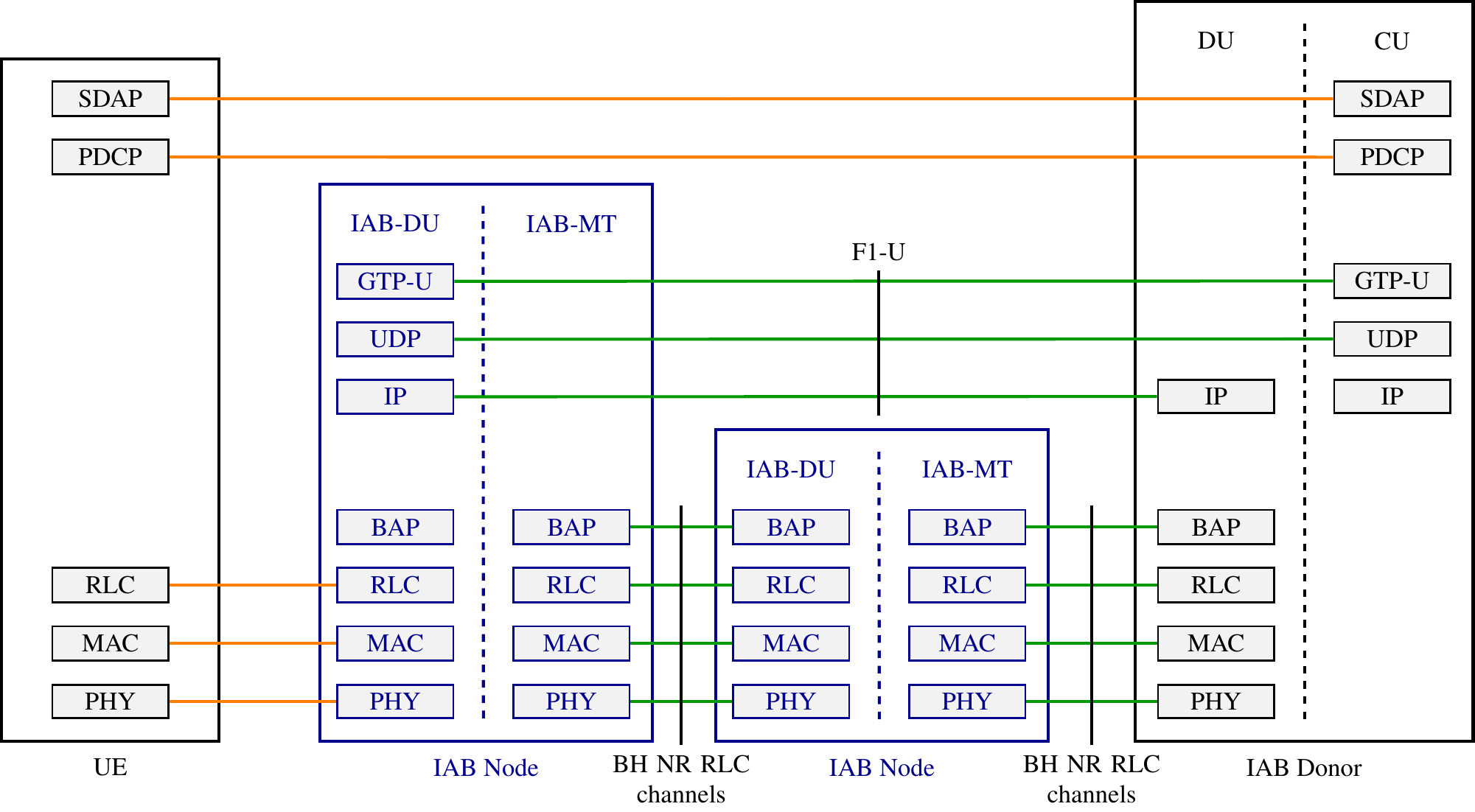} 
		\label{FIG:U-plane-protocols}
	}
	
	\subfloat[Control plane.]{%
		\includegraphics[width=0.95\columnwidth]{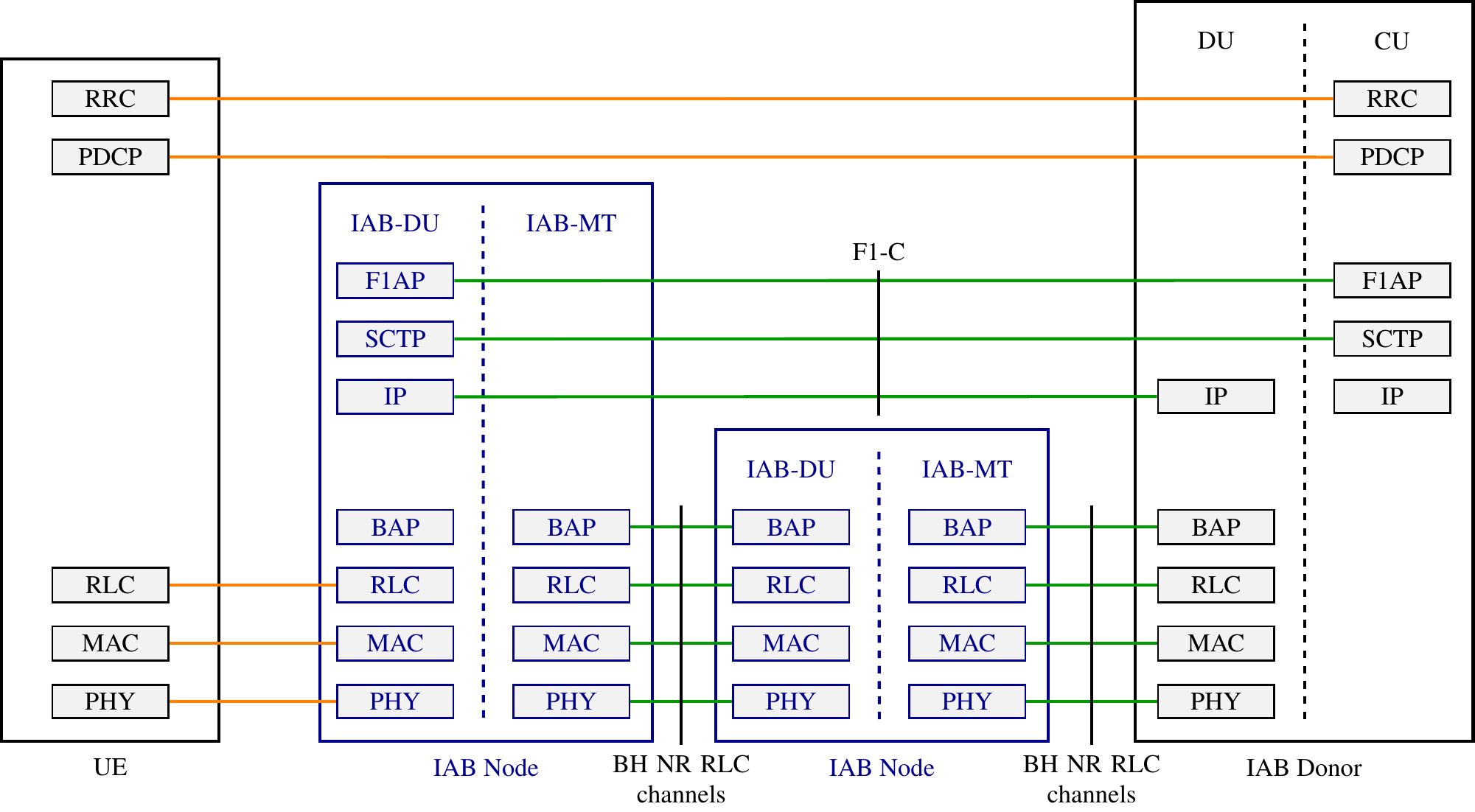} 
		\label{FIG:C-plane-protocols}
	}
	\caption{Protocol stack.}\label{FIG:Protocol-stack}	
\end{figure}

Concerning the \ac{IAB} donor, it is split in \acp{CU} and \acp{DU}~\cite{3gpp.38.300b}. %
This split is transparent to the served nodes and these units can be either collocated or non-collocated. %
As illustrated in~\FigRef{FIG:Protocol-stack}, the \acp{DU} are responsible for lower protocol layers, e.g., \ac{PHY}, \ac{MAC}, and \ac{RLC}, while the \acp{CU} provide upper protocol layers, e.g., \ac{PDCP} and \ac{SDAP}~/~\ac{RRC}. %
The main objective of this split is to allow time-critical functionalities, e.g., scheduling and retransmission, to be performed in \acp{DU} closer to the served nodes, while other functionalities can be performed in \acp{CU} with better processing capacity~\cite{Madapatha2020}. %

\goodbreak
\subsection{Protocol Layers}\label{SUBSUBSEC:PROTOCOL_LAYERS}

As it can be seen in Figure~\ref{FIG:Protocol-stack}, \ac{CU} and \acp{DU} are connected via the F1 interface. %
The open and point-to-point F1 interface supports both the exchange of signaling information by means of F1-C (control plane), and data transmission between end points through F1-U (user plane)~\cite{3gpp.38.470}. %
The main capabilities of F1 interface are the management of radio bearers and backhaul \ac{RLC} channels as well as the transfer of \ac{RRC} messages between \ac{UE} and \ac{gNB}-\ac{CU}. %

According to~\cite{3gpp.38.874}, \ac{IAB} backhaul can operate in both in-band and out-of-band scenarios with respect to access links. %
In in-band mode, there is at least a partial overlap between the used frequency resources by backhaul and access links. %
However, in this case, \ac{IAB} nodes cannot transmit and receive at the same time in order to avoid self-interference, i.e., they operate in \ac{HD}. %
So, the \ac{DU} part of an \ac{IAB} node cannot transmit while its \ac{MT} part is receiving and vice versa. %
In case of in-band scenario, \ac{TDM} or \ac{SDM} should be employed in order to avoid undesirable interferences. %
In out-of-band mode, the backhaul and access links use different frequency resources so as to avoid cross-tier interference. %
\Ac{TDD} should be employed as \ac{NR} operators will typically have access to wide bandwidth at \ac{mmWave}. %
Thus, the time domain should be properly configured to coordinate transmissions in downlink/uplink for access and backhaul links. %
The time share for uplink/downlink transmissions in access and backhaul to optimize system performance can be based on the system load, for example. %

Backhaul \ac{RLC} or BH NR RLC channels are responsible for transporting traffic between \ac{IAB} nodes or between \ac{IAB} donor and \ac{IAB} nodes. %
As the functional \ac{CU}/\ac{DU} split occurs at \ac{RLC} layer, \ac{IAB} nodes are interconnected at this level. %
Hop-by-hop \ac{ARQ} is considered for the \ac{RLC} layer instead of an end-to-end \ac{ARQ}. %
The main disadvantages of hop-by-hop \ac{ARQ} are the higher packet forward latency (due to the \ac{RLC}-state machine on each hop) and the lack of guarantee for lossless uplink packet delivery in certain scenarios, e.g., topology adaptation due to link failure. %
However, end-to-end \ac{ARQ} at \ac{RLC} layer leads to latency due to retransmissions that increases with the number of hops. %
Another drawback of end-to-end \ac{ARQ} is that a packet loss requires retransmission in multiple links including the ones whose transmision was successful~\cite{3gpp.38.874}. %
Multiple backhaul \ac{RLC} channels can be setup on each backhaul link to assure \ac{QoS} guarantees. %
More specifically, there are two types of mapping between \ac{UE} data radio bearers and backhaul \ac{RLC} channels: one-to-one and multiple-to-one mappings. %
In the first case, each \ac{UE} radio bearer is mapped to a separate backhaul \ac{RLC} channel, whereas in the second case several \ac{UE} radio bearers are multiplexed onto a single backhaul \ac{RLC} channel. %
On the one hand one-to-one mapping can ensure strict \ac{QoS} guarantees, but on the other hand multiple-to-one mapping decreases the signaling load and required number of \ac{RLC} channels that should be established. %

One of the premises of \ac{IAB} standardization is to decrease at maximum its impact on \ac{NR} Release 15 specifications. %
However, in order to support routing across \ac{IAB} nodes, i.e., control how packets are forwarded among \ac{IAB} nodes, \ac{IAB} donor, and \acp{UE}, a new protocol sublayer was proposed: the \ac{BAP}~\cite{3gpp.38.340}. %
In this sense, the \ac{BAP} sublayer is responsible for efficient forwarding of \ac{IP} packets between \ac{IAB} nodes on the top of \ac{RLC} backhaul channels. %
In downlink, the \ac{BAP} sublayer of the \ac{IAB} donor \ac{DU} encapsulates \ac{PDCP} packets that, in their turn, are transmitted through \ac{RLC} backhaul channels and, finally, de-encapsulated at the \ac{DU} side of the target \ac{IAB} node. %
In uplink, the \ac{PDCP} packets are encapsulated at the \ac{IAB} node in the origin and de-encapsulated at the \ac{DU} part of the \ac{IAB} donor. %
Each \ac{IAB} node owns a \ac{BAP} address that uniquely identifies it in an \ac{IAB} network. %
The header of a \ac{BAP} packet carries a \ac{BAP} routing \ac{ID} that is setup by the \ac{CU} part of donor \ac{IAB}. %
The routing \ac{ID} is composed of a \ac{BAP} address and a \ac{BAP} path \ac{ID}. %
While the former identifies the destination node where the packet should be delivered, the latter defines the routing path that the packet should follow, if more than one, until reaching the destination node. %
Therefore, at each node, the \ac{BAP} header should be inspected in order to determine if the packet has reached its destination or to determine the next hop that the packet should be forwarded to. %
In downlink, the \ac{DU} part of \ac{IAB} donor is responsible for inserting the header of \ac{BAP} packets, while in uplink this task is performed by the first \ac{IAB} node in the route. %
The routing table used for uplink and downlink can be different from each other. %

\goodbreak
\subsection{Network Procedures}\label{SUBSUBSEC:NETWORK_PROCEDURES}

Regarding integration of new \ac{IAB} nodes to the system, topology adaptation and \ac{IAB} resource configuration, \ac{3GPP} tried to be as much as possible compatible with legacy procedures and adapted the equivalent procedures of a regular \ac{gNB} for the case of an \ac{IAB} node. %
In the following, we briefly present how these procedures work in the context of \ac{IAB}. %

\goodbreak
\subsubsection{IAB Node Integration Procedure}\label{SUBSUBSEC:iab_node_integration}

\begin{figure}
	\centering
	\includegraphics[width=.75\columnwidth]{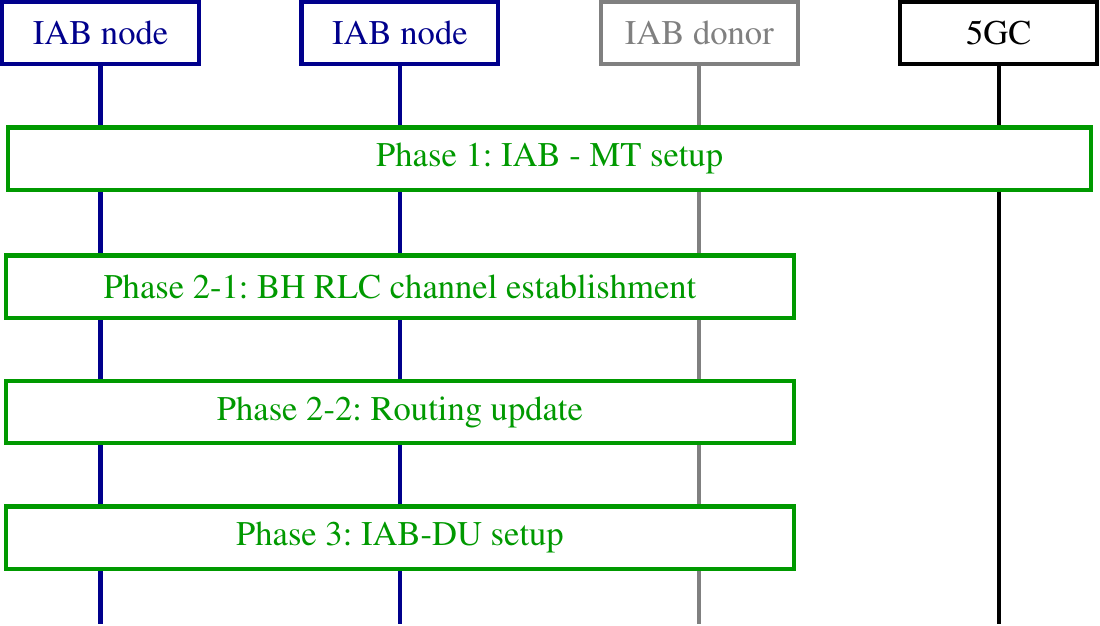}
	\caption{\acs{IAB} node integration procedure~\cite{3gpp.38.401}.}\label{FIG:IAB-initial-access}
\end{figure}

\Ac{3GPP} specified in~\cite{3gpp.38.401} how a new \ac{IAB} node is integrated to the system. %
This procedure is illustrated in~\FigRef{FIG:IAB-initial-access} and is split into three phases. %

In the first phase, the \ac{IAB}-\ac{MT} of the new \ac{IAB} node connects to the network in the same way as a regular \ac{UE} with two differences~\cite{3gpp.38.300b}: 1) the \ac{IAB}-\ac{MT} ignores cell-barring or cell-reservation indications contained in cell system information broadcast; 2) the \ac{IAB}-\ac{MT} only considers a cell as a candidate for cell selection if the cell system information broadcast indicates \ac{IAB} support. %
Except for the mentioned differences, the \ac{IAB}-\ac{MT}, as a regular \ac{UE}, searches the frequency band for \acp{SSB} in order to identify the most suitable cell. %
After a successful random access, the \ac{IAB}-\ac{MT} performs a \ac{RRC} connection setup procedure with the \ac{IAB} donor-\ac{CU}, authentication with the \ac{CN}, and context management and radio bearer configuration with intermediary \ac{IAB} nodes, if any. %

The second phase is split into two main parts. %
The first part concerns the establishment of new BH \ac{RLC} channel or modification of an existing one between the intermediary \ac{IAB} node and the \ac{IAB}-donor-\ac{DU}. %
This part also includes configuring the \ac{BAP} Address of intermediary \ac{IAB} nodes and default \ac{BAP} Routing \ac{ID} for the upstream direction. %
In the second part, the \ac{BAP} sublayer is updated to support routing between the new \ac{IAB} node and the \ac{IAB}-donor-\ac{DU}. %

Finally, the third part concerns the configuration of the \ac{DU} part of the new \ac{IAB} node. %
The F1 link between the \ac{IAB} donor \ac{CU} and the \ac{DU} part of the new \ac{IAB} node is setup with the allocated \ac{IP} address. %
After that, the new \ac{IAB} node is ready to start serving \acp{UE}. %

\goodbreak
\subsubsection{Topology Adaptation}\label{SUBSUBSEC:topology_adaptation}

After being integrated to the system and connected to a parent node, an \ac{IAB} node may need to migrate to a different parent node, e.g., when the link towards its current parent becomes weak due to mobility or obstacles between transmitter and receiver. %
For this, topology adaptation is performed. %

\begin{figure}[t]
	\centering
	\includegraphics[width=.65\columnwidth]{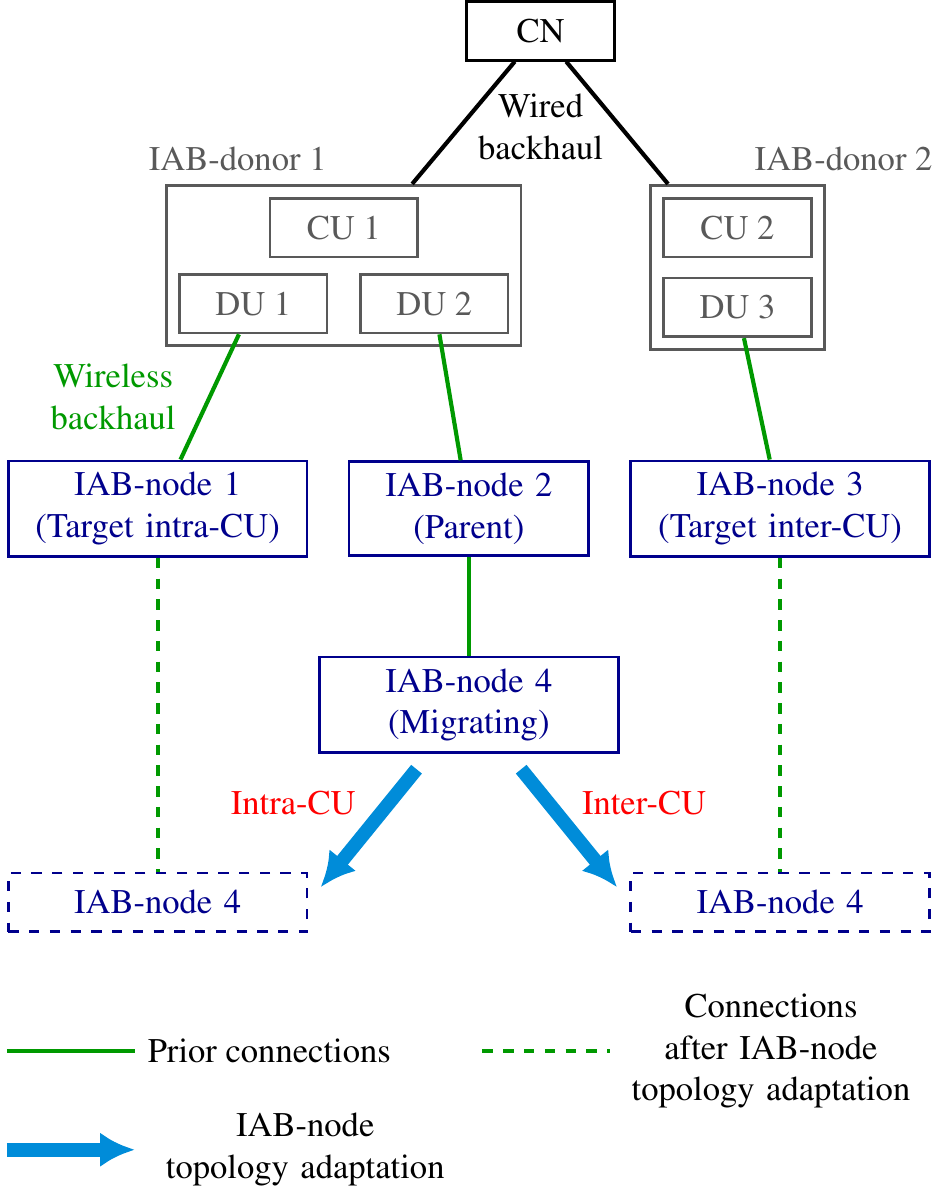}
	\caption{Topology adaptation illustration.}\label{FIG:IAB-topology-adaptation}
\end{figure}

Figure~\ref{FIG:IAB-topology-adaptation} illustrates an example where \ac{IAB} node~$4$ needs to migrate to another parent. %
Its serving \ac{IAB} parent (\ac{IAB} node~$2$) and the target one (\ac{IAB} nodes~$1$ or~$3$) may be served either by the same \ac{IAB} donor-\ac{CU} (intra-\ac{CU}) or by a different \ac{IAB} donor-\ac{CU} (inter-\ac{CU}). %
\Ac{3GPP} defined in~\cite{3gpp.38.401} different procedures for each case. %

In both intra-\ac{CU} and inter-\ac{CU} topology adaptation procedures, first the migrating node (\ac{IAB} node~$4$) sends a measurement report to its parent node (\ac{IAB} node~$2$). %
This measurement report is forwarded from \ac{IAB} node~$2$ towards its \ac{IAB}-donor \ac{CU} (\ac{CU}~$1$) via an \ac{UL} \ac{RRC} transfer. %
The \ac{IAB}-donor \ac{CU} is responsible for evaluating if a topology adaptation is required. %
If so, it chooses a target node to be the new parent of the migrating node. %

If the target node (e.g., \ac{IAB}-node~$1$) is served by this \ac{IAB}-donor
\Ac{CU}, i.e., \ac{CU}~$1$, (intra-\ac{CU} topology adaptation), this \ac{CU}
directly sends a message to the target node with a \verb|CONTEXT_SETUP_REQUEST|.
If \ac{CU}~$1$ receives a positive response, it sends a \ac{RRC} reconfiguration message to \ac{IAB} node~$2$ (the current parent), which forwards this message to \ac{IAB} node~$4$. %
\Ac{IAB} node~$4$ will then perform a random access procedure towards the target parent, i.e., \ac{IAB} node~$1$. %
If everything goes well, the new connection is established, and the migrating \ac{IAB} node sends a \verb|RRC_RECONFIGURATION_COMPLETE| message to its new parent (\ac{IAB} node~$1$), which forwards this message to the \ac{IAB} donor \ac{CU}. %
Finally, the BH \ac{RLC} channels are configured and the \ac{BAP} route and mapping rules are updated. %

Otherwise, if the target node (e.g., \ac{IAB}-node~$3$) is served by another \ac{IAB}-donor \ac{CU} (e.g., \ac{CU}~$2$ in an inter-\ac{CU} topology adaptation), then \ac{CU}~$1$ sends a \ac{HO} request via $X_n$ interface to \ac{CU}~$2$. %
Similar to the previous case, \ac{CU}~$2$ will send a \verb|CONTEXT_SETUP_REQUEST| to the target \ac{IAB} node~$3$ and wait for a response and, if it is positive, \ac{CU}~$2$ will send a \ac{HO} ACK to \ac{CU}~$1$. %
Also similar to the previous case, \ac{CU}~$1$ then sends a \ac{RRC} reconfiguration message to \ac{IAB} node~$2$ (the current parent), which forwards this message to \ac{IAB} node~$4$. %
\Ac{IAB} node~$4$ will then perform a random access procedure towards the target parent, i.e., \ac{IAB} node~$3$. %
If everything goes well, the new connection is established, and the migrating \ac{IAB} node sends a \verb|RRC_RECONFIGURATION_COMPLETE| message to its new parent (\ac{IAB} node~$1$), which forwards this message to its \ac{IAB} donor \ac{CU} (\ac{CU}~$2$). %
Finally, a new F1 is established between \ac{IAB} node~$4$ and \ac{CU}~$2$, the BH \ac{RLC} channels are configured, and the \ac{BAP} route and mapping rules are updated. %

It is worth noting that, for both cases, the parent \ac{IAB}-node never directly communicates with the target parent \ac{IAB}-node. %
The communication between them is performed through the \ac{IAB} donor \ac{CU}. %
Also, according to~\cite{3gpp.38.401}, in upstream direction of intra-\ac{CU} topology adaptation, in-flight packets between the source parent \ac{IAB} node and the \ac{IAB} donor \ac{CU} can be delivered even after the target path is established, while, in-flight downlink data in the source path may be discarded, since the \ac{IAB} donor \ac{CU} can determine unsuccessfully transmitted downlink data over the backhaul link. %

\goodbreak
\subsubsection{IAB Resource Configuration}\label{SUBSUBSEC:iab_rb_config}

The deployment of wireless backhaul is a potential source of interference in the system. %
In order to avoid self-interference and sometimes due to hardware limitations,
in general, \ac{IAB}-\ac{DU} and \ac{IAB}-\ac{MT} operate in \ac{HD}, i.e., \ac{IAB}-\ac{DU} does not transmit at the same time as \ac{IAB}-\ac{MT} receives and vice-versa. %
However, some special cases of \ac{IAB}-\ac{FD} are envisioned, e.g., when the antennas of \ac{IAB}-\ac{DU} and \ac{IAB}-\ac{MT} point towards different directions or the \ac{IAB}-\ac{MT} is outside and the \ac{IAB}-\ac{DU} is inside~\cite{Dhalman2021}. %

Regarding the direction (transmission or reception) in which the frequency resources are used in the time domain by an \ac{IAB}-\ac{MT}, they can be configured via \ac{RRC} signaling by the \ac{IAB} node parent as downlink, uplink or flexible. %
When configured as downlink, it is only used by the parent in the downlink direction (\ac{MT} reception only). %
When configured as uplink, it is only used by the parent in the uplink direction (\ac{MT} transmission only). %
When configured as flexible, it can be used in either direction (but not simultaneously). %
Its instantaneous direction is determined by the parent node scheduler. %

Similar to the \ac{MT} \added{frequency resources}, the \ac{IAB}-\ac{DU} \added{resources }can also be configured as downlink (\ac{DU} transmission only), uplink (\ac{DU} reception only), and flexible. %
However, another configuration is also allowed for the \ac{DU} resources: not available. %
In this case, the \ac{DU} is not allowed to use that resource at all. %

\deleted{Concerning the availability to be scheduled}\added{To coordinate the use of resources by \ac{IAB}-\ac{DU} and \ac{IAB}-\ac{MT} parts}, an \ac{IAB}-\ac{DU} resource can be configured by the parent node as hard, unavailable or soft~\cite{3gpp.38.300b}. %
If configured as hard, the \ac{IAB}-\ac{DU} can use it no matter if it is used by the \ac{IAB}-\ac{MT}. %
In this case, the parent node should avoid transmitting/receiving to/from the \ac{IAB}-\ac{MT} in this resource. %
If configured as unavailable,  it cannot be used by the \ac{IAB}-\ac{DU}, except for some special cases. %
Finally, if configured as soft, the \ac{IAB}-\ac{DU} can use it conditionally either on an explicit indication of availability by the parent node or on an implicit determination of availability by the \ac{IAB}-\ac{DU} based on whether or not the use of that resource impacts the \ac{IAB}-\ac{MT}. %
It is the \ac{CU}'s responsibility to configure both the parent \ac{IAB} node and the \ac{IAB} node such that their respective availability configuration are compatible. %

With respect to \ac{IAB}-\ac{MT} \ac{UL} scheduling, increased latency due to multiple hops can adversely impact the system performance. %
The reason for this is that usually an equipment only requests \ac{UL} data transmission after it actually receives the data to be transmitted, informed via a \ac{BSR}. %
In other words, in a multihop system, the delay is cumulative since an intermediary equipment must first receive the data, before requesting \ac{UL} resources and finally forwarding the data in \ac{UL}. %
To overcome this issue, it was standardized in~\cite{3gpp.38.300b} that an
\Ac{IAB}-node can send a pre-emptive \ac{BSR} to request \ac{UL} resources, i.e., request \ac{UL} resources based on expected data rather than on data already buffered. %
In this case, when data is received by an intermediary \ac{IAB} node, \ac{UL} resources are already reserved to be used for forwarding the data to the \ac{IAB} parent. %

\goodbreak
\subsection{\acs{IAB} in Release 17}\label{SUBSUBSEC:IAB_ON_RELEASE_17}

Due to the restrictions imposed on face-to-face meetings by COVID-19 pandemic, \ac{3GPP} has revised the Release 17 timeline. %
The Release 17 Stage 3 functional freeze is expected to occur on March, 2022. 
A new work item entitled \textit{Enhancements to Integrated Access and Backhaul
	for \ac{NR}} has been approved for Release 17~\cite{3gpp.rp.210758}, the main objectives of which are~\cite{Dhalman2021b}:
\begin{itemize}
	\item Topology adaptation enhancements and routing: Among the main studies in this topic we can mention the specification of inter-donor IAB migration and reduction of the associated signaling load. %
	The main objectives of this topic are the increase of system robustness and load balancing. %
	Moreover, support for CP/UP separation and reduction of service interruption time, e.g., during IAB-node migration, are also envisioned to increase reliability. %
	Remaining studies are focused on multihop latency and congestion mitigation. %
	
	\item Duplexing enhancements: \ac{HD} constraint imposes limitations on \ac{IAB} operation. In this topic, the focus is to improve spectral efficiency and latency by supporting simultaneous operation of \ac{IAB} node's child and parent. %
	In this case, it is possible that the \ac{MT} part of an \ac{IAB} node could transmit while the \ac{DU} part of the same node could be receiving, i.e., \ac{FD} mode. %
	Another possibility is the simultaneous transmission or reception of \ac{MT} and \ac{DU} parts by leveraging spatial multiplexing techniques. %
	Furthermore, dual connectivity scenarios should be considered so as to improve robustness and perform load balancing. %
	
\end{itemize}


\goodbreak
\section{Overview on Fixed IAB}\label{SEC:Survey_fixed_IAB}

\subsection{Taxonomy on Fixed IAB}\label{SUBSEC:Taxonomy_Fix_IAB}

Thanks to the great flexibility of \ac{IAB} networks, the existing literature on fixed \ac{IAB} can be categorized over many different criteria. %
In this section, we provide a taxonomy of the works on \ac{IAB} according to the following criteria: %
\begin{enumerate} %
	\item Studied dimensions and article profile: the system performance on \ac{IAB} networks can be improved by optimizing different aspects. %
	In this criterion, the works can be characterized according to the functionalities that are studied. %
	The considered categories are surveys, overview, testbeds, \ac{IAB} versus fiber-based network performance, network deployment, routing and topology adaptation, resource allocation for access and/or backhaul, power allocation, \ac{UE} association, \ac{MIMO} beamforming; %
	\item System modeling assumptions and constraints: depending on the dimensions that are being optimized in each article, the authors can make some assumptions about the system model that can boost performance gains but, at the same time, can turn the problems harder to solve. %
	In this criterion, we consider the following categories: \ac{FD}, \ac{IRS}, multihop, multiple antennas, \ac{UE}-centric approach, modeling of both uplink and downlink, modeling of access and backhaul links, mesh topology and \ac{NOMA}; %
	\item Problem objectives and \acp{KPI}: the performance of \ac{IAB} networks can be measured over different points of view, therefore, different objectives and metrics are optimized in the reviewed articles. %
	In this criterion, we assume the following categories: fairness, latency, spectral efficiency, energy efficiency, data rate guarantees for \acp{UE}, coverage probability, weighted sum rate, and symbol error rate; %
	\item Solution approaches and mathematical tools: in this criterion, the works can be classified according to the employed tool in order to solve the studied problem(s). %
	Among the common tools we can mention: convex optimization, linear (continuous) optimization, integer linear optimization, heuristics, dynamic programming, machine learning and metaheuristics, stochastic optimization, game theory, stochastic geometry, and statistical analysis. %
\end{enumerate} %

Note that, depending on the assumed criteria, the same work can be classified in more than one category. %
In the following section, we describe some important works on fixed \ac{IAB} clearly identifying the classification according to the established criteria. %

\goodbreak
\subsection{Literature Review on Fixed \ac{IAB}}\label{SUBSEC:Lit_Rev_Fix_IAB}

\begin{figure*}
	\centering
	\includegraphics[width=\textwidth]{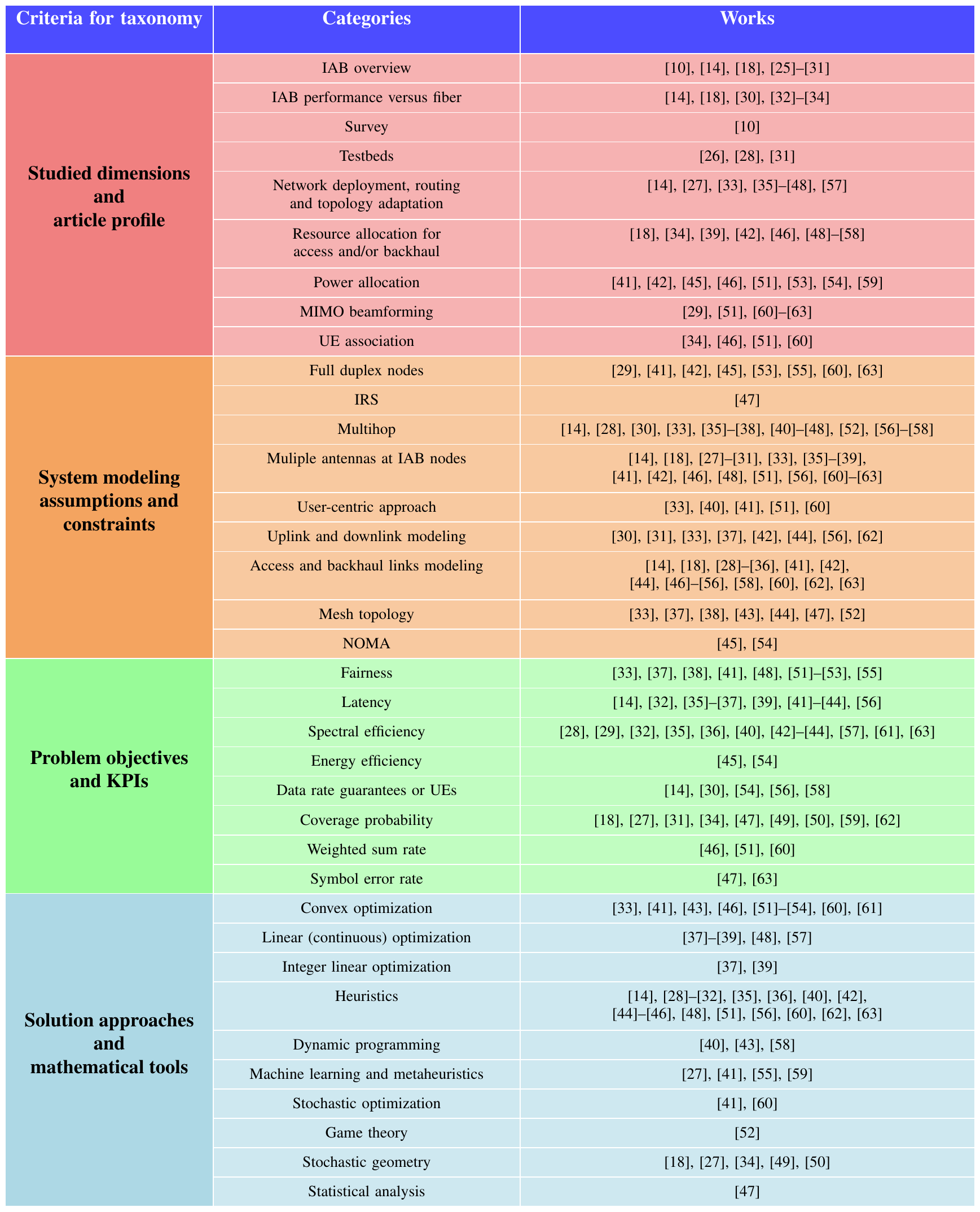}
	\caption{Taxonomy for fixed \ac{IAB}: studied dimensions and article profile, system modeling assumptions and constraints, problem objectives and \acp{KPI}, and solution approaches and mathematical tools.}\label{FIG:FIXED_IAB_TAXONOMY} %
\end{figure*}

In Figure~\ref{FIG:FIXED_IAB_TAXONOMY}, we present the taxonomy for the reviewed articles in fixed \ac{IAB} assuming the four abovementioned criteria.  %
For the sake of organization, the works are presented in the following according to the first criterion. %

\goodbreak
\subsubsection{Surveys, Overview Papers, Testbeds, and \ac{IAB} versus Fiber-Based Network Performance} \label{SUBSUBSEC:SURVEY_OVERVIEW_TESTBEDS_FIBER}

The works~\cite{Teyeb2019,Polese2020,Ronkainen2020,Madapatha2020,Madapatha2021,Zhang2021,Ronkainen2021,Zhang2021_2,Cudak2021,Tian2019} provided an overview of \ac{IAB} technology by describing several aspects regarding architecture, protocol aspects and/or physical layer. %
More specifically, \cite{Zhang2021} provides a survey on fixed \ac{IAB}. %
Its authors classified \ac{IAB} works in the following groups: stochastic geometry-based works, resource allocation, scheduling, cache-enabled \ac{IAB} networks, optical \ac{IAB} networks and non-terrestrial \ac{IAB} networks. %
The works~\cite{Ronkainen2020,Ronkainen2021} presented a proof of concept of a two-hop \ac{IAB} network using two-sectors \ac{IAB} nodes operating at \ac{mmWave}. %
\deleted{The achieved performance was positive showing the potential of \ac{IAB} networks.} %
In \cite{Tian2019}, a testbed was presented composed of a donor \ac{BS}, an \ac{IAB} node and a \ac{UE}. %
\deleted{The experiments took place in a urban macro scenario in Yokohama, Japan.} %
\deleted{According to the trials, \ac{IAB} technology provided a much higher coverage, i.e., \ac{UE}'s throughput higher than 300 Mbps, than the macro-only scenario.} %
\deleted{Results also demonstrated the advantages of a flexible resource allocation between access and backhaul by means of \ac{TDM} and \ac{SDM}.} %
\added{The works~\cite{Ronkainen2020,Ronkainen2021,Tian2019} demonstrated the potential of \ac{IAB} in a practical setup.} %

The works~\cite{Polese2020,Madapatha2020} as well as~\cite{Polese2018,Islam2018,Saha2019,Cudak2021} presented simulation results of \ac{IAB} networks versus fiber-based networks. %
\deleted{In~\cite{Polese2018}, the authors considered a dynamic single-hop \ac{IAB} network with one donor \ac{IAB} node and some \ac{IAB} nodes (relays).} %
\deleted{The \ac{IAB} network was compared to a baseline scenario with only one fiber-connected \ac{gNB}.} %
\deleted{They showed that \ac{IAB} networks benefit cell-edge \acp{UE} in terms of latency compared to a deployment without \ac{IAB} nodes, i.e., only one fiber-connected \ac{gNB}.} %
\deleted{The same authors provided dynamic end-to-end simulations for multihop \ac{IAB} networks in~\cite{Polese2020}.} %
\deleted{They showed that \ac{IAB} performance is close to all-wired networks for low and moderate traffic.} %
\added{Some of the conclusions that can be gathered from those works is that \ac{IAB} performance is close to the one of all-wired networks for low and moderate traffic, and that  the performance of fully-fiber connected networks can be achieved with a small increase in the number of \ac{IAB} nodes in \ac{IAB} networks.} %
\added{Moreover, they showed that \ac{IAB} is a relevant technology for \ac{5G} since it allows to incrementally improve the fiber density in the network.} %
\deleted{In \cite{Cudak2021}, a simulation study was performed by assuming 3D ray tracing model in Lincoln Park neighborhood of Chicago in 39 GHz band.} %
\deleted{The objective with the study was to determine the required number of \acp{BS} to support user data rate requirements for a varying fiber penetration and number of \acp{UE}.} %
\deleted{Although the best performance is achieved by an all-fiber network, at low and moderate loads, \ac{IAB} presented a similar performance to the optimum scenario since more resources are left for backhaul transmission.} %
\deleted{The main outcome of the study was to show that \ac{IAB} is a relevant technology for \ac{5G} since it allows to incrementally improve the fiber density in the network.} %
\deleted{For example, with 4 fiber connected \acp{BS} and 30 \ac{IAB} nodes 90 \acp{UE} can be served; however, over time, by converting 4 \ac{IAB} nodes to fiber \acp{BS}, 160 \acp{UE} could be served.} %

\deleted{The authors in~\cite{Islam2018} showed that \ac{IAB} networks play a fundamental role in the path of networks towards an all-fiber architecture.} %
\deleted{However, the performance of \acp{UE} in good channel state connected to donor \ac{IAB} nodes is degraded in \ac{IAB} networks compared to the non-\ac{IAB} case since the \ac{IAB} donor scheduler has to assign time-frequency resources not only to \acp{UE} but also to backhaul links.} %
\deleted{Coverage probability was studied in~\cite{Madapatha2020} through static simulations using stochastic geometry modeling.} %
\deleted{The authors showed that the performance of fully-fiber connected networks can be achieved with a small increase in the number of \ac{IAB} nodes in \ac{IAB} networks.} %
\deleted{The authors also studied the effect of rain, blockage and tree foliage.} %
\deleted{According to the results, low and moderate rainfall and blockage should not be a problem when appropriate network densities are chosen.} %
\deleted{However, tree foliage may be problematic for networks with low and moderate network densities.} %

\deleted{In~\cite{Saha2019}, the authors compared the rate coverage probability of an \ac{IAB} network with a conventional network without \ac{IAB} nodes or small \acp{BS}, i.e., only \ac{IAB} donors or macro \acp{BS} are assumed, and a network where all \acp{BS}, i.e., small and macro \acp{BS}, have access to a fiber-based backhaul.}  %
\deleted{According to the simulation results, \ac{IAB} outperforms the macro-only network but presents a lower coverage probability compared to the all-fiber scenario.} %
\deleted{Moreover, the authors showed that the offload of \acp{UE} from an \ac{IAB} donor to an \ac{IAB} node does not bring the same benefit of offloading \acp{UE} from a macro to a small \ac{BS} (both with fiber-based backhaul).} %

\goodbreak
\subsubsection{Network Deployment, Routing and Topology Adaptation}\label{SUBSUBSEC:ROUTING_TOPOLOGY}

Network deployment, routing and topology adaptation were studied in~\cite{Ranjan2021, Polese2018b, Polese2020,Islam2018,Rasekh2020,Rasekh2015,Arribas2020,Simsek2021,Vu2019,Li2019,HasanzadeZonuzy2019,Zhai2020,Zhang2020,Lai2020,Jarrah2021,Madapatha2021,Fang2021}. %
In general, for a broad range of coverage constraints/blockage densities, the impact of meshed communication to increase redundancy may be negligible if the network is well planned. %
In~\cite{Islam2018} the authors studied a convex optimization problem with the objective of maximizing the product of \acp{UE}' downlink and uplink data rates (geometric mean) subject to capacity and resource allocation constraints. %
Two \ac{IAB} topologies were considered: mesh and spanning tree based on highest channel quality among \ac{IAB} nodes. %
Simulation results showed that the presented mesh topology improved data rate fairness slightly. %

In~\cite{Rasekh2015} the authors studied a downlink backhaul optimization problem to maximize the minimum data rate among nodes (access links were not considered). %
An alternative formulation was also presented with the objective of penalizing higher delays in order to induce the use of fewer hops. %
The routing problems were solved by using linear optimization techniques. %
The main conclusions obtained from the simulation results were the relevance of taking into account the directional interference among nodes, and that routing complexity can be decreased by splitting the nodes into independent clusters and solving the optimization problem separately for each cluster. %
In~\cite{Rasekh2020}, the authors extended their own work in~\cite{Rasekh2015}. %
\deleted{Firstly,}\added{In this work,} the authors revisited the problem in~\cite{Rasekh2015} where the problem size increases exponentially with the number of links, and provided a new scalable formulation that scales near linearly with the number of links. %
\deleted{Another contribution of the article was the modeling of both downlink and uplink directions.} %
\deleted{The simulation results showed that the solution to the new scalable formulation outperforms the cluster-based solution presented in~\cite{Rasekh2015}.} %
\deleted{Furthermore, the authors compared the achieved throughput for a downlink-only versus a joint downlink and uplink scenario.} %
\deleted{Interestingly, the achieved throughput for downlink and uplink scenario incurred in a performance loss of only 20\% compared to the downlink-only scenario.} %

In~\cite{Vu2019} the authors studied routing and power allocation for a multihop \ac{IAB} network with \ac{FD} \acp{BS} employing hybrid beamforming. %
A utility maximization problem was formulated subject to constraints on the delay violation probability and queue stability.%
\deleted{The original problem was decoupled and, focusing on the routing problem, the authors employed \ac{RL} where the macro \ac{BS} was the agent and the possible paths were the actions.} %
\added{A routing subproblem was solved by employing \ac{RL}.} %
In the results, the authors highlighted the relevance of the path selection functionality in order to assure higher reliability and lower latency. %
\added{Improved latency and better throughput were reported by~\cite{Li2019} where resource allocation for access/backhaul, power allocation and a dynamic routing (based on real network traffic) were studied.} %

\deleted{In~\cite{Li2019} the authors considered a multihop heterogeneous network where resource allocation for access/backhaul, power allocation and routing were studied for downlink and uplink directions.} %
\deleted{Motivated by the high complexity of the formulated mixed continuous/integer optimization problem, the authors firstly decomposed the original problem into three subproblems by assuming a predefined routing solution.} %
\deleted{After solving the three subproblems, the authors presented a dynamic routing solution based on real time network traffic.} %
\deleted{The main idea of the proposed routing solution is to schedule user traffic through the less congested links.} %
\deleted{The dynamic routing solution was able to improve the system throughput and latency when compared to baseline solutions.} %

\deleted{A multihop and mesh \ac{IAB} network was considered in~\cite{HasanzadeZonuzy2019} where broadcast of real time flows (routing) was studied.} %
\deleted{In that scenario, once a packet is created in a given node, this packet should reach every other node in the system.} %
\deleted{The main objective was to maximize the number of packets that arrive at the destination \ac{IAB} nodes within delay requirements.} %
\deleted{Access transmissions are not modeled as the focus is only on backhaul transmissions.} %
\deleted{A utility maximization problem was formulated and an optimal distributed solution (based on dual decomposition technique) and a suboptimal distributed solution were proposed.} %
\deleted{Simulation results showed that the proposed optimal and suboptimal solutions outperformed two centralized throughput-optimal solutons for different utility functions.} %

Mesh topology was studied in~\cite{Zhai2020} for \ac{IAB} networks. %
\deleted{Firstly, the authors claimed that in a \ac{DAG} topology, resource allocation lacks flexibility since \ac{IAB} nodes cannot control their parent backhaul links due to the fixed parent-to-child relationship.} %
\deleted{Motivated by this, t}\added{T}he authors proposed a mesh topology for \ac{IAB} networks where the split \ac{DU}/\ac{MT} in \ac{IAB} nodes is replaced by a peer unit\deleted{, as opposed to \ac{3GPP} considered model}. %
Each peer unit of an \ac{IAB} node can be dynamically configured as master or slave. %
An \ac{IAB} node configured as master can control their links to other \ac{IAB} nodes configured as slaves as well as its own access links. 
Simulation results showed that the proposed mesh topology provides lower packet delays \added{and}\deleted{than the \ac{DAG} topology especially for bursty traffic. While mesh topology achieves small throughput gains compared to \ac{DAG} for constant traffic services,} significant total throughput gains \deleted{can obtained }for bursty traffic \added{when compared to classical topologies}. %
\added{Mesh networks for \ac{IAB} was also studied in \cite{HasanzadeZonuzy2019} where the authors studied the problem of maximizing the number of delivered broadcast packets.} %
\added{Optimal and suboptimal solutions were proposed.} %
\added{Mesh \ac{IAB} with \acp{IRS} was studied in~\cite{Jarrah2021}}. %
\added{The authors reported that the use of \ac{IRS} is capable of improving \ac{SNR} and \ac{SER}, which allows for the use of higher modulation orders, and that outage probability for multiple hops can be reduced if the number of elements in \ac{IRS} is increased}. %
However, it is important to highlight that none of the works presented in this section about mesh topology included in their analysis the increased complexity of enabling a mesh architecture in \ac{IAB} networks in terms of signaling overhead, for example.

The authors in~\cite{Zhang2020} considered a heterogeneous network and studied the problems of power allocation and routing. %
\added{In this work, they assumed cooperation between \acp{BS} that can employ \ac{OMA} and \ac{NOMA} multiple access schemes.} %
\added{The main outcome of this work was the proposal a joint and adaptive greedy routing solution that switches between cooperative \ac{OMA} and \ac{NOMA} in order to maximize energy efficiency subject to data rate and power constraints.} %

\deleted{Two transmission schemes were discussed as background in this work: cooperative \ac{OMA} and cooperative \ac{NOMA}.} %
\deleted{The authors assumed that the small \acp{BS} are split into two groups relatively to the distance to the macro \ac{BS}: near and far small \acp{BS}.} %
\deleted{In cooperative \ac{OMA}, the macro \ac{BS} transmits to the near small \ac{BS} the information desired by the far small \ac{BS}, which in its turn, retransmits the information to the far small \ac{BS} in \ac{FD} mode.} %
\deleted{In the cooperative \ac{NOMA} scheme, the macro \ac{BS} transmits superposed data to a near and a far small \ac{BS} using the same time-frequency resource.} %
\deleted{The near small \ac{BS} performs \ac{SIC} and retransmits in \ac{FD} mode the data to the far small \ac{BS}.} %
\deleted{Since cooperative \ac{OMA} and \ac{NOMA} have its own advantages and disadvantages, a joint and adaptive greedy solution was proposed to improve the system energy efficiency subject to data rate and power constraints.} %
\deleted{In the results, the authors showed that the proposed joint scheme outperforms cooperative and non-cooperative \ac{OMA} and \ac{NOMA} in terms of energy efficiency and average power consumption.} %

In~\cite{Polese2018b}, the authors studied \added{distributed} routing strategies (path selection) for \ac{IAB} networks. %
\deleted{Distributed routing strategies were conceived in the sense that the next hop decision is taken locally at each node.} %
Four routing solutions were presented: highest-quality-first, wired-first, position-aware and maximum-local-rate policies. %
In highest-quality-first solution, the next hop is the one that presents the highest \ac{SINR} whereas in wired-first one, the chosen hop is the one that provides the shortest path to the \ac{IAB} donor. %
The main idea in position-aware policy is to achieve a balance between highest-quality-first and wired-first solutions. %
Finally, the maximum-local-rate solution chooses the next hop with highest data rate that, in its turn, takes into account the number of connected \acp{UE} in the parent \ac{IAB} node, i.e., \ac{IAB} load. %
Performance results showed \added{that}\deleted{the benefits of increasing the number antennas at \ac{IAB} nodes and the percentage of \ac{IAB} donors in the system.} \deleted{W}\added{w}ired-first policy provided a reduced number of hops at the cost of lower worst-\ac{SINR}\deleted{.} \added{while}\deleted{Finally,} the best performance for low-\ac{SINR} regime was achieved by position-aware policy. %
\added{Similar routing solutions and some new variants were evaluated in \cite{Polese2020, Ranjan2021}}. %
\added{In~\cite{Polese2020}, the authors showed that choosing wired-first routing instead of highest-quality-first provides higher throughput and lower latency.} %
\added{In~\cite{Ranjan2021}, the reported results showed that although highest-quality-first routing improved the link quality at backhaul and access links, the number of hops to the core network has increased.} %
\added{Strategies based on path data rate, i.e., end-to-end data rate, were able to reduce the number of hops to the core network.} %

\deleted{In~\cite{Polese2020}, the authors evaluated the impact of routing on \ac{IAB} performance.} %
\deleted{They showed that routing that prioritizes shorter paths instead of the path achieved by choosing the hop with best channel state quality provides higher throughput and lower latency.} %
\deleted{As in \cite{Polese2018b}, in \cite{Ranjan2021}, the authors studied routing for \ac{IAB} networks.} %
\deleted{Different strategies for cell selection of the \ac{MT}-part of \ac{IAB} nodes and \acp{UE} were proposed that take into account link strength, i.e., \ac{RSRP}, local data rate and path data rate, i.e., among all hops until reaching the \ac{IAB} donor.} %
\deleted{Simulation results showed that although the strategy based on signal strength improved the link quality at backhaul and access links, the number of hops to the core network has increased.} %
\deleted{Strategies based on path data rate were able to reduce the number of hops to the core network, i.e., latency, and improve the system load distribution among nodes.} %

A heterogeneous network was assumed in~\cite{Arribas2020} where resource allocation and routing were studied for a two-hop \ac{IAB} network.  
\deleted{The authors assumed that there is a certain amount of data to be delivered from a macro \ac{BS} to each small \ac{BS}.} %
\deleted{Backhaul-only transmissions were evaluated, i.e., access links were not modeled.} %
A mixed integer optimization problem was formulated to minimize the makespan, i.e., the time needed to deliver all packets from macro \ac{BS} to the small \acp{BS}. %
Based on analytical insights, the authors proposed a heuristic solution to the formulated problem. %
The simulation results showed that the makespan increases with the average file and network sizes. %
\deleted{Furthermore, the authors also showed that the makespan decreases with the number of \ac{RF} chains at the macro \ac{BS}, i.e., the spatial multiplexing capacity of the macro \ac{BS}.} %

In~\cite{Lai2020}, the authors assumed a multihop \ac{IAB} network and studied the problems of resource allocation, user association, power allocation as well as \ac{IAB} node positioning. %
Focusing on the \ac{IAB} node positioning problem, long-term channel statistics, \ac{UE} density and \ac{UE}-\ac{BS} association were considered as input to optimize node placement. %
The objective was to maximize weighted sum rate and closed-form solutions were developed for node positioning assuming distributions related to the probability of a \ac{UE} getting assigned resources. %
The proposed node positioning showed to be efficient when compared to exhaustive search solutions. %

Topology formation and adaptation for \ac{IAB} networks were evaluated in~\cite{Simsek2021}. %
\added{Optimal and suboptimal solutions were proposed for a topology formation problem that maximizes the minimum capacity along the system.} %
\deleted{For each node, the authors introduced the concept of score for an \ac{IAB} node that represents the sum of the minimum capacities of the multiple parents of this node towards the serving \ac{IAB} donor.} %
\deleted{The best topology was obtained by maximizing the sum of the \ac{IAB} nodes' scores.} %
\deleted{The optimal solution was obtained by employing dynamic programming at the cost of a high computational complexity.} %
\deleted{Based on the knowledge of the received power for all links in a central node, the authors proposed a heuristic topology formation/adaptation algorithm.} %
\deleted{In the simulation results the authors compared the proposed solution with the optimal one and a baseline solution.} %
\added{As conclusion, t}\deleted{T}hey showed that for few \ac{IAB} nodes, all algorithms achieved a similar performance since the number of options for topology formation is not high. %
\deleted{Also, the proposed solution performed close to the optimal solution for both \ac{ST} and \ac{DAG} topologies.} %

\deleted{In~\cite{Jarrah2021}, the authors studied the impact of the use of \ac{IRS} in wireless mesh backhaul networks.} %
\deleted{By assuming Rician fading channels, the authors derived \ac{SNR}, \ac{SER} and outage probability for single-hop and multihop.} %
\deleted{The analysis for the multihop case was obtained assuming wired-first routing that selects as next node the one that is connected to a wired connection to the core network and that presents a link quality above a given threshold.} %
\deleted{From the results, the authors concluded that the use of \ac{IRS} in backhaul can improve \ac{SNR} and \ac{SER} which allows for the use of higher modulation orders.} %
\deleted{Furthermore, the outage probability for multiple hops can be reduced if the number of elements in \ac{IRS} is increased.} %

In~\cite{Madapatha2021}, the authors assumed \deleted{a hybrid}\added{an} \ac{IAB} network where there are \ac{IAB} donors and small \acp{BS} \added{where}\deleted{.}\deleted{The small \acp{BS} can be either \ac{IAB} nodes connected to the \ac{IAB} donors through wireless backhaul or traditional small \acp{BS} connected to the core network through fiber.}
\deleted{A fixed fraction of the small \acp{BS} can be connected to the core network through fiber.}\deleted{The network deployment was the focus of this article where } two problems were studied: small \ac{BS} positioning and decision if a small \ac{BS} should be directly connected to the core network (small \ac{BS} to donor assignment). %
\deleted{By means of computer simulations based on stochastic geometry, the authors studied the impact of the proposed strategies on system performance, the effect of environmental parameters and nodes capability on coverage probability, and the effect of routing on the performance of \ac{IAB} network.} %
The results showed that the optimization of \ac{IAB} node positioning and \deleted{definition of which small \acp{BS} should be directly connected to the core}\added{small \ac{BS} to donor assignment} \deleted{outperfom}\added{outperform} random positioning\deleted{ and assignment to fiber capabilities}. %
Finally, the authors showed that network planning decreased the need of routing updates during temporal link blocking. %

Resource allocation and routing were the focus of \cite{Fang2021} by assuming a \ac{TDMA}-based multihop \ac{IAB} network. %
An optimization problem of maximizing the minimum node's throughput was formulated subject to time resource constraints. %
Both routing\deleted{, called link scheduling in the article,} and resource allocation were jointly solved by applying an iterative solution based on Simplex and maximum matching theories. %
\deleted{In the simulation results, the authors evaluated the impact of different parameters of the assumed model on the performance of the proposed solution for \ac{IAB} taking macro-only scenario as baseline.} %
\deleted{According to the presented results,}\added{Simulation results showed that}\deleted{ the minimum node's throughput for \ac{IAB} is increased when compared to the one achieved in the macro-only scenario for a broad range of transmit power values. Moreover, the authors showed that} the minimum node's throughput is increased as the number of macro and small \acp{BS} increases as a result of shorter link distances and decreased number of backhaul hops. %
\deleted{Throughput gains are also observed as the main lobe gain of antenna arrays is increased or main lobe beamwidth is reduced.} %

\goodbreak
\subsubsection{Resource Allocation for Access and/or Backhaul}\label{SUBSUBSEC:RESOURCE_ALLOC_ACCESS_BACKHAUL}

As opposed to out-band operation where the access and backhaul operate in different frequencies, in-band backhauling refers to the cases with the access and backhauling sharing the same resources. %
Compared to out-band backhauling, in-band operation leads to higher flexibility at the cost of complexity. %
The state-of-the-art works concentrate mainly on in-band backhauling~\cite{Saha2018,Saha2018b,Saha2019,Kwon2019,Liu2019,Li2019,Zheng2020,Muhammed2020,Lei2020,Madapatha2020,Arribas2020,Lai2020,Fang2021}. %

In~\cite{Saha2018} the authors employed stochastic geometry to model a two-hop heterogeneous \ac{IAB} network. %
The total \ac{mmWave} bandwidth is partitioned into two parts: the first for backhaul and the second for access. %
The access bandwidth is shared among \acp{UE} in a round-robin fashion. %
On the other hand, there are two options on how the backhaul bandwidth is shared among \ac{IAB} nodes: equal partitioning and instantaneous load-based partitioning. %
In the equal partitioning strategy, the bandwidth is equally split among \ac{IAB} nodes. %
In the instantaneous load-based strategy, the allocated bandwidth is proportional to the current load on each \ac{IAB} node. %
Simulation results showed that load-based partitioning outperforms the equal partitioning at the cost of a high signaling rate\deleted{, i.e., the current load on each \ac{IAB} node should be reported to a central controller}. %
\added{In~\cite{Madapatha2020}, the authors studied a similar problem involving bandwidth partitioning between access and backhaul based on a fixed split factor.} %
\deleted{Moreover, the authors showed that the rate coverage probability is maximized for a particular bandwidth partition.} %

\deleted{The results of~\cite{Saha2018} are later extended in~\cite{Saha2018b} where the results are compared with the cases using average}\added{In order to reduce signaling load, in~\cite{Saha2018b}, a} load-based resource partitioning \added{among \ac{IAB} nodes was proposed}. %
Here, the load reported by \ac{IAB} nodes is averaged over long intervals. %
In the simulation results, the authors showed that the average load-based partitioning achieves a good performance-signaling load trade-off.

The work~\cite{Saha2019} is an extension of~\cite{Saha2018}. %
\deleted{Two resource allocation schemes were assumed: \ac{IRA} and \ac{ORA}.} %
\deleted{\Ac{ORA} is similar to the proposal in~\cite{Saha2018} where the bandwidth between access and backhaul is split according to a given factor.} %
\added{The bandwidth split between access and backhaul based on a fixed factor proposed in \cite{Saha2018} is called \ac{ORA} in \cite{Saha2019}.} %
\added{In this work, the authors proposed an alternative solution called \ac{IRA} where the bandwidth assigned by a donor \ac{IAB} to the backhaul for a given \ac{IAB} node depends on the access load of the \ac{IAB} node.} %
\deleted{In \ac{ORA}, the backhaul sharing is proportional to the small \ac{BS} load.} %
\deleted{In \ac{IRA}, the access and backhaul transmissions share the same pool of resources.} %
\deleted{The \ac{IAB} donor shares the bandwidth equally among \acp{UE} (\acp{UE} connected to the donor \acp{IAB} and small \acp{BS}) where the bandwidth for backhaul for a given small \ac{BS} is defined according to the number of \acp{UE} connected to it.} %
\deleted{As \ac{IAB} nodes are \ac{HD}, the bandwidth used for backhaul is not used for access.} %
Simulation results showed that the coverage probability in \ac{ORA} is highly dependent on the bandwidth split factor\deleted{ also as reported by~\cite{Saha2018}}. %
As the number of small \acp{BS} increases, more bandwidth should be reserved to backhaul in \ac{ORA} scheme. %
Coverage probability provided by \ac{IRA} is better than the one achieved with the static \ac{ORA} scheme. %

In~\cite{Kwon2019}, \deleted{an \ac{IAB} network was assumed where macro and small \acp{BS} as well as \acp{UE} are equipped with multiple antennas.}
\deleted{In this article, }the problems of user association, hybrid beamforming design, power allocation and resource allocation were studied. %
A weighted sum rate maximization problem with limited \ac{CSI}\deleted{ subject to beamforming, power and backhaul constraints} was formulated. %
Due to the complexity of the formulated problem, the authors followed a two-stage approach. %
At stage 1, user association and beamforming are designed for access links. %
After that, in stage 2, the decisions for access links are signaled to the macro \ac{BS} that in its turn solves resource allocation  and beamforming problems for backhaul links. %
\deleted{Focusing on resource allocation aspects solved in stage 2, the authors assumed that access and backhaul transmissions take place in different slots within a time frame.} %
Simulation results showed that the time fraction assigned to access links increases as a result of higher transmit power at macro \ac{BS}. %
On the other hand, the increase of the transmit power of small \acp{BS} results in a lower time fraction for access links. %

Resource allocation was studied in~\cite{Liu2019} for a mesh network that combines the concepts of \ac{IAB} and \ac{UPN} where high channel quality \acp{UE} are capable of relay information towards bad channel quality \acp{UE}. %
\deleted{The considered resource allocation in this article was in terms of time sharing for concurrent links.} %
\deleted{Note that access and backhaul links use licensed bands whereas \ac{D2D} communication uses unlicensed bands.} %
A Nash bargaining solution was proposed to \deleted{find the time sharing for the}\added{solve the resource sharing problem among} concurrent links as well as to define how \acp{UE} and the operator should cooperate. %
Simulation results showed that the proposed cooperation scheme for \ac{UPN} and resource allocation for \ac{IAB} resulted in higher network throughput compared to baseline schemes without \ac{UPN} or resource allocation. %

As previously described, resource allocation for access and backhaul was also studied in~\cite{Li2019}. %
In this work, the resource allocation subproblem took the form of defining the spatial multiplexing for the links, i.e., \ac{SDMA} groups, and the number of slots from the frame assigned to each group. %
\Ac{SDMA} groups were chosen by using a greedy algorithm based on graph theory. %
The definition of the number of slots followed a proportional fair policy. %
In the simulation results, the authors showed that the proposed scheme is more robust than baseline solutions when the number of \acp{UE} is increased thanks to the spatial multiplexing capacity. %
\deleted{The authors also showed that the proposed solution has a better capacity of multiplexing downlink and uplink traffic according to \acp{UE}' demands when compared to fixed \ac{TDD} scheme.} %

In~\cite{Zheng2020} the authors assumed a two-hop \ac{IAB} network with \ac{FD}-capable nodes. %
In this work resource allocation for access and backhaul as well as power allocation were studied with the objective of maximizing \deleted{the sum of the logarithm of \acp{UE}'s downlink data rate (}proportional fairness\deleted{)}. %
\deleted{A non-linear and non-convex optimization problem was formulated.} %
Due to the complexity of the formulated problem, the authors resorted to a two-level approach by decomposing the original problem into two problems: backhaul bandwidth allocation and power allocation. %
The authors showed that the bandwidth allocation is a convex subproblem whose solution can be found in closed form. %
In the simulation results, the authors evaluated the impact of the density of connected \acp{UE} to donor \ac{IAB} and \ac{IAB} nodes on the bandwidth allocation. %

Bandwidth and power allocation were studied in~\cite{Muhammed2020} for a heterogeneous network with wireless backhaul where the macro \ac{BS} was equipped with massive \ac{MIMO} and small \acp{BS} employed \ac{NOMA} in access transmissions. %
\deleted{An}\added{A suboptimal solution for the} optimization problem \deleted{with the objective }of maximizing the energy efficiency of the small \acp{BS} subject to minimum \ac{UE} data rate guarantees was formulated. %
\deleted{In order to make the problem tractable, the original problem was transformed in a sequence of linear optimization problems in order to obtain a low-complexity suboptimal solution where bandwidth and power allocation are alternately solved.} %
In the simulation results, the authors showed that there is a specific bandwidth partitioning that leads to the optimal energy efficiency. %
\deleted{Furthermore, energy efficiency is improved for the proposed solution as the number of small \acp{BS} and \acp{UE} connected to small \acp{BS} increases.} %

Downlink subchannel allocation for access and backhaul in a two-hop \ac{IAB} network was studied in~\cite{Lei2020}. %
The nodes are assumed to be \ac{FD}-capable and a single antenna is assumed for \ac{DU} and \ac{MT} parts. %
An optimization problem was formulated with the objective of maximizing a utility function (later set to proportional fairness) subject to data rate guarantees for \acp{UE}. %
\deleted{The formulated problem is a non-convex mixed integer problem and thus complex.} %
\deleted{The authors proposed two deep \ac{RL}-based algorithms to solve the formulated problem: double deep Q-network and actor-critic technique.} %
\deleted{The simulation results showed that the proposed learning-based solutions outperform the static full-reuse solution in terms of fairness.} %
\deleted{Moreover, the actor-critic solution achieves a faster convergence compared to double deep Q-network.}
\added{A learning-based solution was proposed and, according to simulation results, it outperformed the static full-reuse solution in terms of fairness.} %

\deleted{In~\cite{Madapatha2020}, besides providing results comparing \ac{IAB} networks and fiber-based networks as previously described, the authors studied the effect of bandwidth partitioning between access and backhaul in downlink.} %
\deleted{The authors considered that the total bandwidth is split into access and backhaul based on a bandwidth split factor.} %
\deleted{The bandwidth for backhaul assigned by \ac{IAB} donor to \ac{IAB} nodes is based on the supported load by each \ac{IAB} node.} %
\deleted{In access, the bandwidth is equally split among \acp{UE}.} %
\deleted{The simulation results showed that there is an optimal bandwidth split factor depending on the system scenario.} %
\deleted{According to the results, the optimal split factor increases the bandwidth for backhaul as the \ac{IAB} node density increases whereas it increases bandwidth for access as the \acp{UE} density increases.} %

Besides studying routing problems as previously described, the authors in~\cite{Arribas2020} developed a model for resource allocation in backhaul transmissions. %
The fraction of the capacity of each active link to transmit data to small \acp{BS} was also optimized to minimize the makespan. %
One important assumption in this work is that the interference pattern among nodes is known to the macro \ac{BS} or \ac{IAB} donor. %

Considering multihop \ac{IAB} networks, \cite{Lai2020} develops bandwidth allocation schemes for access and backhaul. %
\deleted{Therein, \ac{IAB} nodes were located within a sector of an \ac{IAB} donor.} %
In this work, the authors assumed that orthogonal bands are assigned to each backhaul and access link. %
\deleted{However, frequency reuse is applied to areas that are far apart from each other.} %
In a centralized operation, where most of the resource allocation decisions are taken on \ac{IAB} donors based on channel quality information sent from child nodes, channel aging should be taken into account. %
Channel aging is caused by the fact that the channel quality estimate used by donor \ac{IAB} may change considerably when resource allocation decisions are actually employed by child nodes. %
A subcarrier allocation problem is formulated within each band with the objective of maximizing the weighted downlink spectral efficiency. %
The simulation results showed that channel prediction together with efficient resource allocation lead to a good performance when channel aging and low-mobility \acp{UE} are assumed. %

The work~\cite{Fang2021} studied both routing and resource allocation for access/backhaul in a multihop \ac{TDMA}-based \ac{IAB} network \added{as detailed in Section~\ref{SUBSUBSEC:ROUTING_TOPOLOGY}}. %
\deleted{Both problems are solved jointly by an iterative solution based on Simplex algorithm and maximum matching theory.} %
Focusing on the resource allocation problem, the authors assumed that a frame is divided into a number of time slots with variable lengths. %
\deleted{As \ac{TDMA} and \ac{HD} transceivers are assumed, a given node can only transmit (using the whole bandwidth) to a single node in a given time slot.} %
The formulated problem has the objective of maximizing the minimum nodes's throughput (fairness). %
The proposed scheme outperforms the macro-only scenario over different conditions\deleted{ as detailed in Section~\ref{SUBSUBSEC:ROUTING_TOPOLOGY}}. %

\added{In~\cite{Pagin2022}, the authors focus on how scheduling, i.e., resource allocation, can be performed in a semi-centralized architecture.} %
\added{In \ac{IAB} networks, the scheduling decisions for access links associated with a given node, e.g., \ac{IAB} donor or \ac{IAB} node, are in charge of its local \ac{DU}.} %
\added{Thus, essentially, the proper association among time-frequency resources and links is a distributed task among network nodes.} %
\added{This assumption was not followed by some of the previous described works, such as~\cite{Arribas2020,Li2019,Lai2020,Fang2021} where centralized solutions were proposed.} %

\added{Specifically, in~\cite{Lai2020}, the authors showed that the performance gains of centralized solutions is strongly dependent on the effect of channel aging.} %
\added{Although fully centralized scheduling is not \ac{3GPP} compliant for \ac{IAB}, nodes can exchange information and eventually, it can be assembled in a central node where general recommendations for resource allocation and scheduling can be informed to the other network nodes.} %
\added{This is the key idea of~\cite{Pagin2022}.} %
\added{In this work, a central controller receives general information about network nodes, e.g., \ac{CSI} and buffer status, and provides centralized recommendations for resource partitioning to the distributed schedulers.} %
\added{However, the local \acp{DU} are free to follow or not the indications of the central controller depending on local conditions.} %
\added{One important conclusion of this work is that this semi-centralized framework provides performance gains, however, it strongly depends on the capacity of the system in exchanging information in a timely manner.} %

\added{The previous described works~\cite{Madapatha2020,Saha2018,Saha2019,Saha2018b,Kwon2019} present resource allocation and scheduling solutions that can be classified as a semi-centralized operation where partitioning of bandwidth between access and backhaul is taken in a central node based on the available system information, but the proper association among system resources and links are decided locally.} %

\added{In \cite{Alguafari2022}, the authors proposed a distributed/decentralized solution to scheduling/resource allocation problem.} %
\added{The proposed solution takes advantage of the information spread by network nodes about the bandwidth allocated to their children as well as the
	local perspective obtained about network topology.} %
\added{Based on this information, local optimization problems are solved at each node for resource allocation.} %

\added{Finally, in \cite{Gopalam2022} the authors proposed distributed schedulers for multihop \ac{IAB} assuming a limited number of \ac{RF} chains.} %
\added{An optimal back-pressure scheduling solution as well as a distributed message passing scheme were proposed in order to achieve queue stability.} %
\added{Furthermore, another contribution of \cite{Gopalam2022} was the proposal of local schedulers with a simplified exchanging of messages between nodes.} %

\goodbreak
\subsubsection{Power Allocation}\label{SUBSUBSEC:POWER_ALLOCATION}

Power allocation may be useful for the cases with simultaneous operation at the \ac{IAB} node, especially if the network deployment is not well planned. %
For this reason, power allocation has been discussed in \ac{3GPP} Release 17 work-item on \ac{IAB} enhancements. %
Power allocation was also studied in~\cite{Vu2019,Li2019,Kwon2019,Zheng2020,Muhammed2020,Zhang2020,Lai2020}. %
With exception of \cite{Adare2021}, it is important to note that power allocation was not studied alone in those works; in fact, power allocation was jointly studied with resource allocation for access and/or backhaul, routing/topology adaptation or \ac{UE} association. %
\deleted{The formulated problem in~\cite{Vu2019} was split into routing and power allocation.} %
\deleted{Specifically, the power allocation subproblem for downlink was solved by employing successive convex approximation.} %
\deleted{From the simulation results, although power allocation led to lower delays when compared to a baseline solution without power allocation and routing schemes, the routing scheme showed to be more relevant when delay was concerned.} %
\added{Routing and power allocation was studied in~\cite{Vu2019}.} %
\added{The main outcome of this work was that power allocation leads to only marginal improvements in the delay performance when intelligent routing is already considered.} %
\deleted{In~\cite{Li2019}, power allocation was studied with the objective of maximizing the spectral efficiency.} %
\deleted{The power allocation subproblem for both downlink and uplink was solved by a water-filling-like algorithm for a fixed routing solution.} %
\deleted{The transmit power allocation for each node was rather simplified since the choice of the links in each \ac{SDMA} group guarantees that the interference level are kept at low levels.} %
\added{Power allocation for \ac{SDMA} groups and assuming fixed routing was also studied in~\cite{Li2019} with the objective of maximizing spectral efficiency.} %

Besides user association and resource allocation, power allocation problem was studied in~\cite{Kwon2019}. %
Hybrid beamforming was assumed in this work. %
Therefore, the power allocation problem for downlink was studied together with the baseband beamforming design for access and backhaul links. %
\deleted{For access links, an iterative water-filling algorithm was employed to find a suboptimal solution for power allocation.} %
\deleted{For backhaul links, power allocation and baseband beamforming were found by solving a \ac{SINR} balancing problem.} %
In the simulation results, the authors showed the effect of the available power at macro and small \acp{BS} on the total data rate and time share for access transmissions. %
As expected, the time share for access transmissions increases as the power of macro \ac{BS} increases and/or the power at small \acp{BS} decreases. %

In~\cite{Zheng2020}, the authors decoupled the original optimization problem in resource (bandwidth) allocation and power allocation for maximizing the proportional fairness in the system. %
After solving bandwidth allocation, the transmit power allocation for downlink was solved for access and backhaul independently since there is no coupled constraints. %
\deleted{Both power allocation problems could be optimally solved by applying \ac{KKT} conditions.} %
From the simulation results, the authors showed that throughput can be increased until a certain level as the downlink access transmission \added{power} is augmented. %
For higher access transmission power, the backhaul capacity starts to limit the \acp{UE}' data rates. %

Besides bandwidth allocation, as previously described, power allocation was studied in~\cite{Muhammed2020} with the objective of maximizing energy efficiency. %
The bandwidth and power allocation original problem for downlink was solved by leveraging fractional programming properties and sequential convex approximation. %
The authors showed in the simulation results that energy efficiency increases with the macro \ac{BS}'s transmit power, however, the high transmit power at small \acp{BS} leads to a higher power consumption per data rate which results in poor energy efficiency. %

In~\cite{Zhang2020}, the authors studied routing and power allocation for downlink by proposing a joint cooperative \ac{OMA} and \ac{NOMA} scheme, \deleted{where small \acp{BS} are divided into two groups: near \acp{BS} and far \acp{BS} relative to the macro \ac{BS}}\added{as previously described in Section~\ref{SUBSUBSEC:ROUTING_TOPOLOGY}}. %
\deleted{The joint scheme was compared with the conventional cooperative \ac{OMA} and cooperative \ac{NOMA}.} %
For the proposed and baseline schemes, power allocation for each small \ac{BS} was analytically defined so as to ensure minimum data rate guarantees for backhaul links. %
\deleted{From the simulation results}\added{Simulation results showed that}, \deleted{the authors showed that the energy efficiency and power consumption of the baseline schemes strongly depends on how small \acp{BS} are split into the two groups. As}\added{as} the number of \deleted{small \acp{BS} in the group of near}\added{relaying small} \acp{BS} is gradually increased, the performance of cooperative \ac{OMA} and \ac{NOMA} in terms of energy efficiency and power consumption is improved until a certain point. %
Moreover, \deleted{cooperative \ac{OMA} and \ac{NOMA} present complementary performance depending on how small \acp{BS} are split. Consequently,}\added{results showed that} joint cooperative \ac{OMA} and \ac{NOMA} enjoys the benefits of both \deleted{schemes}\added{cooperative \ac{OMA} and \ac{NOMA}} and presents better performance in different conditions. %

Joint power allocation and subcarrier assignment for downlink were considered in \cite{Lai2020} for a multihop \ac{IAB} network\deleted{ where \ac{IAB} nodes are within a sector of a macro \ac{BS} that plays the role of \ac{IAB} donor}. %
As different subbands were assigned to each backhaul link and access, power allocation was optimally derived for subcarriers within each subband. %
Suboptimal power allocation solutions were also provided, which showed negligible performance loss to the optimal solutions according to simulation results. %

In \cite{Adare2021}, uplink power control for dual-hop \ac{IAB} networks was studied. %
The main objective of the power control solution was to maximize the coverage probability by employing \ac{GA}.
By simulation results, the authors showed that power control is able to substantially improve the coverage probability especially when out-of-band \ac{IAB} is used. %
Also, the authors showed that inter-cell interference does no play an important role since \ac{mmWave} band was assumed and, thus, signal strength quickly decreases with distance. %

\goodbreak
\subsubsection{\ac{MIMO}, Beamforming and \ac{UE} Association}\label{SUBSUBSEC:USER_ASSOC}

The impact of \ac{UE} association is considerably different in \ac{IAB} networks when compared to conventional fiber-based networks. %
The main reason is that the load of \acp{UE} transferred from donor \ac{IAB} to \ac{IAB} nodes comes back to \ac{IAB} donors through the backhaul.
Thus, \ac{UE} association and load balancing should be re-thought for \ac{IAB} networks. %
Moreover, multiple antennas technology, that is a key aspect for \ac{5G}, assumes a relevant role in \ac{IAB} networks since backhaul data rates should be in the order of the total access data rate of \ac{IAB} nodes. %
Spatial multiplexing between access and backhaul links is also another aspect that has the potential to increase \ac{IAB} efficiency. 
In the following, we present some important works related to this topic. %

The proper connection between \acp{UE} and \acp{BS}, i.e., \ac{UE} association, was investigated in~\cite{Saha2019,Kwon2019,Chen2019,Lai2020}. %
In~\cite{Saha2019}, the criterion to define which \ac{BS} a \ac{UE} should connect to is \deleted{the biased average received power. %
	With this criterion, each \ac{BS} (\ac{IAB} or donor \ac{IAB} node) has a constant bias. So, a \ac{UE} connects to the \ac{BS} that has}\added{based on} the highest product between the received power and \deleted{the}\added{a} constant bias. %
In the simulation results, the authors \deleted{studied the impact of increasing the ratio of the bias of \ac{IAB} nodes and \ac{IAB} donors on network performance. By increasing this ratio,}\added{showed that, as the bias factor of \ac{IAB} nodes increases relative to the one of \ac{IAB} donors,} more \acp{UE} are offloaded to the \ac{IAB} nodes, thus increasing the load in backhaul links. %
\deleted{There is an optimal ratio where rate coverage and data rate are maximized for the proposed schemes.} %

The work~\cite{Kwon2019} studied resource allocation and power allocation, as previously described. %
Furthermore, the transmitting \ac{RF} beams of small \acp{BS} and the receiving \ac{RF} beams of the \acp{UE} were optimized using analog beamforming. %
\deleted{Orthogonal beamforming was employed to reduce the channel estimation overhead.} %
In order to solve the baseband beamforming of small \acp{BS}, the weighted maximization problem was converted to a \ac{SLNR} problem in order to reduce the complexity. %
Once access links were designed, beamforming for backhaul links should be designed. %
\Ac{RF} beamforming was defined based on the steering beamforming vectors for the line-of-sight directions of the active small \acp{BS}. %
Then, baseband beamforming was found by solving an \ac{SINR} balancing optimization problem. %
In the simulation results, the proposed hybrid beamforming based solution outperformed baseline solutions for full-\ac{CSI} and achieved an acceptable performance degradation compared to the fully digital beamforming solution. %
\deleted{Also, when limited \ac{CSI} was assumed, the results showed that an acceptable performance degradation is observed for the proposed beamforming solution compared to the full \ac{CSI} case.} %

\Ac{UE} association and beamforming were studied for downlink in~\cite{Chen2019} in a user-centric approach where a \ac{UE} is served by a group of small \acp{BS} (cluster). %
Small \acp{BS}, in \deleted{its}\added{their} turn, receive data intended to the \acp{UE} by means of multicast transmissions\deleted{, i.e., the macro \ac{BS} transmits the data intended to a given \ac{UE} to the small \acp{BS} that compose the cluster serving this \ac{UE}}\added{ from macro \acp{BS}}. %
\deleted{The authors assumed that the macro \ac{BS} employs multiple antennas, \acp{UE} are single-antenna nodes and small \acp{BS} operate in \ac{FD} mode having a single antenna for receiving data in backhaul and multiple antenna for transmitting in access.} %
The authors formulated an optimization problem for maximizing the weighted \ac{UE} data rate under perfect \ac{CSI} through joint design of multicast beamforming at backhaul, access beamforming and small \ac{BS} clustering subject to power constraints at small and macro \acp{BS}. %
\deleted{Due to the high complexity of the formulated problem, a two-step approach was employed where beamforming for access and backhaul were firstly optimized by considering a fixed clustering solution.} %
\deleted{Then, the clustering problem was solved by a heuristic.} %
\added{A heuristic solution was proposed to solve the beamforming design and clustering problems.} %
From the simulation results, for the perfect \ac{CSI} and static clustering assumptions, the authors showed that the proposed solution outperformed baseline solutions especially for large available power at macro \ac{BS}. %
\deleted{The proposed clustering solution presented better performance when compared to static schemes especially when the self-interference suppression capability was improved.} %
\deleted{Finally, the authors also showed that with moderate \ac{CSI}, the proposed solutions achieved a performance close to the perfect \ac{CSI} case.} %

In~\cite{Xue2020}, the authors focused on \ac{MIMO} for backhaul links. %
The expected spectral efficiency for wireless \ac{MIMO} backhaul links is much higher than in \ac{MIMO} access links. %
Therefore, physical impairments such as time offset, which can lead to \ac{ISI}, and phase noise that increases the multi-access interference due to outdated precoder and decorrelator should be carefully mitigated for backhaul links. %
The main contributions of the article were the proposal of a time offset compensation, phase noise estimation and design of a precoder/decorrelator. %
\deleted{From the simulation results, the authors showed the relevance of considering these physical impairments for backhaul links.} %
\deleted{Also, the proposed scheme was able to outperform baseline schemes and to achieve high spectral efficiency.} %
In~\cite{Lai2020}, besides studying other problems such as power/bandwidth allocation and node placement, the authors studied \ac{UE} association. %
The authors showed that \ac{UE} association based on average received power may lead to load imbalance. %
Thus, a load-balancing-based \ac{UE} association was proposed. %
\deleted{Basically, the \ac{UE} association were determined by adjusting the serving area of each node.} %

In~\cite{Sadovaya2021}, the authors investigated the impact of the arrangement of antennas in tri-sectorized \ac{IAB} nodes. %
In order to decrease inter-sector interference and allow the reuse of frequency resources in different sectors of the same \ac{IAB} node, two strategies are employed: spatial and angular diversities. %
In the first one, the arrays of antennas of each sector are disposed in the vertices of an equilateral triangle whereas the \ac{IAB} node equipment are located in the center of mass of the triangle. %
In the second strategy, a minimum angular separation should be kept between \acp{UE} associated to adjacent sectors of the same \ac{IAB} node. %
\deleted{Simulation results showed that the two strategies are capable of reducing the outage probability due to high interference between sectors when compared to the case with no diversity.} %

\Ac{FD}-\ac{IAB} nodes with subarray-based hybrid beamforming scheme was considered in~\cite{Zhang2021_2}. %
By assuming that antenna isolation and \ac{RF} cancellation are perfect, the authors designed digital filters to mitigate residual self-interference (with \ac{MMSE} combiner) at the \ac{IAB} receiver and multi-user interference at transmitter (with \ac{ZF} precoder). %
The proposed scheme is evaluated through simulations by varying channel estimation errors and \ac{RF} insertion losses, i.e., losses in phase shifters, power dividers and power combiners in \ac{mmWave}. %
Simulation results showed the spectral efficiency gains of \ac{FD} versus \ac{HD} \ac{IAB} transceivers as well as showed that the proposed scheme experiences less \ac{RF} insertion loss than the fully-connected hybrid beamforming. %
However, it is important to highlight that the actual implementation of \ac{FD} technology in \ac{IAB} nodes is still very challenging especially due to self-interference issues.
\deleted{However, the subarray-based scheme is more sensitive to channel estimation errors when compared to the fully-connected array.} %
In~\cite{Zhang2021_3}, a further study of the same authors from \cite{Zhang2021_2}, they assumed a fiber Bragg grating-based analog canceler that is placed before \ac{RF} precoding and after \ac{RF} combiner. %
Reduced \ac{BER} were observed in the simulation results when compared to baseline schemes. %


\goodbreak
\section{Mobile IAB}\label{SEC:Survey_Mobile_IAB}
Wireless backhaul may allow the deployment of mobile cells, i.e., \ac{mIAB}, which can hopefully provide uninterrupted cellular services for moving \acp{UE}, e.g., passengers in trains and buses. %

The concept of mobile cells is not new. %
\Acp{MRN} were already studied in the past with a similar purpose. %
It has even been addressed by \ac{3GPP} Release~12 in~\cite{3gpp.36.836}. %
The focus of~\cite{3gpp.36.836} was on high speed trains with known trajectory. %
In the considered topology, outer antennas of access devices, e.g., installed on the top of the train, provide wireless backhaul connection via the \acp{eNB} mounted along the railway, while inner antennas installed inside are responsible for providing wireless connectivity to the passengers. %
Important to highlight that these devices support multi-\ac{RAT} functionalities. %
It means that, while the backhaul link works over \ac{LTE}, the access link can be deployed over other technologies, e.g., \ac{Wi-Fi}. %

A recent technical report from \ac{3GPP} \cite{3gpp.22.839} presents a study on several use cases and requirements for \ac{5G} networks for mobile \ac{BS} relays mounted on vehicles, i.e., \acp{VMR}. %
The study item is part of initial studies for Release 18. %
The use cases cover many aspects including service provision for both onboard \acp{UE} and \acp{UE} in the vicinity of the vehicle as well as seamless connectivity in different scenarios involving mobility of \acp{UE} and relays. %

On the one hand, considering access and backhaul links operating either in different frequency spectrum or even in different technologies simplifies two of the main challenges related to wireless backhaul and moving cells: managing the resources between access and backhaul links and dealing with the dynamic interference between moving cells and crossed fixed cells. %
On the other hand, using the same technology and frequency spectrum for both access and backhaul links may result in efficient system operation and optimized use of the scarce and expensive frequency spectrum, respectively. %

Next subsection presents a literature review of the main topics related to \ac{mIAB}. %

\goodbreak
\subsection{Literature Review}\label{SUBSEC:Survey_Mobile_IAB_Literature}

\begin{figure*}
	\centering
	\includegraphics[width=\textwidth]{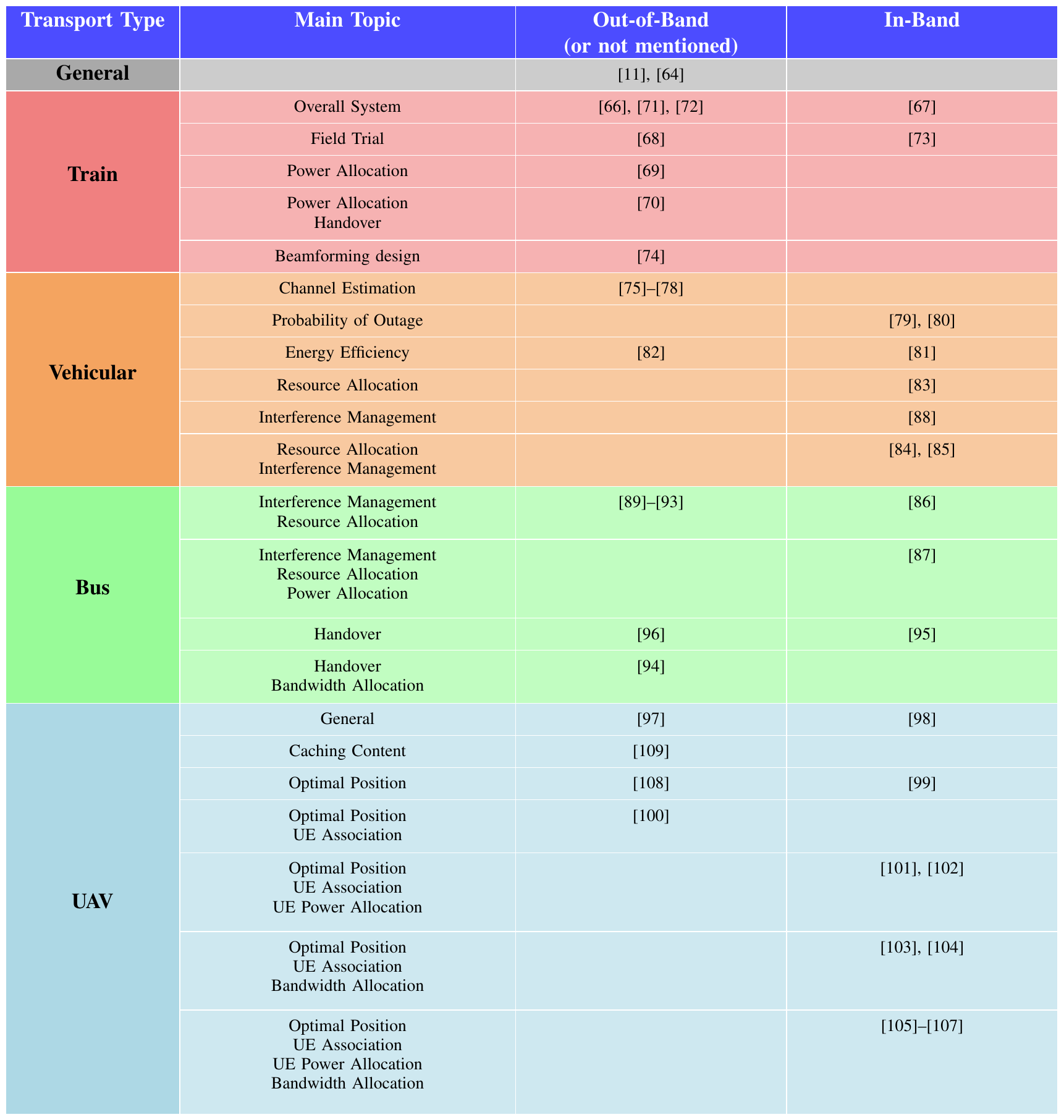}
	\caption{Taxonomy for mobile IAB: types of mobility and studied topics.}\label{FIG:SEC_MOB_IAB_TAXONOMY_1}
\end{figure*}

Works related to \ac{mIAB}, moving relays and moving cells mobile \ac{IAB} can be classified according to different criteria. %
\FigRef{FIG:SEC_MOB_IAB_TAXONOMY_1} presents the classification adopted in the present survey. %
The works are first grouped according to the type of mobility, e.g., train, bus, \ac{UAV}, etc.. %
This is due to fact that, on the one hand, works considering similar type of mobility usually consider similar problems, e.g., \ac{UAV} positioning, and take advantage of specific characteristics of each type of mobility, e.g., previously known mobile trajectory of buses. %
On the other hand, works considering different types of mobility usually consider different environments, e.g., while bus related works try to improve in-vehicle \acp{UE} \ac{QoS} taking into account the presence of out-of-vehicle \acp{UE}, train related works only consider the presence of in-vehicle \acp{UE}. %
For each type of mobility, we present the most recurrent topics and solutions present in the literature. %

\goodbreak
\subsubsection{Train}\label{SUBSUBSEC:Survey_Mobile_IAB_Literature_train}

We start presenting works related to mobile cells deployed at trains. %
This is a typical use case. %
Due to the high speed and shielding effect of the trains, it is challenging to continuously serve onboard \acp{UE} from outside \acp{BS}. %
With the purpose of reducing the penetration losses caused by the Faraday cage characteristics of a \ac{HSR}, state-of-the-art works, e.g.~\cite{Wu2016, Ai2020, Dat2019, Xu2021, Lu2017}, usually consider that the \acp{MRN} are deployed on top of the carriages and connected to in-cabin wireless \acp{AP}. %
The \acp{MRN} communicate with \acp{BS}, playing the role of forwarding data between the passengers on the train and the broadband wireless networks. %
Furthermore, since the trains follow a predefined track and pass through uninhabited areas, a common solution is to deploy the external \acp{BS} along the railway. %

Works~\cite{Wu2016, Ai2020, Plaza2017} provide a review on communication on \ac{HSR}. %
The authors of~\cite{Wu2016}, summarized key challenges and provided a review of techniques used to address the listed challenges. %
They focused primarily on physical layer operations covering the rapidly time-varying channel modeling, estimation of fast time-varying fading channels, Doppler diversity transmissions and non-coherent detections. %
However, they also presented additional discussions on higher layer operations, e.g, handover management, control/user-plane decoupling and network architecture. %
One of the discussed architectures is the one deploying relays on the roof connected to access points installed inside the train through wired links, such as optical fiber. %
The \acp{MRN} act as intermediate nodes between inside \acp{UE} and external \acp{BS}. %
The authors stated that, since the \acp{MRN} are installed outside of the trains, they eliminate the penetration loss. %
In addition, they usually have \ac{LOS} with the \acp{eNB}. %
Furthermore, it was highlighted that \acp{MRN} can also solve the problem of a high number of simultaneous \ac{HO} requests, since the \ac{HO} is performed only between an \ac{MRN} and an \ac{eNB}, instead of between multiple \acp{UE} and a \ac{eNB}. %
That paper did not present either results or more details about the deployment of the \acp{MRN}, e.g., central frequency carrier. %

In~\cite{Ai2020}, current trends of wireless communications for smart railways were presented. %
Besides of state-of-the-art, key challenges and issues, the authors also proposed a network slicing architecture for a \ac{5G}-based \ac{HSR}. %
Similar to~\cite{Wu2016}, in the proposed architecture, \acp{AP} on the train aggregated the data from different \acp{UE}, and relayed them to external \acp{BS} through a wireless backhaul. %
The authors presented simulation results comparing a \ac{MIMO}-\ac{BS}-only architecture with the wireless backhaul architecture. %
However, the simulated scenario was simplified. %
The train traveled a small distance, less then 10 times its length, and there was only one \ac{MIMO}-\ac{BS} positioned in the middle of the path. %
They assumed a \ac{ZF} algorithm to reduce the interference caused by inter-\ac{UE} and inter-\ac{AP} communications. %
The simulations showed that the data rate of the mobile wireless backhaul architecture was higher than that of the \ac{MIMO}-\ac{BS}-only architecture. %
They highlighted that, for velocities larger than 360 km/h, it is a challeging task to deploy \acp{AP} on the train or along the track to make a seamlessly connected heterogeneous network and establish a \ac{UE} association model for service provisioning. %
They also remarked that although \ac{mmWave} shows higher directionality, and thus can be used with more directional antennas making it easier to use \ac{mmWave} simultaneously for mobile access and backhaul, the specifics of the orthogonalization between these two types of transmissions still need to be further studied. %

In~\cite{Choi2021}, the authors advocate that \ac{IAB} nodes are potential candidates to improve service continuity. %
For this, they used real measurement from Gangneung Line in a simulated environment to advocate that \ac{IAB} nodes should be installed in regions where the signal of \ac{LTE} and \ac{LTE-R} are weak in order to avoid \acp{RLF}.
According to their results, \ac{IAB} can improve service reliability by $10\%$, in average. %
The work~\cite{Plaza2017} describes the main characteristics and requirements for critical and non-critical communications in railways that must be addressed by \ac{5G}. %
The authors highlighted the shielding effect of the train to radio signals, due to its metallic construction. %
They mentioned that this effect has been quantified in $20$-\SI{30}{dB} and therefore, it is necessary to use moving cells for internal network provisioning of reliable communications. %
They also highlighted the importance of using satellites since high speed lines can travel across large uninhabited areas. Therefore, the trains must be equipped with antennas with mechanical or electrical control of azimuth and elevation. %

In~\cite{Ichinose2018, Dat2019} results from real measurements are presented. %
In~\cite{Ichinose2018}, the authors presented an overview of field trials to assess technical and operating conditions of a \ac{5G} mobile communication system capable of supporting high-speed \SI{2}{Gbps} throughput for \ac{HSR}. %
The trials were performed on February 19--23, 2018  in Japan. %
Two \ac{5G} \acp{eNB} were deployed operating at center frequency of \SI{27.875}{GHz}, bandwidth of \SI{700}{MHz}, and $2$ antenna units each, with $96$ antenna elements in each unit. %
Concerning the mobile equipment, it was mounted inside the train car, on the front windshield in front of the driver's seat with $64$ antenna elements. %
The network was split into two sub-networks: an inter-server network between a master server at the \ac{BS} and a cache server inside the train, and an end-user network between the cache server and the \acp{UE} inside the train. %
The \acp{UE} were downloading multiple 4K/8K video files. %
A maximum throughput of \SI{2.08}{Gbps} was verified when the train was moving at a speed of \SI{90}{km/h}. %

In~\cite{Dat2019}, the authors deployed an experimental setup for a proof-of-concept. %
In order to implement a handover-free communication of up to several tens of kilometers, they installed \acp{RAU} along the railway track and connected them to a central station. %
The \acp{RAU} were responsible for providing a wireless backhaul to a mobile \ac{BS} installed on a train. %
The backhaul was deployed at W-band ($75$-\SI{100}{GHz}). %
\Acp{UE} on the train communicated with the \ac{BS} inside the train. %
They assumed that the train information, including location, velocity, and list of identified \acp{BS}, was available at the train operation center. %
This information was utilized to control and switch radio cells instead of using control signals as in cellular networks. %
They exploited transmission diversity with the Alamouti coding scheme and cell coordination between adjacent cells for interference cancellation. %
They successfully transmitted approximately \SI{20}{Gbps} and \SI{10}{Gbps} in \ac{DL} and \ac{UL}, respectively. %

Since the trains' internal network is almost considered as an independent network, state-of-the art works usually focus on the backhaul link between the train antenna and external \acp{BS}, as is the case in~\cite{Xu2021, Lu2017, Gao2020}. %
More specifically, in~\cite{Xu2021}, the authors investigated the power allocation problem between \acp{BS} and \acp{MRN} deployed on trains with the objective of maximizing system achievable sum rate. %
The proposed power allocation algorithm was based on \ac{MADRDPG}, which is capable of learning power decisions from past experience instead from an accurate mathematical model. %
Two constraints were considered. %
First, the power allocated to each \ac{MRN} should not only be non-negative, but also be no more than the maximum transmitting power. %
Second, the minimum \ac{QoS} requirements of each \ac{MRN} should be met. %
They modeled the \ac{MADRDPG} as: the state was represented by the \acp{MRN} channel gain of current time step, beamforming design of the current time step, and achievable rate and emitting power of the last time step; the action indicated the power allocated to each MR;\@ and the reward was equal to the system achievable sum rate when the \acp{MRN} \ac{QoS} contraints were satisfied, otherwise, it was zero. %
Simulation results showed that the proposed solution outperformed other state-of-the-art solutions based on machine learning. %

As in~\cite{Xu2021}, the authors of~\cite{Lu2017} also investigated the power allocation but focused on \ac{HO} performance of \ac{HSR} with \ac{MRN}. %
This work also adopted the two-hop communication architecture, where the train passengers were served through \acp{MRN}. %
With such a two-hop transmission architecture, all passengers within the train were treated as a ``big user'' of the \ac{BS}. %
This avoided multiple simultaneously \ac{HO} requests. %
Another addressed challenge related to \ac{HO} was the impact of fading on~\acp{HO}. %
Aiming at making the signal from different \acp{BS} more distinguishable, which results in better \ac{HO} performance, the authors adjusted the power in the \acp{BS}' overlapping region by one parameter without knowing the precise \ac{CSI}. %
More precisely, the overlapping region was divided into two sub-regions, one closer to the \ac{sBS} and other closer to the \ac{tBS}. %
Within the first subregion, the authors proposed to increase the transmit power of the \ac{sBS} and decrease the transmit power of the \ac{tBS}, while, within the second subregion, they proposed the opposite, i.e., increase the transmit power of the \ac{tBS} and decrease the transmit power of the \ac{sBS}. %
For this, they assumed perfect knowledge of trains' position. %
Mathematical analysis and simulation results showed that the proposed solution decreased the \ac{HO} failure occurrence probability. %

Finally, in~\cite{Gao2020}, the authors investigated efficient \ac{HBF} design for train-to-ground communications in \ac{mmWave}. %
They developed a two-stage \ac{HBF} algorithm in blockage-free scenarios. %
In the first stage, the \ac{MMSE} method was adopted for optimal \ac{HBF} design with low complexity and fast convergence. %
In the second stage, the orthogonal matching pursuit method was utilized to approximately recover the analog and digital beamformers. %
Furthermore, in blocked scenarios, they designed an anti-blockage scheme by adaptively invoking the proposed \ac{HBF} algorithm, which can deal with random blockages. %
Simulation results showed that the proposed solution outperformed state-of-the-art solutions in terms of sum rate for different speed values and blockage probability. %

\goodbreak
\subsubsection{Vehicular}\label{SUBSUBSEC:Survey_Mobile_IAB_Literature_vehicular}

Regarding \ac{mIAB} deployed in vehicles, before presenting works that addressed system level problems, e.g., resource allocation and interference management, we present a more fundamental article that proposed a way to deploy a set of antennas in a vehicle to allow transmissions with good signal quality to and from mobile nodes. %

The authors of~\cite{Guo2020} proposed an \ac{MRN} setup configured with two groups of antennas on the roof of a vehicle: 1) the front predictor antennas; and 2) the behind receive antennas. %
In this way, the \ac{CSI} acquired by front predictor antennas is exploited for data transmission to the behind receive antennas when they reach the same position, and the channel aging effect is minimized. %
If the receive antennas end up in the same position, it will experience the same radio environment, and the \ac{CSI} will be almost perfect. %
If the receive antennas do not reach the same point as the predictor ones, due to, e.g., the processing delay is not equal to the time needed until the receive antennas reach the same point as the predictor ones, the receive antennas may receive the data in a place different from the one where the predictor antennas sent the pilots. %
Such spatial mismatch may lead to \ac{CSI} inaccuracy, which deteriorates the system performance. %
Different methods based on transmission delay adaptation or rate adaptation are proposed to reduce the effect of spatial mismatch. %

Similar to~\cite{Guo2020}, other works from the literature have also already investigated the topic of predictor antennas from a theoretical perspective, e.g,~\cite{Guo2021, Guo2021b}, and as experimental testbeds, e.g., \cite{Huy2018}. %

The authors of~\cite{Sui2012, Sui2012b} analyzed the end-to-end outage probability and capacity at a vehicular \ac{UE} of single-hop direct transmission (baseline case), and dual-hop transmission via an \ac{MRN} as well as a fixed relay node. %
They assumed that the downlinks of different cells were synchronized in time, where the backhaul links were only active in the first time slot and access links were only active in the following slot. %
Thus, backhaul links and access links did not interfere with each other. %
By theoretical analyses and computational simulations the authors showed that in the cases of moderate to high vehicle penetration loss, \acp{MRN} deployed on top of vehicles enhanced \ac{QoS} of the vehicular \acp{UE} compared to the scenarios with only direct transmission and with fixed relay nodes. %

In~\cite{Sui2013}, the authors of~\cite{Sui2012, Sui2012b} extended their study and compared the required average transmit power in similar scenarios under an outage probability constraint. %
They showed that dual-hop transmission via an \ac{MRN} can also significantly reduce the power consumption of the network. %

As the authors of~\cite{Sui2013}, the authors of~\cite{Gui2021} also addressed the topic of energy efficiency. %
They proposed a Q-learning-based scheme to stabilize the energy efficiency of \ac{IAB} nodes backhaul. %
For this, each \ac{mIAB} node, i.e., vehicle-mounted \acp{AP}, periodically reported its location and velocity to an access controller. %
The access controller was responsible for predicting the most suitable \ac{BS} or \ac{D2D} node to serve the \ac{IAB} nodes in the future and for configuring them to be connected to the predicted \ac{BS} or \ac{D2D} node. %
The \ac{IAB} nodes, in their turn, were responsible for selecting a suitable transmission beamwidth and transmission power that kept the energy efficiency above a given threshold. %
The selection was performed by using a Q-learning based solution. %

The resource allocation for macrocell \acp{UE} and the backhaul of the existing and newly arrived moving small cells deployed in vehicles was studied in~\cite{Jangsher2017}. %
The authors formulated the problem of maximizing the downlink data rate of the newly arrived moving small cells in the macrocell such that the data rate of the macrocell \acp{UE} and existing moving small cells \acp{UE} is protected. %
They proposed an adjustable-power-based resource allocation to allocate RBs and power in the macrocell. %

The problem of resource sharing between moving cells and access links was also studied in~\cite{Jaffry2019a}. %
More precisely, they considered a typical mobile cell scenario was considered, where in-vehicle cellular \acp{UE} are served by an in-vehicle antenna over the access link. %
A separate external antenna connects the mobile cells to the nearest \ac{eNB} over the backhaul link, and to the neighboring mobile cells over the sidehaul link. %
The authors proposed algorithms to enable resource sharing between a mobile-cell, access link, sidehaul link and conventional cellular users taking into account interference caused by the wireless backaul of mobile cells. %
The algorithms' performance was compared with the optimal, yet time-consuming, brute-force method and presented good results based on Monte-Carlo simulations. %

The work~\cite{Jaffry2019a} was extended in~\cite{Jaffry2019b}. %
The authors proposed new resource sharing and user scheduling algorithms that jointly aimed at ensuring that the access link shared the sub-channel either with backhaul link or with the out-of-vehicles macrocell \acp{UE}. %
They took into account results from~\cite{Jaffry2018} and~\cite{Jaffry2018b}, such as the minimization of interference to backhaul link and provisioning of high \ac{QoS} for in-vehicle users by considering the use of directional access link antenna even at low transmit power. %
The authors also exploited the vehicular penetration loss along with self-interference-coordination to further reduce interference between the resource sharing links. %
They pointed as interesting direction for future research the deployment of the backhaul link for mobile cells through \acp{UAV}. %

\deleted{Regarding specifically the}\added{The} problem of mutual interference between macro and moving small cell \acp{UE}\deleted{, it} is highlighted in~\cite{Jaziri2016}. %
Its authors studied the impact of deploying moving small cells in a Manhattan grid scenario in the presence of stationary traffic hotspot inside a macro cell. %
Simulation results showed that deploying moving small cells to offload traffic in the congested macro cell can be beneficial. %
However, when the small cells were moving far away from the traffic hotspot, the system performance was degraded compared to a network composed of only macro cells due to the high mutual interference between macro and small cell users. %

\goodbreak
\subsubsection{Bus}\label{SUBSUBSEC:Survey_Mobile_IAB_Literature_bus}

In the context of \ac{mIAB} considering buses, the majority of the works available in the literature focus on how to allocate resources in order to avoid the interference between access link of inside and outside \acp{UE}. %
A common adopted strategy is to split \acp{UE} into inside and outside \acp{UE} and allocate them in different parts of the spectrum. %
Another topic also addressed in the literature is related to \ac{HO} from two perspectives: how to avoid multiple \ac{HO} due to the mobility of the \acp{UE} and how to optimize the \ac{HO} process considering that multiple \acp{UE} might change its \ac{sBS} at the same time since they are moving together. %
In the following, we briefly discuss some papers. %

The authors of~\cite{Mastrosimone2015, Mastrosimone2015b, Mastrosimone2017} considered a macrocell with several public vehicles embedded with \acp{MFAP}. %
They proposed to deploy \ac{mmWave} communications at \SI{60}{\GHz} inside vehicles for access links while the backhaul link of \acp{MFAP} and the direct link between an external \ac{UE} and the macro cell use \ac{LTE}. %
The authors analyzed three critical cases: a) the bus was near the \ac{eNB}; b) two buses were close together; and c) an outside \ac{UE} was close to a bus. %
They verified that, in case a), the \ac{eNB} was a big interferer for inside \acp{UE}, because its transmission power was higher than that of the \acp{MFAP}. %
\Acp{UE} served by the \acp{MFAP} measured a low \ac{SINR}, and were forced to disconnect from \acp{MFAP} and connect to the \ac{eNB}. %
This resulted in a high number of simultaneous \acp{HO}. %
However, as the bus moved, as soon as the signal of the \acp{MFAP} was acceptable, \acp{UE} had to re-engage to the \acp{MFAP}. %
In the case c), they showed that an outside \ac{UE} in the proximity of a bus, especially when it was away from the \ac{sBS}, could connect to the \acp{MFAP}, but its connection only lasted for a few minutes, depending on the speed of the bus. %

Similar to~\cite{Mastrosimone2015, Mastrosimone2015b, Mastrosimone2017}, the authors of~\cite{Chae2012} considered moving femtocell deployed inside buses. %
The authors proposed a dynamic frequency allocation scheme that allocates the same frequency band for inside \acp{UE} even if they are in different buses, while the outside \acp{UE} are allocated in a different frequency band. %

In~\cite{Jangsher2014}, the authors also considered small cells deployed on the buses to serve the passengers. %
They studied the interference between moving small cells and proposed a \ac{PGRA} algorithm considering that the path of the moving small cell is fixed but it moves at non-uniform speed. %

Different of~\cite{Mastrosimone2015, Mastrosimone2015b, Mastrosimone2017, Chae2012, Jangsher2014}, \cite{Jaffry2018} proposed a model to allow the downlink backhaul sub-channels to be shared by in-vehicle downlink access link. %
This model is based on vehicular penetration effect and \ac{LOS} communication. %
While vehicular penetration depends upon the material and construction properties of the transport vehicle, the \ac{LOS} communication can be enabled using directional antennas. %
This paper also advocated that construction parameters and antenna positioning inside vehicles should be considered for future transport vehicles to increase the spectral efficiency for cellular network. %

The authors of~\cite{Jaffry2018}, extended their study in~\cite{Jaffry2018b}. %
They presented new results regarding the exploitation of vehicular penetration effect and \ac{LOS} communication to allow the downlink backhaul sub-channels to be shared by in-vehicle downlink access link transmission. %
Furthermore, a technique for access link power control was also proposed to reduce the interference to the backhaul link, while maintaining high link quality for in-vehicle users. %

Regarding \ac{HO} aspects of \ac{mIAB}, the authors of~\cite{Chowdhury2012}~addressed the problem of group \ac{HO} of \acp{UE} inside buses or trains. %
They proposed a resource management scheme that contains bandwidth adaptation policy and dynamic bandwidth reservation policy. %
More specifically, some instants before the moment when a bus or train is expected to arrive at a station, a given amount of resources at the station are reserved for the coming \acp{UE}. %
These resources remain reserved for a predefined interval of time. %
Simulation results showed that the \ac{HO} call dropping probability considerably reduced compared to the case where no pre-reservation was performed without much impacting the bandwidth utilization. %

The works~\cite{Raheem2014, Tayyab2020} also investigated the topic of \ac{HO} in the context of \ac{mIAB}. %
The authors of~\cite{Raheem2014} analyzed the system performance of three different scenarios: in the first, there were only macro cells, in the second one there were macro cells and fixed femtocells and in the third one there were macro cells and mobile femtocells deployed in buses. %
The \ac{LTE} technology was used for all links. %
The \ac{QoS} of cell edge \acp{UE} improved after adding fixed and mobile femtocells. %
However, mobile \acp{UE} enjoyed better performance after adding the mobile femtocells, since these cells could reach areas that fixed femtocells could not. %
A result comparing the number of \acp{HO} considering three different \ac{HO} strategies (namely, proactive, normal and reactive) was presented. %
In the presented results, the reactive strategy presented a lower number of \acp{HO} compared to the other strategies. %
However, it was not presented how other metrics were impacted by these three strategies, e.g., link failure rate and \ac{UE} throughput. %

In~\cite{Tayyab2020}, the authors considered a scenario with \acp{MRN} deployed at the top of buses and serving onboard \acp{UE}. %
They provided an analysis of the HO performance and the associated power consumption in \ac{LTE}. %
Through computational simulations, the authors compared two cases. %
The first case was without \acp{MRN} deployed on the buses, meaning that all onboard \acp{UE} performed their individual \ac{HO} procedure with macro \acp{eNB}. %
The second case was with \acp{MRN}, meaning that only the relays performed the \ac{HO} procedure to the macro \acp{eNB}. %
Their results showed that the deployment of \acp{MRN} reduced the \ac{HO} rate, the \ac{HO} failure rate, the ping-pong rate, the \ac{UE} power consumption and \ac{eNB} power consumption. %
Furthermore, the simulation results also showed that \ac{UL} transmission errors were the most dominant cause of \acp{MRN} \ac{HO} failure. %
They highlighted that when deploying the relays, the \ac{HO} procedure played an even more important role in keeping \acp{UE} connectivity.  %
With the relays there was a single point of failure, i.e., when the relay \ac{HO} failed, all \acp{UE} connected to the relay were dropped. %

\goodbreak
\subsubsection{UAV}\label{SUBSUBSEC:Survey_Mobile_IAB_Literature_UAV}

\Ac{UAV}-based \ac{MRN} will not be part of the \ac{3GPP} Release~18 discussions on \ac{mIAB}, and it is left for future uses in beyond \ac{5G}. %
However, it is still interesting to study its performance from an academic point of view. %
Compared to the previous types of mobility, e.g., trains and buses, \acp{UAV} have important differences. %
Firstly, in, e.g., buses, the \ac{MT} and \ac{DU} are installed outside and inside the buses, respectively, which gives good separation between them, while in \acp{UAV} this cannot be accomplished. %
Besides, \acp{UAV} are usually used to serve outside \acp{UE} so, for instance, the access links are also moving, but for buses the main purpose is to serve onboard \acp{UE} with stationary access links. %
More precisely, the main use-case for \acp{UAV} is coverage extension while for buses is capacity increment. %
Furthermore, a key point of \acp{UAV} is the chance for playing with their height while taking into account their energy consumption and lifetime, which are the main bottleneck for prolonged use of \acp{UAV}. %

The integration of drones into cellular networks was addressed in~\cite{Yaliniz2019}. %
Two different perspectives were presented: %
1) how wireless networks can support personal or professional use of drones (called mobile-enabled drones (MEDs)); %
and 2) how drones can support wireless network performance, i.e., boosting capacity on demand, increasing, coverage range, enhancing reliability and agility as an aerial node called  \acp{WID}.  %
Regarding \acp{WID}, different possibilities were addressed. %
One of them was the use of drones as \ac{IAB} nodes. %
According to the authors, \acp{WID} as intermediate \ac{IAB}-nodes can reduce the number of hops, and provide topology flexibility thanks to \ac{LOS} and mobility. %
More specifically, \acp{WID}' mobility enables flexibility in topology design and alleviates the problem of coverage holes by following the crowd at the cell edge. %
They highlighted that co-channel interference may be a limiting factor. %

Other advantages of the integration of \acp{UAV} were presented in~\cite{Gapeyenko2018}. %
They considered a scenario where \acp{UE} could connect to \acp{UAV} operating in \ac{mmWave}, which co-existed with \ac{5G} \ac{NR} \acp{BS}. %
Through mathematical analyses, they demonstrated that under certain speed, intensity, and service capacity, the use of \ac{UAV}-based relays enabled significant gains for the system performance. %
In particular, outage probability and outage duration in the considered scenario became notably reduced, while spectral efficiency increased substantially. %

The optimization of the \ac{UAV} 3D-position in order to improve coverage is one of the most recurrent problems in the literature related to the deployment of \acp{UAV} as relay-nodes. %
The authors of~\cite{Perez2019} considered a ray tracing software called WinProp to study coverage of \ac{UAV}-\ac{IAB} nodes operating at \ac{mmWave}. %
They considered two modes: \ac{AF} and \ac{DF}. %
Out-of-band and in-band \ac{IAB} was assumed for \ac{AF} and \ac{DF}, respectively. %
For the \ac{AF} case, the authors defined the power allocation based on the received backhaul link \ac{SINR}. %
The \acp{UAV} were positioned so as to maximize the coverage. %
In the \ac{DF} relaying mode, if the received \ac{SINR} was above a certain threshold, a \ac{UAV} forwarded its received packet. %
Trivial results were presented: \acp{UAV} improved coverage in the system compared with the case of \ac{IAB}-donor only. %

In~\cite{Kalantari2017}, besides the optimal \ac{UAV} positioning, the optimal \ac{UE}-\ac{UAV} association was also investigated. %
Their objective was to maximize the number of \acp{UE} served by the \acp{UAV}. %
The formulated problem was subject to a limited peak aggregate rate supported by the wireless backhaul link, a limited bandwidth available for access between \acp{UE} and \ac{UAV} and a \ac{UE} \ac{QoS} requirement in terms of maximum path loss that a \ac{UE} could tolerate before outage. %
Besides, the \acp{UE} could have different priorities to be served by the \acp{UAV}. %
The authors proposed a centralized solution, where for each candidate coordinate of the \ac{UAV}-\ac{BS} placement, the problem was transformed into a binary integer linear program, which was then solved through the branch-and-bound method. %

In~\cite{Fouda2018}, the authors went even further. %
In order to maximize the overall instantaneous sum-rate, they tried to optimize 3D locations of the \acp{UAV}, \ac{UE}-\ac{UAV}/\ac{gNB} association, \ac{UE}-power allocation and precoder design at backhaul links. %
The authors considered a heterogeneous scenario with a macro \ac{BS} and \acp{UAV}. %
\Ac{DL} was assumed. %
\Ac{gNB} had multiple antennas (massive \ac{MIMO}) and \acp{UAV} had one receiving antenna and multiple transmit antennas. %
Terrestrial and aerial \acp{UE} had only one antenna. %
Aerial \acp{UE} received interference from \ac{DL} signals to other aerial users, from \ac{gNB} to terrestrial \acp{UE}, and from backhaul links (\ac{gNB} to \acp{UAV}). %
\Acp{UAV} are assumed \ac{FD}. %
The following simplifying assumptions were made: \ac{ZF} canceled out the intra tier interference (perfect \ac{CSI}) and self-interference was completely eliminated at \acp{UAV}. %
The studied optimization problem was the maximization of the sum rate (instantaneous) at aerial and terrestrial \acp{UE} subject to \ac{SINR} constraints and total available power at nodes. %
One of the main conclusions of the article was that the performance gains of using \acp{UAV} was increased when \acp{UE} were distributed in multiple hotspots far from each other in contrast to the case where \acp{UE} were in a single hotspot. %

In~\cite{Fouda2019}, the authors of~\cite{Fouda2018} extended their work. %
They proposed an interference management algorithm based on \ac{PSO} to jointly optimize \ac{UE}-\ac{UAV}/\ac{BS} association, downlink power allocations and the 3D deployment of \acp{UAV} in in-band \ac{UAV}-assisted \ac{IAB} networks. %
The \acp{UAV} were \ac{FD}-capable. %
Two \acp{UAV} spatial configurations were compared. %
In the first one, the \acp{UAV} were configured as a single \ac{DAA} to jointly serve ground \acp{UE} that were spatially distributed in a single hotspot. %
In that configuration, the \acp{UAV} did not interfere with each other. %
In the second configuration, the \acp{UAV} served the \acp{UE} in an independent way. %
The numerical results showed that, on one hand, when the ground cellular \acp{UE} were normally distributed into multiple bad coverage areas, the spatial configuration of distributed \acp{UAV} outperformed that of the \ac{DAA} in terms of the overall network sum rate. %
On the other hand, when the ground cellular \acp{UE} were concentrated in a single bad coverage area, the spatial configuration of the \ac{DAA} outperformed that of distributed \acp{UAV}. %
Numerical results confirmed that \acp{UAV} can be used for coverage improvement and capacity boosting in the \ac{IAB} cellular networks. %

In~\cite{Qiu2019}, besides of \ac{UAV} placement and \ac{UE}-\ac{UAV} association, the authors also investigated resource allocation in an in-band \ac{IAB} scenario. %
Orthogonal allocation was considered to avoid interference. %
Besides, in order to mitigate interference between macro-\acp{BS} and \acp{UAV},
reverse \ac{TDD} between them is adopted, e.g., while the macro-\acp{BS} are in the \ac{DL} mode with respect to their serving \acp{UE}, \acp{UAV} are in \ac{UL} mode. %
The number of \ac{UL} and \ac{DL} slots are the same. %
\Ac{HD} is assumed at \acp{UAV}. %
\Acp{UAV} can be seen as macro \acp{UE}. %
From the ground \acp{UE}' point of view, only interference from other \acp{UAV} is modeled. %
More specifically, the authors formulated an optimization problem to maximize the \ac{UE} utility (log of data rate) subject to fronthaul and backhaul resources and user assignment. %
The problem was complex and a suboptimal approach was adopted. %
First, it was assumed fixed \ac{UAV} placement and \ac{UE} association and then the resource allocation was optimally solved. %
After that, user association and \ac{UAV} placement were solved. %
Results showed the advantage of \acp{UAV} in \acp{HetNet} and the relevance of the joint optimization considering fronthaul and backhaul constraints. %

The problem of optimal \ac{UAV} 3D positioning, \ac{UE}-\ac{UAV} association and resource allocation was also investigated in~\cite{Kalantari2017b}. %
A scenario with one macro \ac{BS} and multiple \acp{DBS} with in-band wireless backhaul was considered. %
As in~\cite{Qiu2019}, to reduce interference, reverse \ac{TDD} was employed and access and backhaul links used orthogonal frequency channels. %
Besides, as the transmit power of \ac{DBS} \acp{UE} was lower than that of a macro \ac{BS}, the interference from the \ac{DBS} \acp{UE} at macro \ac{BS} \acp{UE} was assumed negligible. %
It was assumed that \ac{URLLC} \acp{UE} could only be connected to the macro \ac{BS} in order to decrease latency, while \ac{eMBB} \acp{UE} could be either associated directly to the macro \ac{BS} or to a \ac{DBS}. %
The formulated problem was decomposed into two sub-problems and solved as: 1) \ac{UE}-\ac{BS} association and wireless backhaul bandwidth allocation were found through a primal decomposition method; and %
2) the locations of \acp{DBS} were updated using the \ac{PSO} algorithm. %
The authors concluded that clustered \acp{UE} could be covered with \acp{DBS} at lower altitudes, which can increase \ac{SNR}. %
Furthermore, they also concluded that it was necessary to develop efficient interference cancellation methods for dense deployments of \acp{DBS}, since preventing overlapped areas among \acp{DBS} prevented \acp{DBS} to be deployed in its optimal locations. %

In~\cite{Zhang2019, Zhang2019b, Zhang2019c} the authors added another dimension to the list of optimized parameters of~\cite{Kalantari2017b}. %
They also optimized the \ac{UE} power allocation in addition to \ac{UAV} 3D positioning, \ac{UE}-\ac{UAV}/\ac{BS} association and resource allocation. %
In~\cite{Zhang2019, Zhang2019b, Zhang2019c} the \acp{DBS} had in-band \ac{FD} capabilities, while the \acp{UE} were \ac{HD} capable. %
The focus was on \ac{DL}. %
However, while in~\cite{Zhang2019}, the objective was to minimize the number of \acp{DBS} while maximizing the overall transmission rate, in~\cite{Zhang2019b, Zhang2019c}, the objective was only to maximize the total system throughput. %
The \acp{UE} could be either associated directly to a macro \ac{BS} or to a \acp{DBS}. %
Neither different \acp{UE} nor different \acp{DBS} were scheduled at the same resources. %
However, it was assumed that \acp{DBS} could cancel part of their self-interference and due to this the backhaul link of a \ac{DBS} reused the frequency spectrum of its access link. %
Due to the complexity of the problem (NP-hard), it was decomposed into two sub-problems: %
1) \acp{DBS} placement problem (vertical and horizontal positions); and %
2) Joint \ac{UE}-\ac{BS} association, power and bandwidth assignment problem. %
Approximation algorithms were proposed to solve the sub-problems. %
In~\cite{Zhang2019b}, simulation results demonstrated gains of deploying \acp{DBS} compared to a system without them. %
In~\cite{Zhang2019c}, the authors identified, for the considered scenario, a \ac{DBS} altitude for which \ac{DBS} position lower than that the path loss was dominated by the \ac{NLOS} component, and \ac{DBS} position higher than that the path loss was dominated by the \ac{LOS} component. %

Different of the previously mentioned works, \cite{Tafintsev2020a} considered \ac{UL} performance and a constraint on the \ac{UAV} buffer. %
The authors of~\cite{Tafintsev2020a} also provided a short review of \ac{3GPP} \acp{TR} and decisions about \ac{UAV} and \ac{IAB}. %
They listed attractive \ac{UAV} applications for mobile operators, e.g, wireless connection in disaster-affected regions, and network densification during temporary mass events. %
\Ac{PSO} was used to optimize \ac{UAV} positioning. %
The authors concluded that as the number of \ac{UAV}-nodes increased, the performance difference between static grid deployment of \acp{UAV} and dynamic positioning decreased, since a grid based deployment densely spanned across the entire service area, and all of the \acp{UE} were always within coverage of at least one \ac{UAV}-\ac{BS}. %
However, they highlighted that with an extremely high number of \ac{UAV}-\ac{BS}, the interference started to play an important role for the achievable throughput by not allowing the algorithm to position the \ac{UAV}-\ac{BS} as close as needed. %
Furthermore, they also showed that ideal backhaul scheme significantly overestimates the actual throughput, since all of the traffic generated by the \acp{UE} in the \ac{UAV}-\ac{BS} coverage area is usually assumed to be delivered to the \acp{AP} successfully. %

Besides the applications listed in~\cite{Tafintsev2020a}, another use of \acp{UAV} was presented in~\cite{Khuwaja2020}. %
In that paper, \acp{UAV} and ground small cells were deployed to cache\footnote{The use of cache has not been standardized in 3GPP.} content close to the \acp{UE} in order to reduce traffic congestion in backhaul. %
\Ac{UE} association probability for the \acp{UAV} and the ground small-cells was derived using stochastic geometry. %
Besides, the successful content delivery probability was also derived by considering both the intercell and intracell interference. %
The successful content delivery performance has been improved by $26.6\%$ on average by caching popular content in the \acp{UAV}. %


\goodbreak
\section{Performance Evaluation}\label{SEC:Perf_Eval}

This section presents a performance comparison between a scenario with \ac{mIAB} and two benchmark scenarios: one with only macro \acp{gNB}, called here as \textit{only macros} scenario, and other with macro and pico \acp{gNB} fiber-connected to the \ac{CN}, called here as \textit{macros-picos} scenario. %
The details concerning the considered simulation modeling are presented in \SecRef{SUBSEC:Simulation_Assumptions} and the results are discussed in \SecRef{SUBSEC:Numerical_Results}. %

\goodbreak
\subsection{Simulation Assumptions}\label{SUBSEC:Simulation_Assumptions}

\begin{figure}[!t]
	\centering
	\includegraphics[width=0.9\columnwidth]{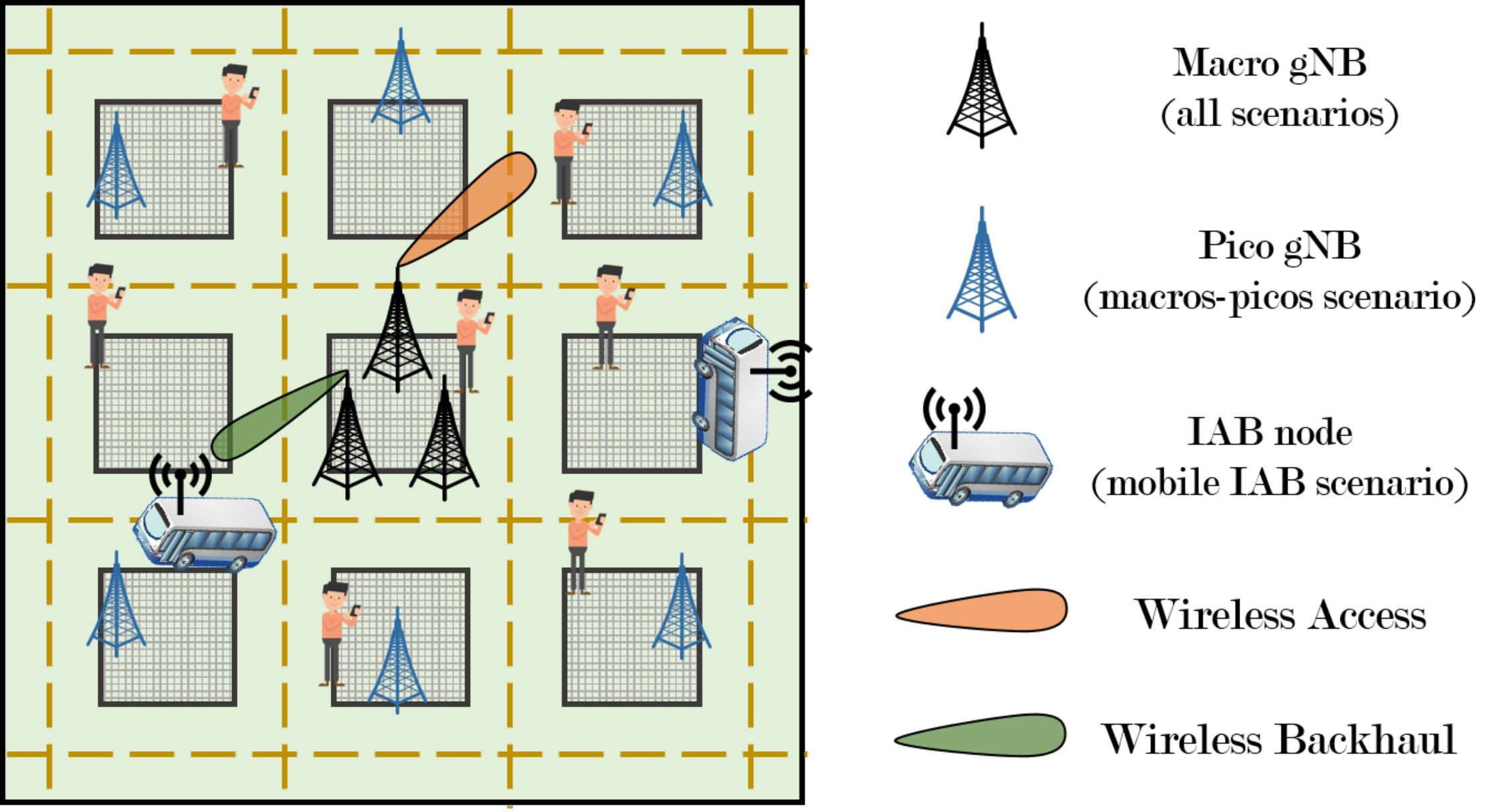} 
	\caption{Simplified Madrid grid. }\label{FIG:Madrid-grid}
\end{figure}

It was considered a simplified version of the Madrid grid~\cite{METIS:D6.1:2013} as in~\cite{Sui2015}. %
As illustrated in~\FigRef{FIG:Madrid-grid}, in this scenario, there were nine square blocks, with dimension of \SI{120}{m}~$\times$~\SI{120}{m}. %
The blocks were surrounded by \SI{3}{\meter}~wide sidewalks and separated of each other by \SI{14}{m}~wide streets with four lanes, two in each direction. %
In the central block there were 3 not co-located macro \acp{gNB} deployed as in~\FigRef{FIG:Madrid-grid}. %
Pedestrians and buses were randomly placed in the sidewalks and in the streets, respectively. %
In the intersections, they had a probability of \SI{60}{\%} to continue straight ahead and \SI{40}{\%} to turn left or right with equal probability to each side. %
The pedestrians walked in the sidewalks and were allowed to cross the roads only in the intersections. %
Passengers were randomly located inside the buses at any available seat. %
During the simulation, their position relative to their bus did not change. %
In the \ac{mIAB} scenario, \ac{mIAB} nodes were deployed at the buses. %
The \ac{DU} and \ac{MT} were placed at the back of the buses; however, the \ac{DU} was inside and the \ac{MT} was outside at the roof. %
Important to mention that an \ac{mIAB} node could not connect to another \ac{mIAB} node. %
It could only be served by an \ac{IAB} donor, i.e., a \acp{gNB}. %
Other parameters related to the entities present in the system, e.g., \ac{IAB} donors, \ac{mIAB} nodes, pedestrians and passengers, are presented in~\TabRef{TABLE:Entities-characteristics}. %

\begin{figure}[!t]
	\centering
	\includegraphics[width=0.85\columnwidth]{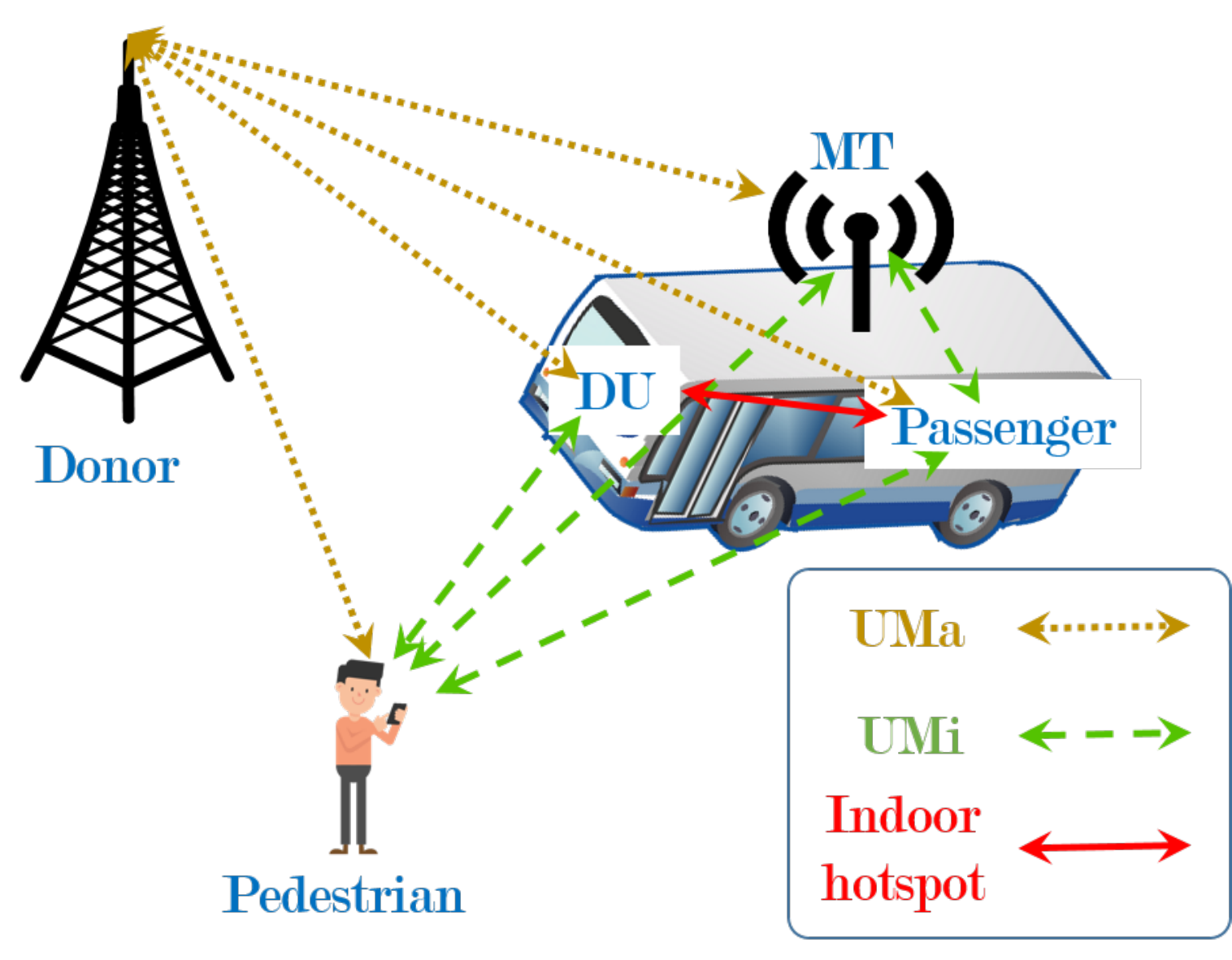}
	\caption{Channel types.}\label{FIG:Channel-types}
\end{figure}

\begin{table*}[!t]
	\centering
	\caption{Entities characteristics.}
	\label{TABLE:Entities-characteristics}
	\begin{tabular}{llllll}
		\toprule
		\textbf{Parameter} & \textbf{\ac{IAB} Donor} & \textbf{\ac{mIAB} node - \ac{DU}} & \textbf{\ac{mIAB} node - \ac{MT}} & \textbf{Pedestrian} & \textbf{Passenger} \\
		\midrule
		Height & \SI{25}{\meter} & \SI{2.5}{\meter} & \SI{3.5}{\meter} & \SI{1.5}{\meter} & \SI{1.8}{\meter} \\
		Transmit power & \SI{35}{\decibel m} & \SI{24}{\decibel m} & \SI{24}{\decibel m} & \SI{24}{\decibel m} & \SI{24}{\decibel m} \\
		Antenna tilt & $12^{\circ}$ & $4^{\circ}$ & $0^{\circ}$ & $0^{\circ}$ & $0^{\circ}$ \\
		Antenna array & URA $8\times 8$ & URA $8\times 8$ & ULA $64$ & Single antenna & Single antenna \\
		Antenna element pattern & \ac{3GPP} 3D~\cite{3gpp.38.901} & \ac{3GPP} 3D~\cite{3gpp.38.901} & Omni & Omni & Omni \\
		Max. antenna element gain & \SI{8}{\decibel i} & \SI{8}{\decibel i} & \SI{0}{\decibel i} & \SI{0}{\decibel i} & \SI{0}{\decibel i} \\
		Speed & \SI{0}{km/h} & \SI{40}{km/h} & \SI{40}{km/h} & \SI{3}{km/h} & \SI{40}{km/h} \\
		\bottomrule
	\end{tabular}
\end{table*}

\begin{table*}[!t] 
	\centering
	\small
	\caption{\acs{TDD} scheme adopted in only macros and macros-picos scenarios.}
	\label{TABLE:macro-pico-TDD}
	\resizebox{\textwidth}{!}{
		\begin{tabular}{l|llllllllll|ll|l}
			\toprule
			Slot                & 1  & 2      & 3  & 4  & 5  & 6  & 7      & 8  & 9  & 10 & DL usage & UL usage & Total usage \\ 
			\midrule
			Macro and Pico gNBs & DL & S (DL) & UL & UL & UL & DL & S (DL) & UL & UL & DL & 50\%     & 50\%     & 100\% \\
			\bottomrule
		\end{tabular}
	}
\end{table*}

\begin{table*}[!t] 
	\centering
	\small
	\caption{\acs{TDD} scheme adopted in the \ac{mIAB} scenario.}
	\label{TABLE:IAB-TDD}
	\resizebox{\textwidth}{!}{
		\begin{tabular}{l|llllllllll|cc|c}
			\toprule
			Slot      & 1             & 2             & 3             & 4             & 5             & 6             & 7             & 8             & 9             & 10            & DL usage & UL usage & Total usage \\  
			\midrule
			IAB donor & DL            & UL            & $\varnothing$ & DL            & $\varnothing$ & UL            & DL            & $\varnothing$ & UL            & DL            & 40\%     & 30\%     & 70\%  \\
			Backhaul  & DL            & $\varnothing$ & UL            & DL            & UL            & $\varnothing$ & $\varnothing$ & UL            & $\varnothing$ & DL            & 30\%     & 30\%     & 60\%  \\
			IAB node  & $\varnothing$ & UL            & DL            & $\varnothing$ & DL            & UL            & DL            & DL            & UL            & $\varnothing$ & 40\%     & 30\%     & 70\% \\
			\bottomrule
		\end{tabular}
	}
\end{table*}

The simulations were conducted at \SI{28}{\GHz} with \SI{50}{\MHz} of system bandwidth. %
\deleted{It was considered a slot duration of 0.25~ms and subcarrier spacing of 60~kHz. }%
The channel link between the entities present in the simulations was modeled using the 5G-SToRM channel model described in~\cite{Pessoa2019}, where~\cite{Pessoa2019} implements its channel model according to~\cite{3gpp.38.901}. %
The adopted channel model from~\cite{Pessoa2019,3gpp.38.901} is spatially and time consistent. %
It considers a distance-dependent path-loss, a lognormal shadowing component and a small-scale fading. %
Moreover, as illustrated in~\FigRef{FIG:Channel-types}, all the links with the donor, i.e., donor-pedestrian, donor-\ac{MT}, donor-\ac{DU} and donor-passenger, were modeled as \ac{UMa}. %
All links involving the \ac{MT}, except the link donor-\ac{MT}, were modeled as \ac{UMi}, as well as the links \ac{DU}-pedestrian and pedestrian-passenger. %
The link \ac{DU}-passenger was modeled as an indoor hotspot one. %
The bus body had a penetration loss of \SI{40.1}{dB}~\cite{Mastrosimone2017}. %
Thus, the links with one entity inside the bus and other outside, such as donor-\ac{DU}, \ac{DU}-pedestrian, \ac{MT}-passenger and pedestrian-passenger, suffered this penetration loss. %
Furthermore, the link between a \ac{DU} of a bus and a passenger of another bus suffered twice this penetration loss. %
Besides, all the links that crossed the bus body were considered \ac{NLOS}. %
The other links could be either in \ac{LOS} or \ac{NLOS} with a transitional state between them as described in~\cite{3gpp.38.901}. %


Regarding resource scheduling, a slot was the minimum scheduling unit. %
In the time domain, a slot consisted of $14$~\ac{OFDM} symbols with total time duration of \SI{0.25}{ms}, while in the frequency domain it consisted of one \ac{RB} with subcarrier spacing of \SI{60}{kHz}, where one \ac{RB} corresponded to $12$~consecutive subcarriers. %
	
In the frequency domain, the \ac{RR} scheduler was adopted to schedule the \acp{RB}. %
The \ac{RR} iteratively allocated the \acp{RB}, scheduling in a given \ac{RB} the \ac{UE} bearer waiting the longest time in the queue. %
In the access links, \ac{MISO} transmissions were performed with a matched filter used as precoder. %
On the other hand, in the backhaul, \ac{MIMO} transmissions were performed. %
To take advantage of the multiple antennas at both sides of a backhaul link, the transmission within an \ac{RB} was multiplexed into streams. %
For each stream, it was computed a pair of precoder/decoder using the \ac{SVD} of the channel between an \ac{IAB} donor and the \ac{MT} part of an \ac{mIAB} node. %
At most the eight streams with the highest singular values were scheduled, subject to having an estimated \ac{MCS} higher than one. %
	
In the time domain, \ac{TDD} schemes were used to separate transmission and reception of the signals, i.e., in a given slot, data could traverses a link in just one direction, either \ac{UL} or \ac{DL}. %
More specifically, the \ac{TDD} scheme presented in~\TabRef{TABLE:macro-pico-TDD} was considered in the only macros and macros-picos scenarios, while the \ac{TDD} scheme in~\TabRef{TABLE:IAB-TDD} was considered in the \ac{IAB} scenario\added{, where backhaul \ac{UL} means that the \ac{mIAB} \ac{MT} part transmits data to the \ac{IAB} donor, while backhaul \ac{DL} means that the \ac{mIAB} \ac{MT} part receives data from the \ac{IAB} donor}. %
The \ac{TDD} scheme in~\TabRef{TABLE:macro-pico-TDD} is standardized by \ac{3GPP} in \cite{3gpp.36.211b}, while the one in ~\TabRef{TABLE:IAB-TDD} is based on the one proposed in \cite{Ronkainen2020} for three hops (in our case we are considering just two hops). %
Remark that on the one hand, in the scheme of~\TabRef{TABLE:macro-pico-TDD}, macro and pico \acp{gNB} are active in $100\%$ of the time, transmitting either in \ac{DL} or in \ac{UL} with equal time usage for both \ac{DL} and \ac{UL}. %
On the other hand, in the scheme of~\TabRef{TABLE:IAB-TDD}, \ac{IAB} donor and \ac{mIAB} node are active only in $70\%$ of the time and with different time usage between \ac{DL} and \ac{UL}. %
One of the reasons for this is the introduction of silence intervals to reduce interference in the system. %
Specifically regarding the buses, to avoid self-interference, they could not simultaneously receive and transmit data, e.g., receive in the \ac{MT} while transmitting in the \ac{DU}. %

As a measure of signal strength, the \acp{UE} and the \ac{mIAB} nodes were measuring the \ac{RSRP} of candidate serving cells, where the \ac{RSRP} is the linear average over the power contributions (in Watts) of the resource elements confined within a \ac{SSB} transmitted by a candidate serving cell, as defined in~\cite{3gpp.38.215}. %
The topology adaptation was based on the highest measured \ac{RSRP}. %
\deleted{Concerning the link adaptation, it was adopted the}\added{The} \ac{CQI}/\ac{MCS} mapping curves standardized in~\cite{3gpp.38.214} \added{were used for link adaptation }with a target \ac{BLER} of \SI{10}{\%}. %
It was also considered an outer loop strategy to avoid the increasing of the \ac{BLER}. %
According to this strategy, on the one hand, when a transmission error occurred, the estimated \ac{SINR} used for the \ac{CQI}/\ac{MCS} mapping in the link adaptation was subtracted by a back-off value of \SI{1}{dB}. %
On the other hand, when a transmission occurred without error, the estimated \ac{SINR} had its value added by \SI{0.1}{dB}. %

As previously mentioned, three scenarios were considered. %
The first with only three macro \acp{gNB} deployed at the central block, called here as only macros scenario. %
The second with the three macro \acp{gNB} (acting as \ac{IAB} donors) plus \ac{mIAB} nodes deployed in the buses, called here as \ac{mIAB} scenario. %
The third with the three macro \acp{gNB} plus $6$ pico \acp{gNB} deployed at the positions indicated in~\FigRef{FIG:Madrid-grid} (at the vertices of a hexagon) called here as macros-picos scenario. %
In these three scenarios there were $6$ buses (as many as pico \acp{gNB}) and $72$ \acp{UE}, i.e., passengers plus pedestrians. %
The \acp{UE}' traffic, in both \ac{DL} and \ac{UL}, was modeled as \ac{CBR} flows with packet-inter-arrival time equal to four slots. %
In the simulations, it was analyzed the impact of the packet size and the impact of the percentage of passengers considering the $72$ \acp{UE} in the system. %
For this, nine possible combinations of packet size and percentage of passengers were tested, where the packet size could be equal to $1024$, $2048$ or $3072$ bits and the percentage of passengers considering the $72$ \acp{UE} in the system could be equal to $25\%$, $50\%$ or $75\%$ meaning that in each bus there were $3$, $6$ or $9$ passengers. %
\TabRef{TABLE:Simul_Param} summarizes the main simulation parameters. %

In the following, the simulation results are presented. %
Firstly, we compare the downlink performance of passengers and pedestrians in the three considered scenarios, focusing on their throughput, latency and \ac{MCS} usage. %
Secondly, we analyze their performance on the uplink, highlighting the main differences compared to the downlink. %
Then, we focus on the wireless backhaul quality in the \ac{mIAB} scenario. %
Finally, we present the statistics of links served by the \ac{IAB} donors. %
More specifically, we analyze the percentage of time that the \ac{IAB} donors served backhaul and access links. %

\begin{table}[!t]
	\centering
	\small
	\setlength{\tabcolsep}{1ex}
	\caption{Simulation parameters.}
	\label{TABLE:Simul_Param}
	\begin{tabularx}{0.99\columnwidth}{>{\raggedright\arraybackslash}X>{\raggedright\arraybackslash}X}
		\toprule
		\textbf{Parameter} & \textbf{Value} \\
		\midrule
		Layout & Simplified Madrid grid~\cite{METIS:D6.1:2013,Sui2015} \\
		Carrier frequency & \SI{28}{\GHz}\\
		System bandwidth & \SI{50}{\MHz}\\
		Subcarrier spacing & \SI{60}{\kHz}\\
		Number of subcarriers per \acs{RB} &  $12$\\
		Number of \acsp{RB} & $66$\\
		Slot duration & \SI{0.25}{\ms} \\
		OFDM symbols per slot & $14$ \\
		Channel generation procedure & As described in~\cite[Fig.7.6.4-1]{3gpp.38.901}\\
		Path loss  & Eqs. in~\cite[Table 7.4.1-1]{3gpp.38.901}\\
		Fast fading & As described in~\cite[Sec.7.5]{3gpp.38.901} and \cite[Table7.5-6]{3gpp.38.901} \\
		AWGN power per subcarrier & \SI{-174}{dBm}\\
		Noise figure &  \SI{9}{\decibel}\\
		Mobility model & Pedestrian and Vehicular \cite{3gpp.37.885}\\
		Number of buses & $6$ \\
		Passengers~$+$~pedestrians & 72 \\
		Percentage of passengers & \{$25\%, 50\%, 75\%$\} \\
		Number of passengers per bus & \{$3, 6, 9$\} \\
		\acs{CBR} packet size & \{$1024, 2048, 3072$\} bits \\
		\acs{CBR} packet inter-arrival time & \SI{4}{slots}\\
		\bottomrule
	\end{tabularx}
\end{table}

\goodbreak
\subsection{Simulation Results}\label{SUBSEC:Numerical_Results}

\subsubsection{Downlink}\label{SUBSUBSEC:Simulations-DL}

\subsubsection*{Downlink Throughput}\label{SUBSUBSEC:Simulations-DL-throughput}
\begin{figure}[!t]
	\centering
	\subfloat[75\% of passengers.]{%
		\includegraphics[width=\columnwidth]{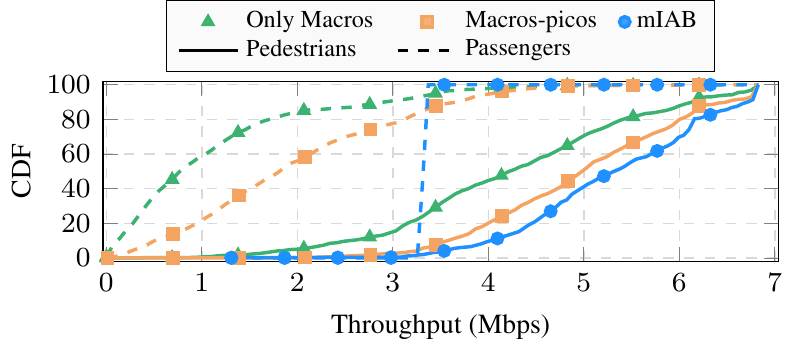}
		\label{FIG:Simulation-results-DL-throughput-impact-of-perc-passengers-75}
	}
	
	\subfloat[50\% of passengers.]{%
		\includegraphics[width=\columnwidth]{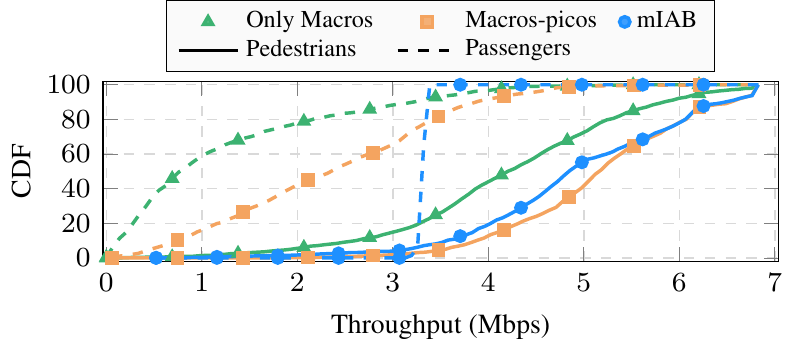}
		\label{FIG:Simulation-results-DL-throughput-impact-of-perc-passengers-50}
	}	
	
	\subfloat[25\% of passengers.]{%
		\includegraphics[width=\columnwidth]{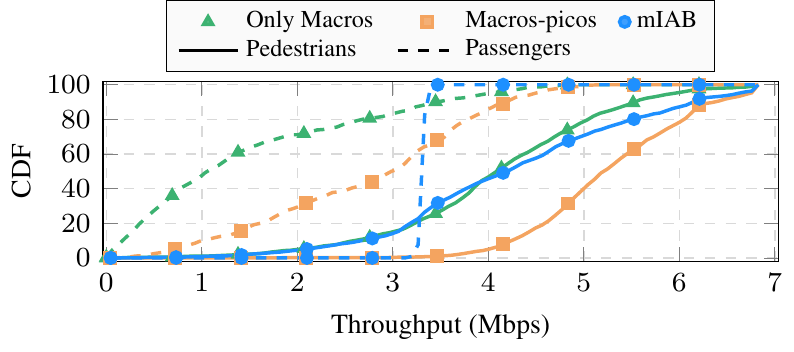}
		\label{FIG:Simulation-results-DL-throughput-impact-of-perc-passengers-25}
	}			
	\caption{Downlink throughput – Impact of percentage of passengers, considering packet size equal to $3072$~bits.}\label{FIG:Simulation-results-DL-throughput-impact-of-perc-passengers}	
\end{figure}

Firstly, let's analyze the impact of introducing \ac{mIAB} on the \ac{DL} throughput of passengers and pedestrians. %
Figure~\ref{FIG:Simulation-results-DL-throughput-impact-of-perc-passengers}	presents the \ac{DL} throughput of passengers (dashed lines) and pedestrians (solid lines) considering a packet size of $3072$~bits and for three different percentages of passengers. %

First, notice that in the three figures of~\FigRef{FIG:Simulation-results-DL-throughput-impact-of-perc-passengers}, the \ac{DL} throughput of passengers in the \ac{mIAB} scenario (blue dashed lines with circles as markers) was similar for almost $100$\% of them. %
This was due to their vicinity to the \ac{mIAB} node, i.e., they were next to the \ac{DU} antennas inside the buses. %
Besides, compared to the only macros scenario (green dashed lines with triangles as markers), in the \ac{mIAB} scenario, the \ac{DL} throughput of the majority of the passengers was remarkably improved. %
One could argue that in the \ac{mIAB} scenario there were more \acp{gNB}, i.e., the ones in the buses, than in the only macros scenario. %
However, even when compared to the macros-picos scenario (orange dashed lines with squares as markers), where the number of \acp{gNB} was equal to the one in the \ac{mIAB} scenario, the passenger \ac{DL} throughput of \ac{mIAB} scenario was higher than the one for macros-picos scenario in most of the percentiles. %
Besides, while the \ac{DL} throughput of all passengers in the \ac{mIAB} scenario was higher than \SI{3}{Mbps}, more than half of the passengers in the macros-picos scenario were not able to achieve this throughput. %
The main reason for this is the fact that signals serving passengers in the only macros and macros-picos scenarios suffered a penetration loss of \SI{40.1}{dB}, due to the signal crossing the bus body, while, in the \ac{mIAB} scenario, the passengers were connected to antennas deployed inside the bus that did not suffer this loss. %

Regarding the pedestrians \ac{DL} throughput in the \ac{mIAB} scenario, we can see in~\FigRef{FIG:Simulation-results-DL-throughput-impact-of-perc-passengers}, that it varied with the percentage of passengers in the system, thus, with the number of pedestrians. %
Decreasing the percentage of passengers, i.e., increasing the number of pedestrians, their \ac{DL} throughput decreased. %
One of the reasons was the adopted fixed \ac{IAB} \ac{TDD} scheme, \TabRef{TABLE:IAB-TDD}, which partitioned the slots between backhaul and \acp{UE} directly connected to \ac{IAB} donors without taking into account the real necessity of the system. %
Thus, it could be better if one adopted a dynamic resource scheduling taking into account the load of backhaul and \acp{UE} directly connected to the \ac{IAB} donors. %
Besides, compared to the other scenarios, in the \ac{IAB} scenario, the \ac{IAB} donors, which served most part of the pedestrians, were active in the \ac{DL} only in $40\%$ of the time, while in the other scenarios, according to the adopted \ac{TDD} scheme in~\TabRef{TABLE:macro-pico-TDD}, the macro \acp{gNB} were active in the \ac{DL} during $50\%$ of the time. %
However, we remark that even if the pedestrians \ac{DL} throughput decreased in the \ac{mIAB} scenario compared to the others, it still achieved values high enough to keep a good connectivity. %

\begin{figure}[!t]
	\centering
	\subfloat[Packet size = 3072 bits.]{%
		\includegraphics[width=\columnwidth]{figs/Throughput-DL-50-passenger-3072-packet.pdf}
		\label{FIG:Simulation-results-DL-throughput-impact-of-packet-size-3072}
	}	
	
	\subfloat[Packet size = 2048 bits.]{%
		\includegraphics[width=\columnwidth]{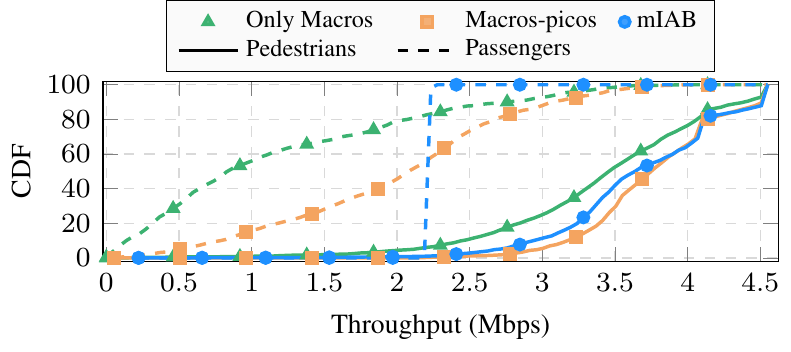}
		\label{FIG:Simulation-results-DL-throughput-impact-of-packet-size-2048}
	}	
	
	\subfloat[Packet size = 1024 bits.]{%
		\includegraphics[width=\columnwidth]{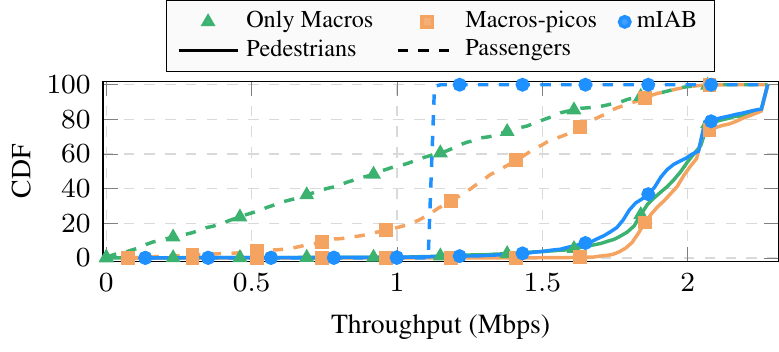}
		\label{FIG:Simulation-results-DL-throughput-impact-of-packet-size-1024}
	}			
	\caption{Downlink throughput – Impact of packet size, considering percentage of passengers equal to~$50\%$.}\label{FIG:Simulation-results-DL-throughput-impact-of-packet-size}	
\end{figure}

Figure~\ref{FIG:Simulation-results-DL-throughput-impact-of-packet-size} also presents the \ac{DL} throughput of passengers (dashed lines) and pedestrians (solid lines), but for three different packet sizes and considering $50\%$ of the \acp{UE} in the system as passengers. %
Similar to the analyses of~\FigRef{FIG:Simulation-results-DL-throughput-impact-of-perc-passengers}, we can conclude here that the deployment of \ac{mIAB} nodes inside the buses improved the passengers \ac{DL} throughput, while keeping the pedestrians with a good \ac{DL} throughput. %
Furthermore, we highlight that for higher percentages of passengers in the system and higher packet size, i.e., more loaded systems, the gains of using \ac{mIAB} relative to the only macros and macros-picos scenarios were higher. %

\goodbreak
\subsubsection*{Downlink Data Transmission Quality}\label{SUBSUBSEC:Simulations-DL-Data-Transmission-Quality}

\begin{figure}[!t]
	\subfloat[Only macros scenario.]{%
		\includegraphics[width=\columnwidth]{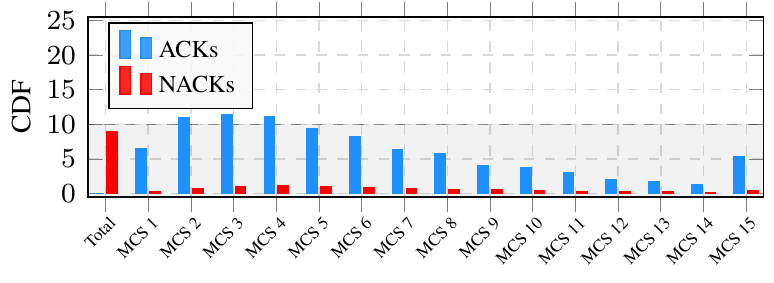}
		\label{FIG:Simulation-results-DL-MCS-passengers-macro-only}
	}
	
	\subfloat[\ac{mIAB} scenario.]{%
		\includegraphics[width=\columnwidth]{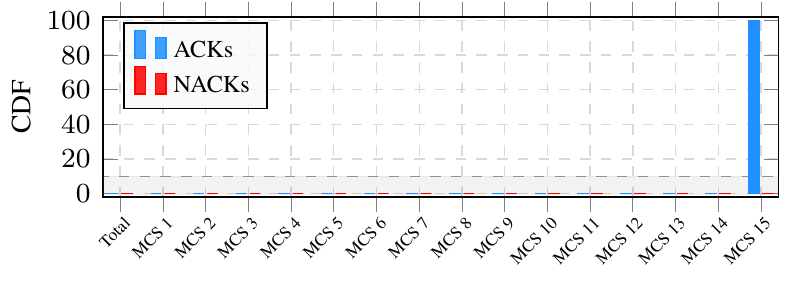}
		\label{FIG:Simulation-results-DL-MCS-passengers-IAB}	
	}			
	\caption{Histogram of passengers' \ac{MCS} usage in \ac{DL} in a scenario with $50\%$ of passengers and packet size of $3072$~bits.}\label{FIG:Simulation-results-DL-MCS-passengers}	
\end{figure}

Figures~\ref{FIG:Simulation-results-DL-MCS-passengers-macro-only} and \ref{FIG:Simulation-results-DL-MCS-passengers-IAB} present the passengers' \ac{MCS} usage in  \ac{DL} for the only macros and \ac{mIAB} scenarios, respectively for the configuration with $50\%$ of passengers in the system and packet size of $3072$~bits. %
The red bars represent the percentage of transmissions with error and the blue bars represent the percentage of successful transmissions. %
First, we remark that \FigRef{FIG:Simulation-results-DL-MCS-passengers-IAB} does not refer to the passengers end-to-end link quality (\ac{IAB} donor - passengers), it only refers to the quality of the link between the \ac{mIAB} nodes and their served passengers. %
The quality of the link between the \ac{IAB} donors and the \ac{mIAB} nodes will be analyzed in~\SecRef{SUBSUBSEC:Simulations-backhaul}. %

On the one hand, notice that in \FigRef{FIG:Simulation-results-DL-MCS-passengers-macro-only}, in the only macros scenario, the passengers received data mainly with lower \acp{MCS}. %
This was mainly due to the bus penetration loss, which degraded the link between passenger and macro \acp{gNB}. %
On the other hand, notice that in \FigRef{FIG:Simulation-results-DL-MCS-passengers-IAB}, in the \ac{mIAB} scenario, the passengers received data from the \ac{mIAB} node with the highest \ac{MCS} and with almost zero transmission error. %
This is a benefit of deploying the \ac{DU} part of the \ac{mIAB} node inside the buses. %

\begin{figure}[!t]
	\centering
	\subfloat[Only macros scenario.]{%
		\includegraphics[width=\columnwidth]{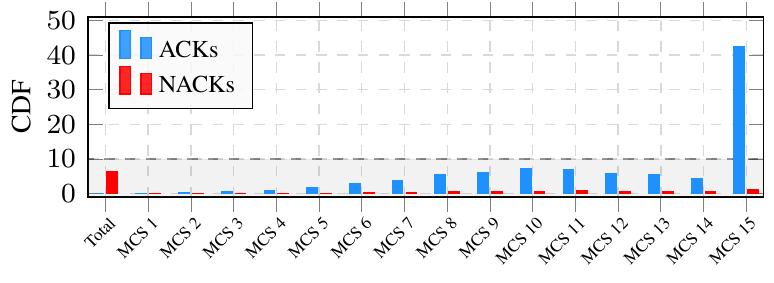}
		\label{FIG:Simulation-results-DL-MCS-pedestrians-macro-only}
	}	
	
	\subfloat[\ac{mIAB} scenario - pedestrians connected to an \ac{IAB} donor.]{%
		\includegraphics[width=\columnwidth]{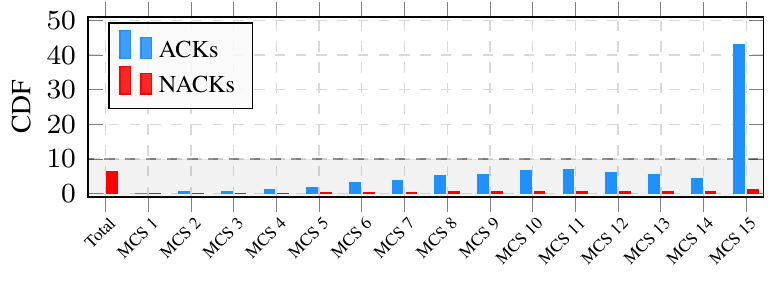}
		\label{FIG:Simulation-results-DL-MCS-pedestrians-IAB-serving-Donor}
	}	
	
	\subfloat[\ac{mIAB} scenario - pedestrians connected to an \ac{mIAB} node.]{%
		\includegraphics[width=\columnwidth]{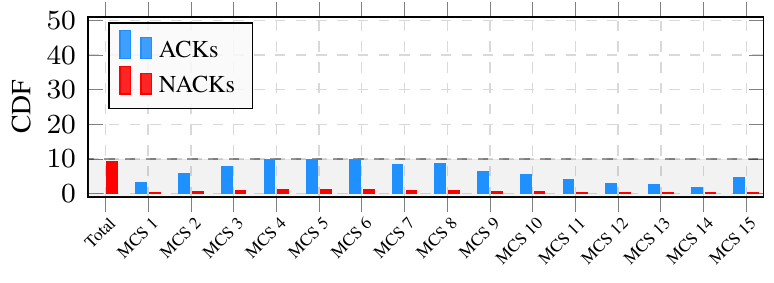}
		\label{FIG:Simulation-results-DL-MCS-pedestrians-IAB-serving-Node}
	}			
	\caption{Histogram of pedestrians' \ac{MCS} usage in \ac{DL} in a scenario with $50\%$ of passengers and packet size of $3072$~bits.}\label{FIG:Simulation-results-DL-MCS-pedestrians}
\end{figure}

The pedestrians' \ac{MCS} usage in \ac{DL} is presented in~\FigRef{FIG:Simulation-results-DL-MCS-pedestrians}. %
Figure~\ref{FIG:Simulation-results-DL-MCS-pedestrians-macro-only} is related to the only macros scenario, while \FigRef{FIG:Simulation-results-DL-MCS-pedestrians-IAB-serving-Donor} concerns pedestrians connected to an \ac{IAB} donor in the \ac{mIAB} scenario and \FigRef{FIG:Simulation-results-DL-MCS-pedestrians-IAB-serving-Node} concerns pedestrians connected to an \ac{mIAB} node also in the \ac{mIAB} scenario. %
First, notice that the histograms in \FigRef{FIG:Simulation-results-DL-MCS-pedestrians-macro-only} and \FigRef{FIG:Simulation-results-DL-MCS-pedestrians-IAB-serving-Donor} are similar with small differences, meaning that the \ac{SINR} of a pedestrian connected to a macro \ac{gNB} in the only macros scenario and the \ac{SINR} of a pedestrian connected to an \ac{IAB} donor in the \ac{mIAB} scenario were similar. %
Furthermore, considering that the pedestrian signal strength was also similar in both cases, since the macro \ac{gNB} and the \ac{IAB} donor had similar characteristics, we conclude that the interference in both cases was also similar. %

\begin{figure}
	\centering
	\includegraphics[width=\columnwidth]{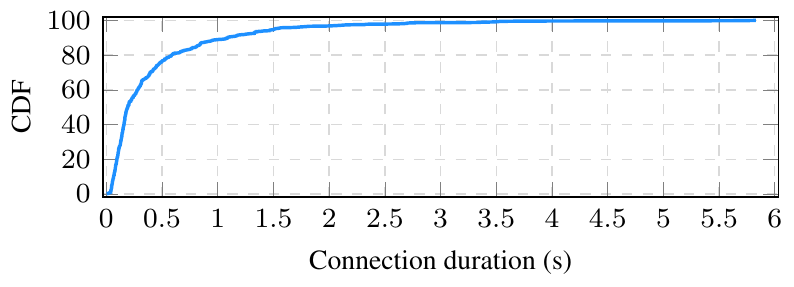}
	\caption{\ac{CDF} of connection duration between a pedestrian and a \ac{mIAB} node.}\label{FIG:IAB-Simulation-results-pedestrians-time-connected2Bus}
\end{figure}

In the \ac{mIAB} scenario, $15\%$ of the pedestrians were served by an \ac{mIAB} node for at least a couple of \acp{TTI}. %
Particularly, their connection lasted less than \SI{1}{\s} in $90\%$ of the cases, as illustrated in~\FigRef{FIG:IAB-Simulation-results-pedestrians-time-connected2Bus}, which is not good in practice. %
They connected to a \ac{mIAB} node due to a higher \ac{RSRP}. %
However, when they were receiving data from the \ac{DU} part of the \ac{mIAB} node, the \ac{MT} part was transmitting upstreaming data to the \ac{IAB} donor in the backhaul, as described in \TabRef{TABLE:IAB-TDD} slots $3$, $5$ and $8$. %
Thus, the \ac{MT} part of the \ac{mIAB} node caused a high interference to the pedestrians connected to an \ac{mIAB} node, since both were active at the same time and both were outside the bus, meaning that the interfering link did not suffer attenuation due to the crossing of the bus body. %
This inference caused low values of \ac{SINR} and so the usage of low \acp{MCS} as shown in~\FigRef{FIG:Simulation-results-DL-MCS-pedestrians-IAB-serving-Node}. %
Important to remark that this interference was not a problem for the passengers, since they were inside the bus and the signal from the \ac{MT} part suffered from the bus body penetration loss. %

Based on these results, one can conclude that pedestrians should avoid connecting to an \ac{mIAB} node, unless the signal between them is strong enough to compensate the interference from the \ac{MT} part and the connection is expected to last longer than a given threshold. %
For this, one could consider an admission policy to allow a \ac{UE} to connect to a \ac{mIAB} cell. %
The admission criteria could be: a minimum measured \ac{RSRP} value; a maximum measured interference level (\ac{RSRQ} instead of \ac{RSRP}); the relative \ac{UE} and bus geographical position in a given time interval; etc.. %

\begin{figure}[!t]
	\centering
	\subfloat[Only macros scenario.]{%
		\includegraphics[width=\columnwidth]{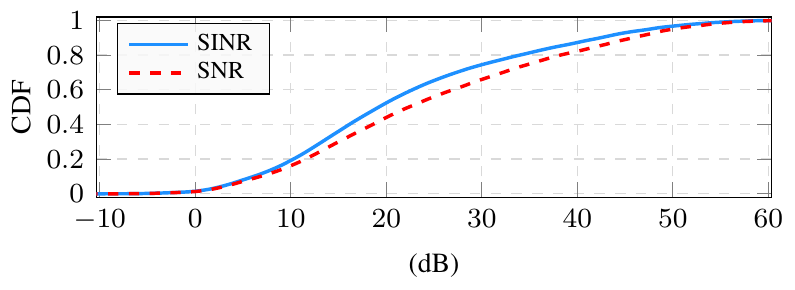}
		\label{FIG:SINR-SNR-pedestrians-macro-only-macro-DL}
	}	
	
	\subfloat[\ac{mIAB} scenario - pedestrians connected to an \ac{IAB} donor.]{%
		\includegraphics[width=\columnwidth]{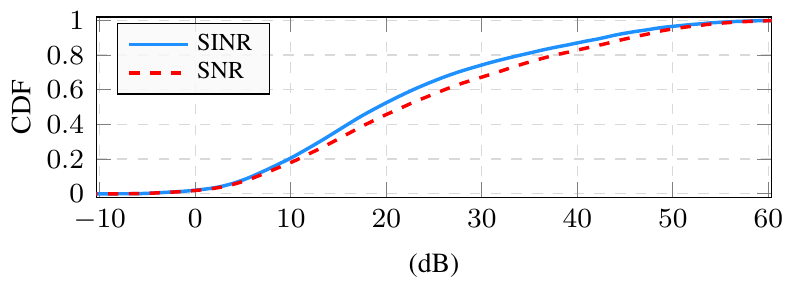}
		\label{FIG:SINR-SNR-pedestrians-iab-macro-DL}
	}			
	\caption{\ac{SINR} and \ac{SNR} of pedestrians in a scenario with $50\%$ of passengers and packet size of $3072$~bits.}
	\label{FIG:Simulation-results-DL-SINR-SNR-pedestrians}
\end{figure}

\added{Figures~\ref{FIG:SINR-SNR-pedestrians-macro-only-macro-DL} and~\ref{FIG:SINR-SNR-pedestrians-iab-macro-DL} present the \ac{CDF} of \ac{SINR} and \ac{SNR} in the \ac{DL} of pedestrians connected to a macro \ac{gNB} in the only macros and \ac{mIAB} scenarios, respectively. %
	Comparing these two figures, notice that the difference between the \ac{SINR} and \ac{SNR} curves, i.e., the interference, is similar (the difference neither increased nor decreased when changing the scenario). %
	Thus, we can conclude that the deployment of \ac{mIAB} nodes may not impact the interference that the pedestrians connected to a macro \ac{gNB} suffered. %
	This is an important result, since the possible dynamic interference caused by \ac{mIAB} nodes in the \acp{UE} served by \ac{IAB} donors is one of the concerns regarding the deployment of \ac{mIAB} cells. %
	The interference avoidance was obtained due to the adopted \ac{TDD} scheme (\TabRef{TABLE:IAB-TDD}), which did not allow transmissions in the access links of \ac{mIAB} nodes when \ac{IAB} donors were operating in \ac{DL} in their access links. %
	The topic of exploiting a \ac{TDD} scheme for interference handling in \ac{mIAB} networks is addressed in~\cite{Monteiro2022}. }%

\begin{figure}[!t]
	\centering
	\subfloat[Only macros scenario.]{%
		\includegraphics[width=\columnwidth]{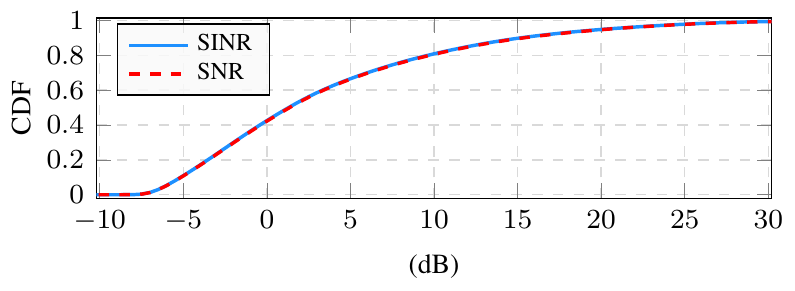}
		\label{FIG:SINR-SNR-passengers-macro-only-macro-DL}
	}	
	
	\subfloat[\ac{mIAB} scenario - passengers connected to a \ac{mIAB}.]{%
		\includegraphics[width=\columnwidth]{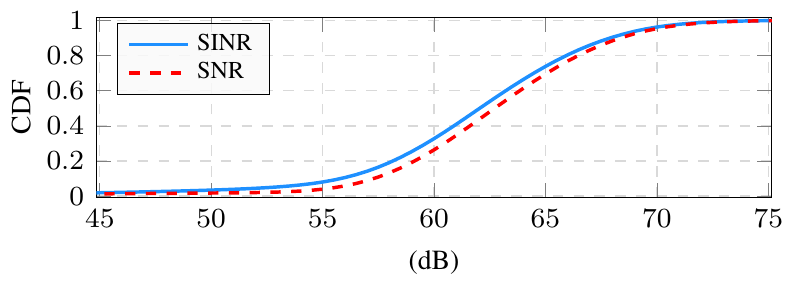}
		\label{FIG:SINR-SNR-passengers-iab-iab-DL}
	}			
	\caption{\ac{SINR} and \ac{SNR} of passengers in a scenario with $50\%$ of passengers and packet size of $3072$~bits.}
	\label{FIG:Simulation-results-DL-SINR-SNR-passengers}
\end{figure}

\added{Figures~\ref{FIG:SINR-SNR-passengers-macro-only-macro-DL} and~\ref{FIG:SINR-SNR-passengers-iab-iab-DL} present the \ac{CDF} of \ac{SINR} and \ac{SNR} in the \ac{DL} of passengers connected to a macro \ac{gNB} in the only macros scenario and connected to an \ac{mIAB} node in the \ac{mIAB} scenario, respectively. %
	In \FigRef{FIG:SINR-SNR-passengers-macro-only-macro-DL}, the \ac{SINR} and \ac{SNR} curves are superposed, meaning that in the only macros scenario the interference was not a problem for the passengers. %
	This was a consequence of the bus body loss that filtered possible interference links coming from neighbor \acp{gNB}. %
	The drawback of the bus body loss is that it filters not only the interference but also the signals serving the passengers, as we can see by the low values of the \ac{SNR}. %
	In this scenario, $80\%$ the passengers had an \ac{SNR} lower than \SI{10}{dB}. %
	On the other hand, in the \ac{mIAB} scenario, their \ac{SNR} had a boost when deploying the \ac{mIAB} nodes in the buses (\FigRef{FIG:SINR-SNR-passengers-iab-iab-DL}). %
	In this scenario, when the backhaul (\ac{MT} part of the \ac{mIAB} nodes) was operating in the \ac{UL}, they interfered with the passengers receiving in the \ac{DL} from the \ac{DU} part of the \ac{mIAB} nodes. %
	However, as we can see in \FigRef{FIG:SINR-SNR-passengers-iab-iab-DL}, the \ac{SINR} was still high enough to allow transmissions with the highest \ac{MCS}. }%

\goodbreak
\subsubsection*{Downlink Latency}\label{SUBSUBSEC:Simulations-DL-Latency}

\begin{figure}[!t]
	\centering
	\subfloat[75\% of passengers.]{%
		\includegraphics[width=\columnwidth]{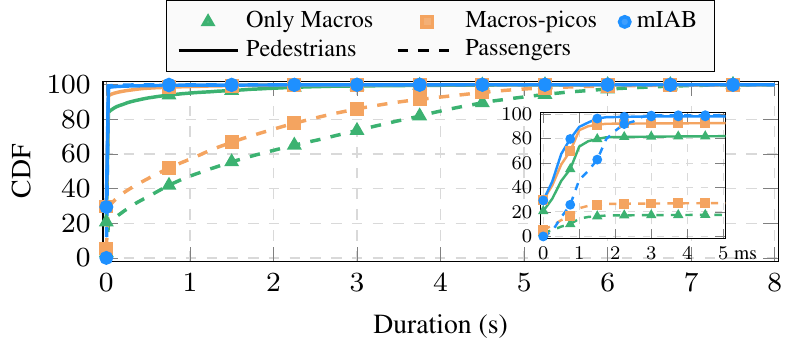}
		\label{FIG:Simulation-results-DL-latency-impact-of-perc-passengers-75}
	}	
	
	\subfloat[50\% of passengers.]{%
		\includegraphics[width=\columnwidth]{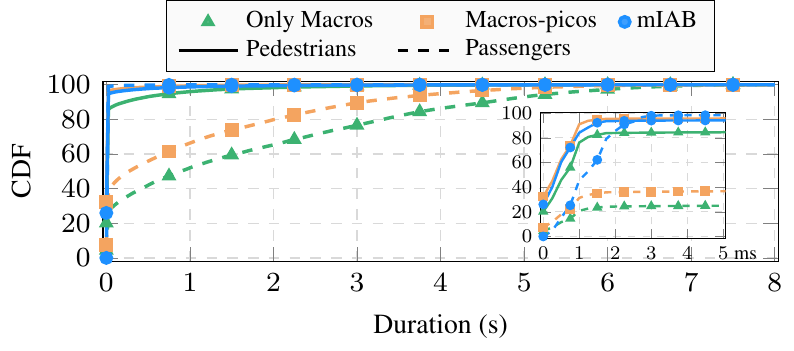}
		\label{FIG:Simulation-results-DL-latency-impact-of-perc-passengers-50}
	}
	
	\subfloat[25\% of passengers.]{%
		\includegraphics[width=\columnwidth]{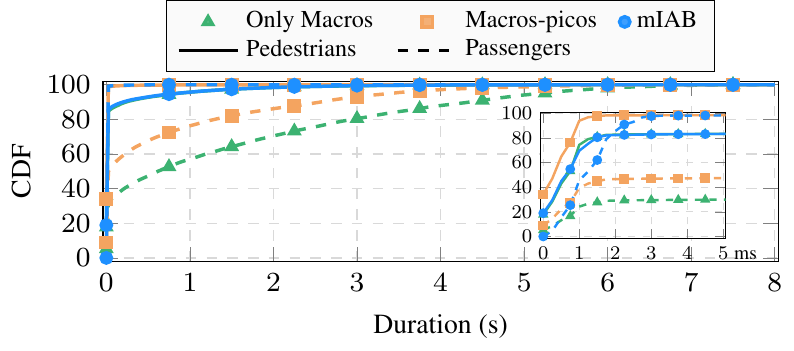}
		\label{FIG:Simulation-results-DL-latency-impact-of-perc-passengers-25}
	}			
	\caption{Downlink latency – Impact of percentage of passengers, considering packet size equal to $3072$~bits.}\label{FIG:Simulation-results-DL-latency-impact-of-perc-passengers}	
\end{figure}

\begin{figure}[!t]
	\centering
	\subfloat[Packet size = 3072 bits.]{%
		\includegraphics[width=\columnwidth]{figs/Latency-DL-50-passenger-3072-packet.pdf}
		\label{FIG:Simulation-results-DL-latency-impact-of-packet-size-3072}
	}	
	
	\subfloat[Packet size = 2048 bits.]{%
		\includegraphics[width=\columnwidth]{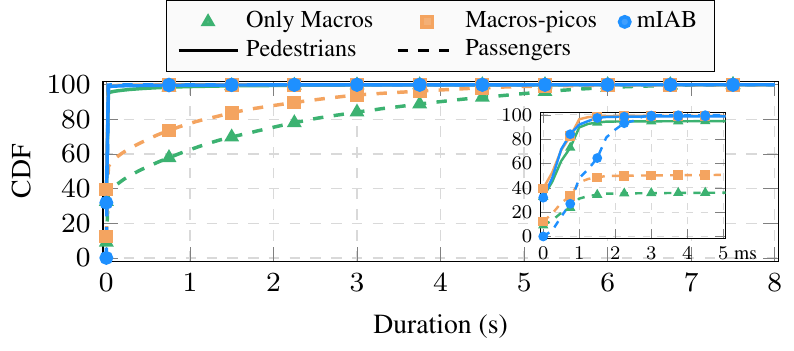}
		\label{FIG:Simulation-results-DL-latency-impact-of-packet-size-2048}
	}	
	
	\subfloat[Packet size = 1024 bits.]{%
		\includegraphics[width=\columnwidth]{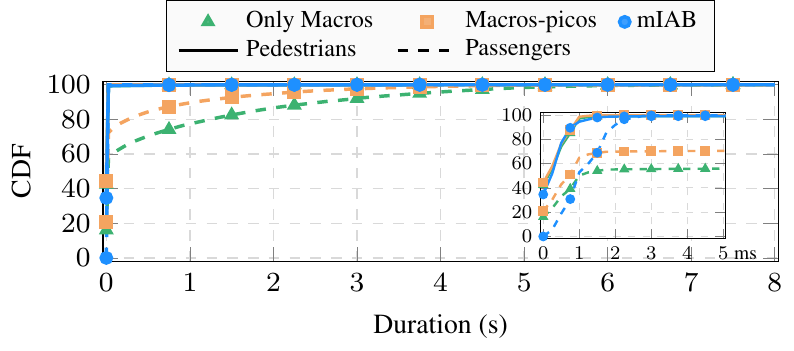}
		\label{FIG:Simulation-results-DL-latency-impact-of-packet-size-1024}
	}			
	\caption{Downlink latency – Impact of packet size, considering percentage of passengers equal to~$50\%$.}\label{FIG:Simulation-results-DL-latency-impact-of-packet-size}	
\end{figure}

\Ac{DL} transmissions with low \ac{MCS}, thus low \ac{DL} throughput, to the passengers in the only macros and macros-picos scenarios impacted the passengers \ac{DL} latency as it can be seen in \FigRef{FIG:Simulation-results-DL-latency-impact-of-perc-passengers} and \FigRef{FIG:Simulation-results-DL-latency-impact-of-packet-size}. %
These figures present the \ac{CDF} of the \acp{UE} end-to-end \ac{DL} latency for different values of the percentage of passengers and of the packet size, respectively. %
Dashed and solid curves represent the \ac{DL} latency of passengers and pedestrians, respectively. %
On the one hand, one can see that passengers using delay sensitive services will suffer in the only macros and macros-picos scenarios, since in the simulations their \ac{DL} latency achieved values up to \SI{6}{\s}. %
On the other hand, in the \ac{mIAB} scenario, \ac{DL} packet latency was negligible. %
In that scenario, for the majority of the passengers, in the worst case, the latency was equal to \SI{2}{\ms}. %
It corresponded to the case where packets were generated in the \ac{IAB} donor at slot~$5$ of \TabRef{TABLE:IAB-TDD} and needed to wait until slot~$10$ to be transmitted in the backaul from the \ac{IAB} donor to the \ac{mIAB} node and only at the next slot~$3$ of the next frame it was transmitted from the \ac{mIAB} node to the passengers, totalizing $8$~slots, i.e., \SI{2}{\ms}, of delay. %
Furthermore, similar to the analyses of \ac{DL} throughput, we can see that for higher percentages of passengers in the system and higher packet size, i.e., more loaded systems, the gains in the \ac{mIAB} scenario for the passengers \ac{DL} latency were higher compared to the other scenarios. %

\goodbreak
\subsubsection{Uplink}\label{SUBSUBSEC:Simulations-UL}

Until now, we have analyzed the \ac{mIAB} performance in the \ac{DL}. %
Now, let's focus on the \ac{UL}. %
In general, in the \ac{UL} we came to similar conclusions as the ones drawn in the \ac{DL} analyses. %
However, due to different transmit power and interference pattern in \ac{DL} and \ac{UL}, some points were different and will be highlighted in the following. %

\begin{figure}[!t]
	\centering
	\subfloat[Only macros scenario.]{%
		\includegraphics[width=\columnwidth]{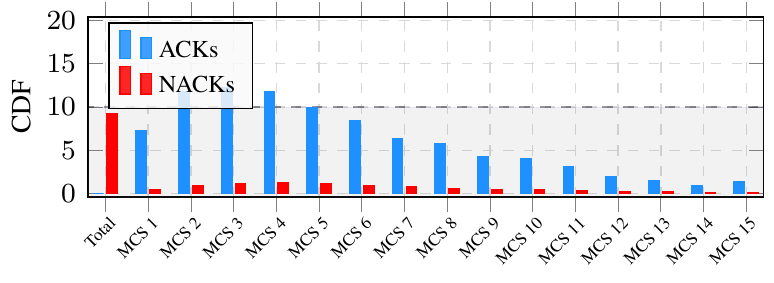}
		\label{FIG:Simulation-results-UL-MCS-passengers-macro-only}
	}	
	
	\subfloat[\ac{mIAB} scenario.]{%
		\includegraphics[width=\columnwidth]{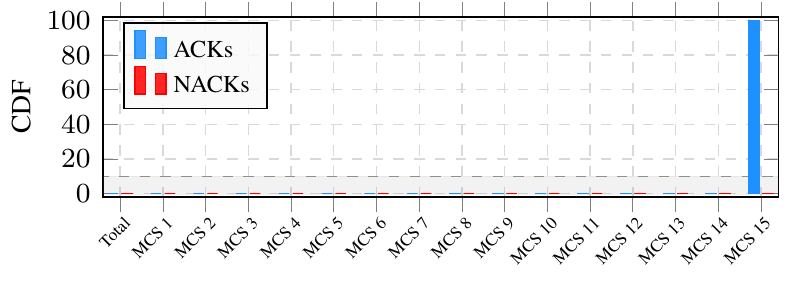}
		\label{FIG:Simulation-results-UL-MCS-passengers-IAB}	
	}		
	
	\caption{Histogram of passengers' \ac{MCS} usage in \ac{UL} in a scenario with $50\%$ of passengers and packet size of $3072$~bits.}\label{FIG:Simulation-results-UL-MCS-passengers}	
\end{figure}

Figure~\ref{FIG:Simulation-results-UL-MCS-passengers} presents the passengers' \ac{MCS} usage in \ac{UL} for the only macros and the \ac{mIAB} scenarios for the configuration with $50\%$ of passengers in the system and packet size of $3072$~bits. %
It is equivalent to \FigRef{FIG:Simulation-results-DL-MCS-passengers}, but for the \ac{UL}. %
The red bars represent the percentage of transmissions with error and the blue bars represent the percentage of successful transmissions. %
Comparing \FigRef{FIG:Simulation-results-DL-MCS-passengers-macro-only} and \FigRef{FIG:Simulation-results-UL-MCS-passengers-macro-only}, one can notice that in the \ac{DL} of the only macros scenario, even if the majority of the transmissions were with lower \acp{MCS}, $5\%$ of the transmissions were successful with the highest \ac{MCS}. %
However, in the \ac{UL}, less than $1.5\%$ of the transmissions were successful with the highest \ac{MCS}. %
The main reason for this was the lower transmit power in the \ac{UL} compared to the \ac{DL}. %
As a consequence, the \ac{SINR} of the passengers was lower, allowing them to mostly use only the lowest \acp{MCS}. %

\begin{figure}[!t]
	\centering
	\subfloat[Only macros scenario.]{%
		\includegraphics[width=\columnwidth]{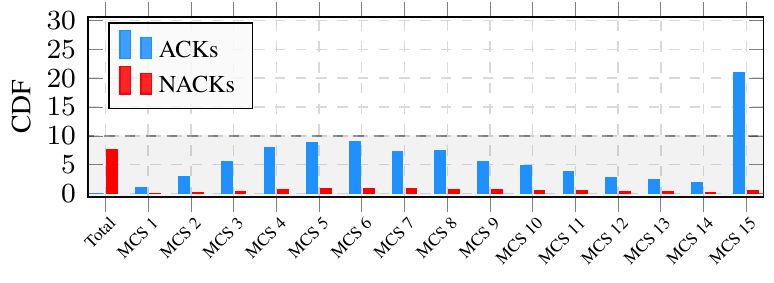}
		\label{FIG:Simulation-results-UL-MCS-pedestrians-macro-only}
	}	
	
	\subfloat[\ac{mIAB} scenario - pedestrians connected to an \ac{IAB} donor.]{%
		\includegraphics[width=\columnwidth]{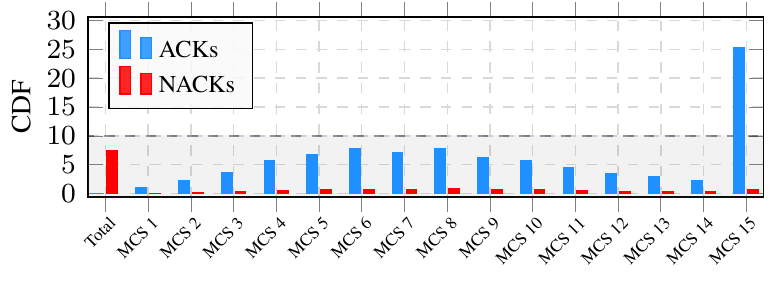}
		\label{FIG:Simulation-results-UL-MCS-pedestrians-IAB-serving-Donor}
	}	
	
	\subfloat[\ac{mIAB} scenario - pedestrians connected to an \ac{mIAB} node.]{%
		\includegraphics[width=\columnwidth]{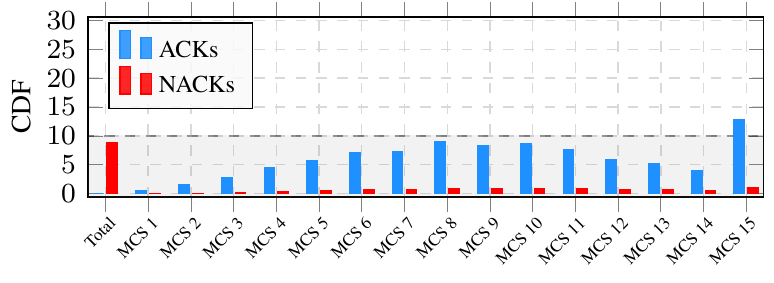}
		\label{FIG:Simulation-results-UL-MCS-pedestrians-IAB-serving-Node}
	}			
	\caption{Histogram of pedestrians' \ac{MCS} usage in \ac{UL} in a scenario with $50\%$ of passengers and packet size of $3072$~bits.}\label{FIG:Simulation-results-UL-MCS-pedestrians}
\end{figure}

The pedestrians' \ac{MCS} usage in  \ac{UL} is presented in \FigRef{FIG:Simulation-results-UL-MCS-pedestrians}. %
Figure~\ref{FIG:Simulation-results-UL-MCS-pedestrians-macro-only} is related to the only macros scenario, while \FigRef{FIG:Simulation-results-UL-MCS-pedestrians-IAB-serving-Donor} concerns pedestrians connected to an \ac{IAB} donor in the \ac{mIAB} scenario and \FigRef{FIG:Simulation-results-UL-MCS-pedestrians-IAB-serving-Node} concerns pedestrians connected to an \ac{mIAB} node also in the \ac{mIAB} scenario. %
Figure~\ref{FIG:Simulation-results-UL-MCS-pedestrians} is equivalent to \FigRef{FIG:Simulation-results-DL-MCS-pedestrians}, but for the \ac{UL}. %
Comparing \FigRef{FIG:Simulation-results-DL-MCS-pedestrians-IAB-serving-Node} and \FigRef{FIG:Simulation-results-UL-MCS-pedestrians-IAB-serving-Node}, one can notice that, for the pedestrians connected to an \ac{mIAB} node, in the \ac{DL}, only $5\%$ of the transmissions were successful with the highest \ac{MCS}, while, in the \ac{UL}, $13\%$ of the transmissions were successful with the highest \ac{MCS}. %
This difference is explained by the fact that, even though the transmit power in the \ac{UL} was lower, the interference was even lower. %
More precisely, analyzing \TabRef{TABLE:IAB-TDD}, we can see that a pedestrian connected to an \ac{mIAB} node receiving data in the \ac{DL} suffered interference from the \ac{MT} part of the \ac{mIAB} node, which was transmitting in the \ac{UL} to the \ac{IAB} donor and which did not suffer from the bus penetration loss since it was placed outside the bus. %
However, when the same pedestrian was transmitting in the \ac{UL} to the \ac{DU} part of an \ac{mIAB} node, the \ac{DU} part suffered interference from other pedestrians connected to the \ac{IAB} donor and transmitting data to it in the upstreaming. %
Thus, the interference that they caused in the \ac{DU} also suffered from the bus penetration loss. %

\begin{figure}[!t]
	\centering
	\includegraphics[width=\columnwidth]{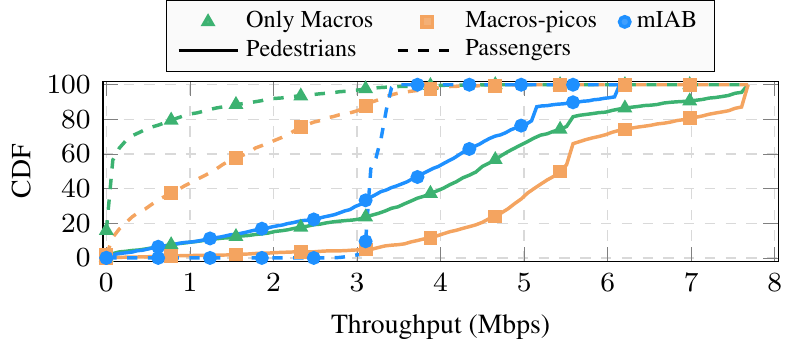}
	\caption{Uplink throughput considering packet size equal to $3072$~bits and percentage of passengers equal to~$50\%$.}\label{FIG:Simulation-results-Throughput-UL-50-passenger-3072-packet}
\end{figure}

The highlighted differences regarding the \ac{MCS} distribution impacted the \ac{UL} throughput. %
Figure~\ref{FIG:Simulation-results-Throughput-UL-50-passenger-3072-packet} presents the \ac{UL} throughput of passengers (dashed lines) and pedestrians (solid lines) considering packet size equal to $3072$~bits and percentage of passengers equal to~$50\%$. %
It is equivalent to \FigRef{FIG:Simulation-results-DL-throughput-impact-of-perc-passengers-50}, but for the \ac{UL}. %
As we have seen, the link quality of passengers in the only macros scenario was worse in the \ac{UL} than in the \ac{DL} due to the lower transmit power. %
As a consequence, in the only macros scenario, passengers \ac{UL} throughput was worse than their \ac{DL} throughput, as one can see comparing \FigRef{FIG:Simulation-results-Throughput-UL-50-passenger-3072-packet} and \FigRef{FIG:Simulation-results-DL-throughput-impact-of-perc-passengers-50}. %
Thus, for the passengers, the gains of deploying \ac{mIAB}, in terms of throughput, were even higher in the \ac{UL}. %
Concerning the pedestrians, in the \ac{mIAB} scenario, their \ac{UL} throughput was lower than their corresponding \ac{DL} throughput. %
It is explained by the fact that, in the adopted \ac{TDD} scheme for the \ac{mIAB} scenario, less time slots are used for \ac{UL} transmissions than for \ac{DL} transmissions. %

\begin{figure}[!t]
	\centering
	\includegraphics[width=\columnwidth]{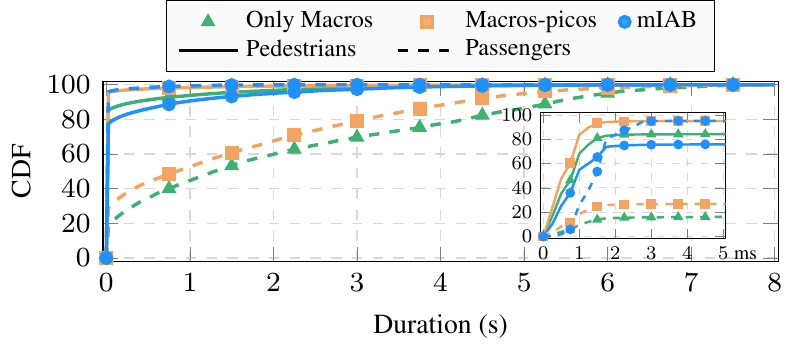}
	\caption{Uplink latency considering packet size equal to $3072$~bits and percentage of passengers equal to~$50\%$.}\label{FIG:Simulation-results-latency-UL-50-passenger-3072-packet}
\end{figure}

Figure~\ref{FIG:Simulation-results-latency-UL-50-passenger-3072-packet} presents the \ac{UL} latency results. %
It is equivalent to \FigRef{FIG:Simulation-results-DL-latency-impact-of-perc-passengers-50}, which presents the \ac{DL} latency results. %
Conclusions similar to the ones obtained frrom the \ac{UL} throughput can be drawn regarding the \ac{UL} latency. 

\goodbreak
\subsubsection{Backhaul Quality}\label{SUBSUBSEC:Simulations-backhaul}

\begin{figure}[!t]
	\centering
	\subfloat[Downlink.]{%
		\includegraphics[width=\columnwidth]{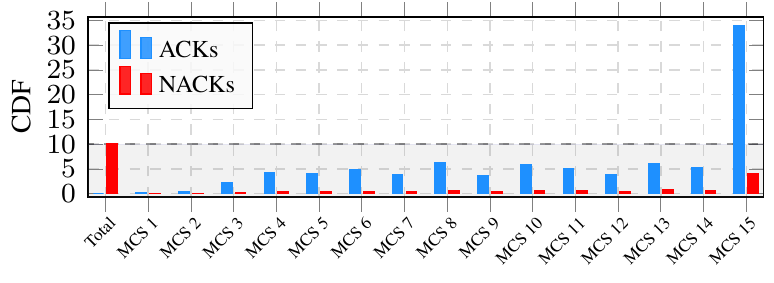}
		\label{FIG:Simulation-results-DL-MCS-backhaul}
	}	
	
	\subfloat[Uplink.]{%
		\includegraphics[width=\columnwidth]{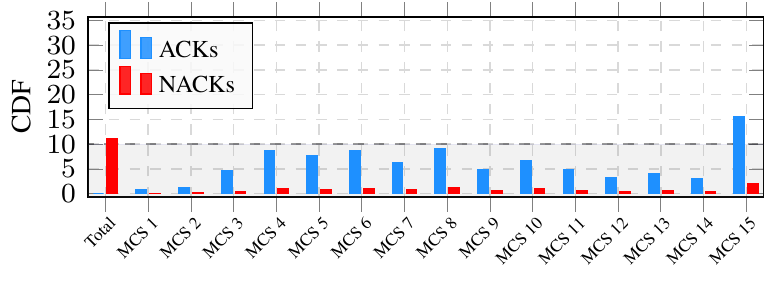}
		\label{FIG:Simulation-results-UL-MCS-backhaul}	
	}			
	\caption{Histogram of backhaul's \ac{MCS} usage in a scenario with $50\%$ of passengers and packet size of $3072$~bits.}\label{FIG:Simulation-results-MCS-backhaul}	
\end{figure}

In a \ac{mIAB} scenario, besides the quality of the links of passengers and pedestrians, one should also be careful with the quality of the backhaul link, since when a bus suffered from a bad link to their serving \ac{IAB} donor, all the passengers connected to it also suffered. %
Figure~\ref{FIG:Simulation-results-MCS-backhaul} presents the histogram of \ac{DL} and \ac{UL} backhaul's \ac{MCS} usage in a scenario with $50\%$ of passengers and packet size of $3072$~bits. %
Comparing \FigRef{FIG:Simulation-results-DL-MCS-backhaul} and \FigRef{FIG:Simulation-results-UL-MCS-backhaul} with \FigRef{FIG:Simulation-results-DL-MCS-passengers-macro-only} and \FigRef{FIG:Simulation-results-UL-MCS-passengers-macro-only}, respectively, one can again conclude that it was better for a passenger to communicate with the \ac{IAB} donor through an \ac{mIAB} node instead of being directly connected to it. %
The backhaul better link quality was mainly due to the deployment of the \ac{mIAB} node \ac{MT} part outside the bus, thus overcoming the bus penetration loss, and also due to the utilization of a \ac{ULA} with $64$~antenna elements instead of a single antenna as the passengers did. %

\goodbreak
\subsubsection{Links Served by \ac{IAB} Donors}\label{SUBSUBSEC:Simulations-links-served-by-macro}

We have seen that the adopted \ac{TDD} scheme impacts the results. %
Thus, it is interesting to try to understand how to better allocate the resources. %
For this, it is important to understand the profile of the links served by the \ac{IAB} donors. %
Hence, in this part we analyze the number of links served by the \ac{IAB} donors, the percentage of them that are access links and how many links are served by the \ac{IAB} donors in the backhaul of \ac{mIAB} nodes. %

\begin{figure}[!t]
	\centering
	\subfloat[75\% of passengers.]{%
		\includegraphics[width=\columnwidth]{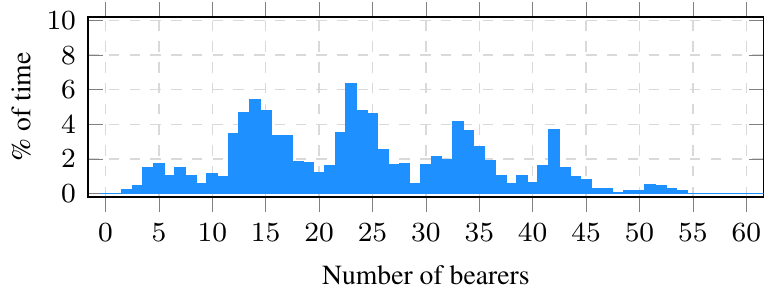}
		\label{FIG:Simulation-links-served-by-macro-75-passengers}
	}	
	
	\subfloat[50\% of passengers.]{%
		\includegraphics[width=\columnwidth]{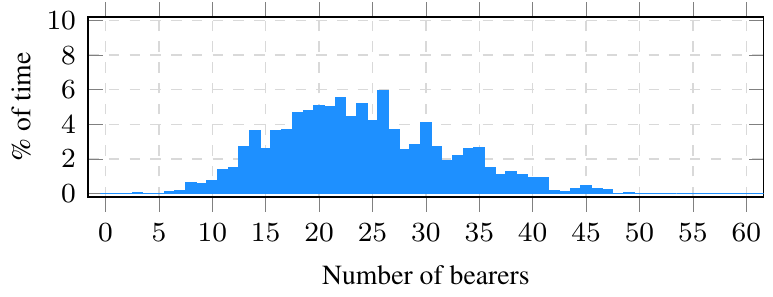}
		\label{FIG:Simulation-links-served-by-macro-50-passengers}
	}
	
	\subfloat[25\% of passengers.]{%
		\includegraphics[width=\columnwidth]{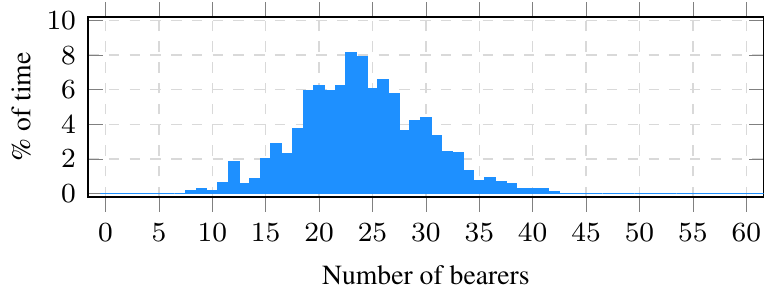}
		\label{FIG:Simulation-links-served-by-macro-25-passengers}
	}			
	\caption{Histogram of the number of links served by macro \acp{gNB}.}\label{FIG:Simulation-links-served-by-macro}	
\end{figure}

Figure~\ref{FIG:Simulation-links-served-by-macro} presents the histogram of the number of links served by \ac{IAB} donors. %
When serving a bus, i.e., an \ac{mIAB} node, an \ac{IAB} donor indirectly served through the backhaul the \acp{UE} served by the bus (almost always, only passengers). %
Thus, when a bus disconnected from the \ac{IAB} donor, the load in the donor suddenly changed a lot, since the passengers went away together with the bus. %
Due to this, as we can see in \FigRef{FIG:Simulation-links-served-by-macro} the number of links served by an \ac{IAB} donor could be quite different from one donor to another, since if they served a different number of buses, the difference in the number of served links was approximately multiple of the number of passengers per bus. %

This effect is well visible in~\FigRef{FIG:Simulation-links-served-by-macro-75-passengers}, which refers to the case of $75\%$ of passengers in the system, i.e., nine passengers per bus. %
Notice that, in this figure, there are six peaks (six was the number of buses in the simulations), each one centered in a multiple of nine, which is the number of passengers in a bus for this case. %
When the number of passengers was reduced from nine to six and three as in \FigRef{FIG:Simulation-links-served-by-macro-50-passengers} and \FigRef{FIG:Simulation-links-served-by-macro-25-passengers}, respectively, the load variability of the \ac{IAB} donors decreased, since, in those cases, moving a bus from one \ac{IAB} donor to another did not heavily impact the load of those donors. %

Based on these results, one should consider adopting topology adaptation strategies that take into account the \ac{IAB} donor load. %
This may avoid overloading \ac{IAB} donors when changing the serving donor of a given bus. %

\begin{figure}[!t]
	\centering
	\subfloat[75\% of passengers.]{%
		\includegraphics[width=\columnwidth]{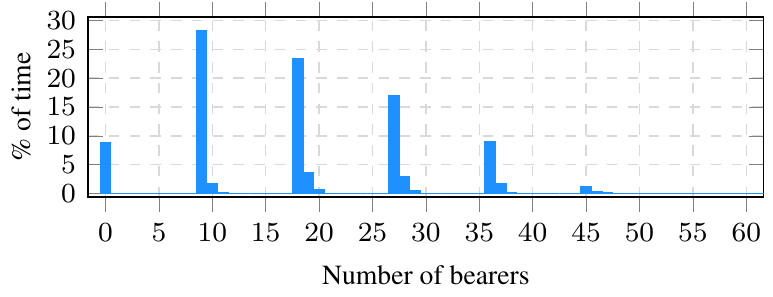}
		\label{FIG:Simulation-backhaul-links-served-by-macro-75-passengers}
	}	
	
	\subfloat[50\% of passengers]{%
		\includegraphics[width=\columnwidth]{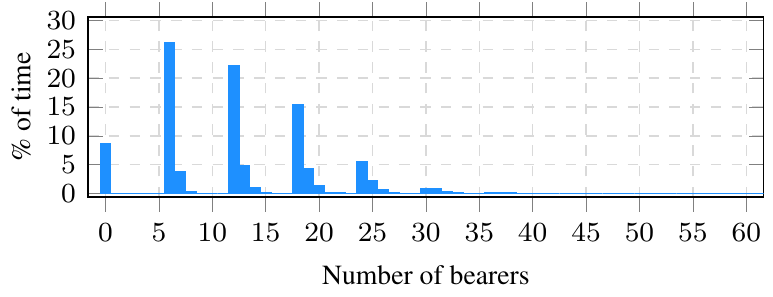}
		\label{FIG:Simulation-backhaul-links-served-by-macro-50-passengers}
	}	
	
	\subfloat[25\% of passengers]{%
		\includegraphics[width=\columnwidth]{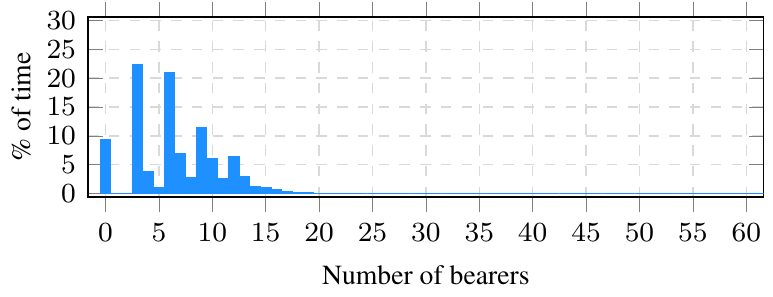}
		\label{FIG:Simulation-backhaul-links-served-by-macro-25-passengers}
	}			
	\caption{Histogram of the number of links in the backhaul served by macro \acp{gNB}.}\label{FIG:Simulation-backhaul-links-served-by-macro}	
\end{figure}

Figure~\ref{FIG:Simulation-backhaul-links-served-by-macro} focuses only on the number of backhaul links served by \ac{IAB} donors and presents the histogram of this number. %
Since, in the backhaul, the \ac{IAB} donors were serving links from the buses, one could expect that through the backhaul the donors would serve a number of links multiple of the number of passengers in the buses. %
However, \FigRef{FIG:Simulation-backhaul-links-served-by-macro} shows that sometimes the backhaul served a number of links a little bit higher than a number multiple of the number of passengersper bus. %
For example, in~\FigRef{FIG:Simulation-backhaul-links-served-by-macro-75-passengers}, which refers to the case of $75\%$ of passengers in the system, i.e., nine passengers per bus, there are six peaks, each one in multiples of nine, plus some bars corresponding to numbers a little bit higher that the closest multiple of nine. %
This is explained by the fact that buses could serve not only passengers, but also pedestrians, totalizing a number of served links equal to the number of passengers plus a small number of pedestrians. %

\begin{figure}[!t]
	\centering
	\subfloat[75\% of passengers.]{%
		\includegraphics[width=\columnwidth]{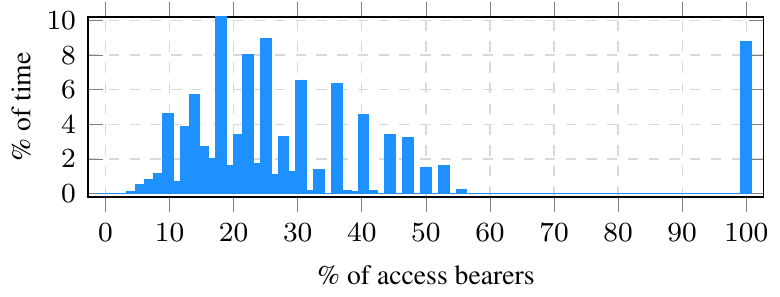}
		\label{FIG:Simulation-access-links-served-by-macro-75-passengers}	
	}	
	
	\subfloat[50\% of passengers.]{%
		\includegraphics[width=\columnwidth]{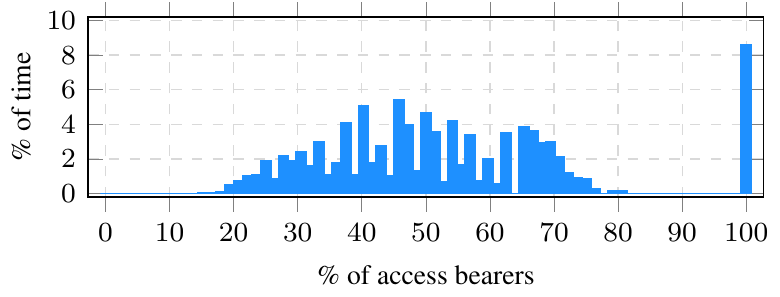}
		\label{FIG:Simulation-access-links-served-by-macro-50-passengers}
	}	
	
	\subfloat[25\% of passengers.]{%
		\includegraphics[width=\columnwidth]{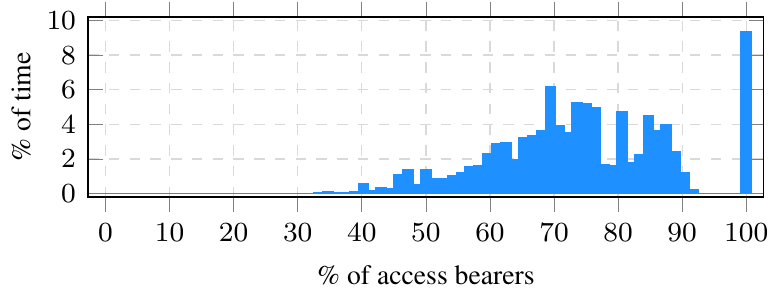}
		\label{FIG:Simulation-access-links-served-by-macro-25-passengers}
	}			
	\caption{Histogram of the percentage of links served by the macro \acp{gNB} that are access links.}\label{FIG:Simulation-access-links-served-by-macro}	
\end{figure}

Finally, \FigRef{FIG:Simulation-access-links-served-by-macro} presents the histograms of the percentage of links served by the \ac{IAB} donors that correspond to access links. %
The complementary of these numbers corresponds to backhaul links. %
Remark that the percentage of links that are access links varies a lot. %
As a consequence, one can conclude that adopting fixed time/frequency resources multiplexing schemes between donors and nodes may waste resources. %
For example, when the percentage of access links served by the \ac{IAB} donor was low, it could be better to schedule more resources to the backhaul of connected \ac{mIAB} nodes. %
In other words, one should consider adopting dynamic time/frequency resource multiplexing between access and backhaul based on the type of served links. %

Based on the conclusions drawn in this section and on the concepts presented until now, next section summarizes the main learned lessons and points out open issues and possible directions to address them. %


\goodbreak
\section{Lessons Learned, Open Issues and Future Directions}\label{SEC:Survey_Issues_Future}

\subsection{Cell Planning}\label{SUBSEC:Cell_Planning}
The \ac{mIAB} state-of-the-art works which consider transports with previously known behavior, e.g., trains and buses, usually take advantage of that and deploy \ac{IAB} donors in strategic places, e.g., along \added{the }railway track. %
For systems without a well-defined behavior, e.g., vehicles \added{moving in a city}, it is more challenging to \added{design the system in advance, e.g., to define where the \ac{IAB} donors should be deployed and how resource allocation and interference management can be handled}\deleted{prepare the system in advance for the dynamic changes}. %
A promising direction is to \added{adopt}\deleted{consider} \ac{AI}-based solutions \deleted{in order }to \added{learn possible \ac{mIAB} node trajectories and how the traffic varies along the day in order to be able to deploy \ac{IAB} donors in strategic positions}\deleted{predict the requirements of the \ac{mIAB} and prepare the system in advance}. %

\goodbreak
\subsection{Mobility Management}\label{SUBSEC:Mobility_Management}
\added{A challenge imposed by the mobility of an \ac{mIAB} node is that it can suddenly overload its target \ac{IAB} donor if the donor is not prepared to receive this new \ac{mIAB} node.} %
Similar to the previous topic, one adopted approach to deal with \added{the mobility of \ac{mIAB} nodes}\deleted{multiple \acp{UE} performing \ac{HO} at the same time} is to exploit the knowledge of the transport position and \added{its }trajectory, e.g., where a train or bus is and when they are going to arrive at the station. %
It allows the system to configure in advance the next \ac{IAB} donor to receive the incoming \acp{UE}. %
However, this strategy is more challenging when we deal with vehicles without a well-defined trajectory. %
A promising solution to investigate is also to use prediction-based methods in those cases. %
\added{For example, if the system is able to predict the arrival of an \ac{mIAB} node, and its corresponding additional traffic, instants before it arrives, nearby \acp{BS} can perform load balancing in advance in order to reserve resources for the coming node.} %

\goodbreak
\subsection{Dynamic Interference Management}\label{SUBSEC:Interference_Management}
Whether interference management is required or not depends on different parameters such as the \ac{mIAB} transmit power, scheduling, power adaptation capability of the nodes, etc. %
Many works considered the access link provided by \ac{mIAB} nodes \deleted{as }deployed in a frequency spectrum different from the one used by the rest of the system in order to eliminate the interference caused by the mobility of those \acp{UE} through the system. %
This is not an efficient solution since the spectrum is expensive and scarce and might stay unused in a given area as long as no \ac{mIAB} node is there. %
\added{On the other hand, it is still a challenge to deploy access and backhaul links at the same frequency band due to the dynamic interference caused by the mobility of the \ac{mIAB} nodes.} %
\added{Thus, solutions}\deleted{Solutions} for in-band interference management must be further investigated\deleted{, e.g., considering power control and/or dynamic spectrum allocation}. %
\added{For example, one can perform power control in the \ac{mIAB} nodes access links in order to reduce/avoid interfering with nearby \acp{UE} connected to the \ac{IAB} donor.} %
\added{Another option is to insert slots of silence on the \ac{TDD} frame pattern, as investigated in~\cite{Monteiro2022}. }%
\goodbreak
\subsection{Bandwidth Allocation for Access and Backhaul}\label{SUBSEC:Bandwidth_Allocation}
Most of the works considering in-band deployment of \ac{mIAB} adopted a fixed split of the resource (either in frequency or in time) between access and backhaul link. %
A strategy for sharing resources between backhaul link and access links of inside and outside \acp{UE} is still a challenge and must be further investigated, since it can lead to more efficient use of the spectrum. %
\added{For example, dynamic spectrum partitioning could be performed taking into account the load of data to be transmitted in the backhaul and in the access links of \ac{IAB} donor and \ac{mIAB} nodes.} %
\added{This solution guarantees fairness and avoids the waste of resources in case part of the spectrum is reserved for a given type of link, e.g., backhaul, but there is not enough data to be transmitted in that link.} %

\goodbreak
\subsection{Antenna Deployment}\label{SUBSEC:Antenna_Deployment}
It is almost a consensus that at least for trains and buses antennas are going to be deployed outside and at the top of these transports being connected to inside \acp{AP} which will provide connectivity for inside \acp{UE}. %
However, practical considerations must still be further evaluated, e.g., each operator will need to deploy its own equipment or will the infrastructure be shared, e.g., computational resources, antennas and frequency spectrum. %
\added{Moreover, since a possible use case of \ac{mIAB} nodes is to serve not only onboard \acp{UE} but also surrounding \acp{UE}, it is important to evaluate and compare pros and cons of deploying \ac{DU} antennas either only inside or both inside and outside of \ac{mIAB} nodes.} %

\goodbreak
\subsection{Role of \acs{UAV}}\label{SUBSEC:Role_UAV}
\Ac{UAV}-based communication is not considered in \ac{3GPP} Release~18 work-item on \ac{mIAB}. %
However, from an academic perspective, it is still interesting to investigate its performance. %
Works which use \acp{UAV} to enhance the coverage of the network usually first define an optimal position for the \acp{UAV} and fix them there. %
It is still under evaluated how \acp{UAV} perform when continually moving either to provide access link for \acp{UE} or to provide a wireless backhaul for other \acp{IAB} nodes. %

\goodbreak
\subsection{Improved Reliability for C-plane}\label{SUBSEC:Improved_reliability_CPlane}
\Ac{3GPP} standard documents do not yet cover \ac{mIAB} aspects. %
Different aspects should still be investigated, e.g., the possibility of splitting user and control planes as in \ac{DC} scenarios. %
In the context of \ac{mIAB} U-plane could be multi-hop while C-plane would be one-hop~\cite{3gpp.RP-192709}. %
This approach aims at reducing latency and failure probability. %
Furthermore, U/C-planes could even be deployed at different frequencies, where U-plane would be deployed at \ac{mmWave} for enhanced throughput while C-plane would be deployed at C-band for higher reliability and coverage. %

\goodbreak
\subsection{Topology Adaptation on mobile IAB}\label{SUBSEC:Topology_Adap_Mobile_IAB}
With inter-\ac{CU} topology adaptation, an \ac{IAB} node migrates from an old parent to a new parent where both parent nodes are served by different \ac{IAB} donor \acp{CU}. %
The current discussions are about the need for both \ac{MT} and \ac{DU} parts of the migrating \ac{IAB} node to handover. %
The main point is that \ac{DU} migration requires that all child nodes and \acp{UE} should perform handover to the new target donor-\ac{CU}. %
This is not the case when only \ac{MT}-part handover is executed. %
\Ac{DU} migration involves context transfer of the child \ac{IAB} nodes and \acp{UE}, update of their security keys and causes a large amount of signaling to be exchanged via the Xn, F1 and radio interfaces~\cite{3gpp.R3-212413}. %
This flood of signaling is called signaling storm. %
The main concern is that this procedure may cause service interruption as, currently, each \ac{DU} can be connected simultaneously to only one \ac{CU}. %
One of the envisaged solutions is the possibility of the support of two logical \acp{DU}, but this may increase the complexity of \ac{IAB}. %
Although these discussions on \ac{3GPP} are focused on fixed \ac{IAB}, they certainly are of utmost relevance for future discussions on \ac{mIAB} where topology adaptation should be more frequent. %

\goodbreak
\subsection{Signaling Optimization for Group Handover}\label{SUBSEC:Signal_opt_group_handover}

Another important aspect that has been discussed on \ac{3GPP} meetings is about the optimization of group handover when the \ac{DU} part changes its parent in inter-\ac{CU} topology adaptation. %
If standard handover procedure is followed, all child nodes and \acp{UE} would perform Xn handover procedure including the \ac{RACH} procedure. %
However, note that the \ac{RACH} procedure is not necessary in these cases since there is no change in the \ac{UE}'s strongest cell and child \ac{IAB} nodes' strongest parent. %
Finally, group signaling could be tailored for this case where handover messages of several \acp{UE} could be concatenated/compressed~\cite{3gpp.R3-206332}. %
As an example, each \ac{UE}'s \ac{RRC} reconfiguration message could be significantly reduced by updating only few parameters. %


\goodbreak
\section{Conclusions}\label{SEC:Survey_Conclusions}
As presented in this work, \ac{mIAB} is a candidate solution to address at least two challenges of current networks. %
Firstly, it enhances the service of moving \acp{UE}, e.g., passengers of trains and buses, guaranteeing that they are always connected and avoiding overloading the network with \ac{HO} messages when a group of moving \acp{UE} enters simultaneously the coverage area of a new cell. %
Secondly, it allows a quick deployment of new cells to provide/improve connectivity in high demanding areas with the deployment of \ac{mIAB} nodes. %

The basis for the deployment of \ac{mIAB} were presented. %
They are: current status of \ac{IAB} standardization in \ac{5G} \ac{NR} and state-of-the-art works regarding fixed \ac{IAB} and \ac{mIAB}. %
In order to help readers interested in this subject, the state-of-the-art works were classified following different criteria. %
On the one hand, fixed mobile related works were classified according to the studied dimensions, system modeling assumptions/constraints, considered problem objectives and \acp{KPI}, solution approaches and adopted mathematical tools. %
On the other hand, \ac{mIAB} related works were grouped according to the type of mobility, e.g., train, bus, \ac{UAV}, etc.. %
It was noticed that works considering similar type of mobility usually consider similar problems, e.g., \ac{UAV} positioning, and take advantage of specific characteristics of each type of mobility, e.g., previously known mobile trajectory of buses. %

After the literature review, we have performed an extensive analysis of \ac{mIAB} based on computational simulations. %
It was considered an urban macro scenario where \ac{IAB} nodes were deployed in buses in order to improve the passengers' connection. %
It was presented passengers and pedestrians throughput, latency and link quality in both downlink and uplink transmission directions. %
Results related to the wireless backhaul were also presented as well as the profile of links served by the \ac{IAB} donors. %
It was concluded that the deployment of the \ac{DU} part of an \ac{IAB} node inside a bus and its \ac{MT} part outside the bus remarkably improves the passengers throughput and latency. %
This is due to the fact that this deployment overcomes the significant bus penetration loss that a link between a \ac{BS} outside the bus  and a passenger suffers. %
Moreover, it was also concluded that an admission policy to allow a \ac{UE} to connect to a \ac{mIAB} cell should be adopted. %
The admission criteria could be: a minimum measured \ac{RSRP} value; a maximum measured interference level (\ac{RSRQ} instead of \ac{RSRP}); the relative \ac{UE} and bus geographical position in a given time interval; etc.. %
Furthermore, it was concluded that the adopted topology adaptation and \ac{TDD} scheme play an important role directly impacting the interference management strategy. %
Thus, it would be better if dynamic topology adaptation and time/frequency resource scheduling were adopted taking into account the \ac{IAB} donor load, e.g., amount of served traffic from access and backhaul links. %

Finally, we have summarized lessons learned, open issues and future directions related to \ac{mIAB}. %
One important challenge is how to deal with the dynamic interference caused by mobile cells when moving around the system, specially in the case with in-band deployment of backhaul and access. %
Besides, a promising solution for this and other challenges is the use of predictive solutions, e.g., \ac{AI}-based, in order to anticipate what is going to happen and prepare the system in advance. %


\ifCLASSOPTIONcaptionsoff
  \newpage
\fi

\printbibliography

\begin{IEEEbiography}[{\includegraphics[width=1in,height=1.25in,clip,keepaspectratio]{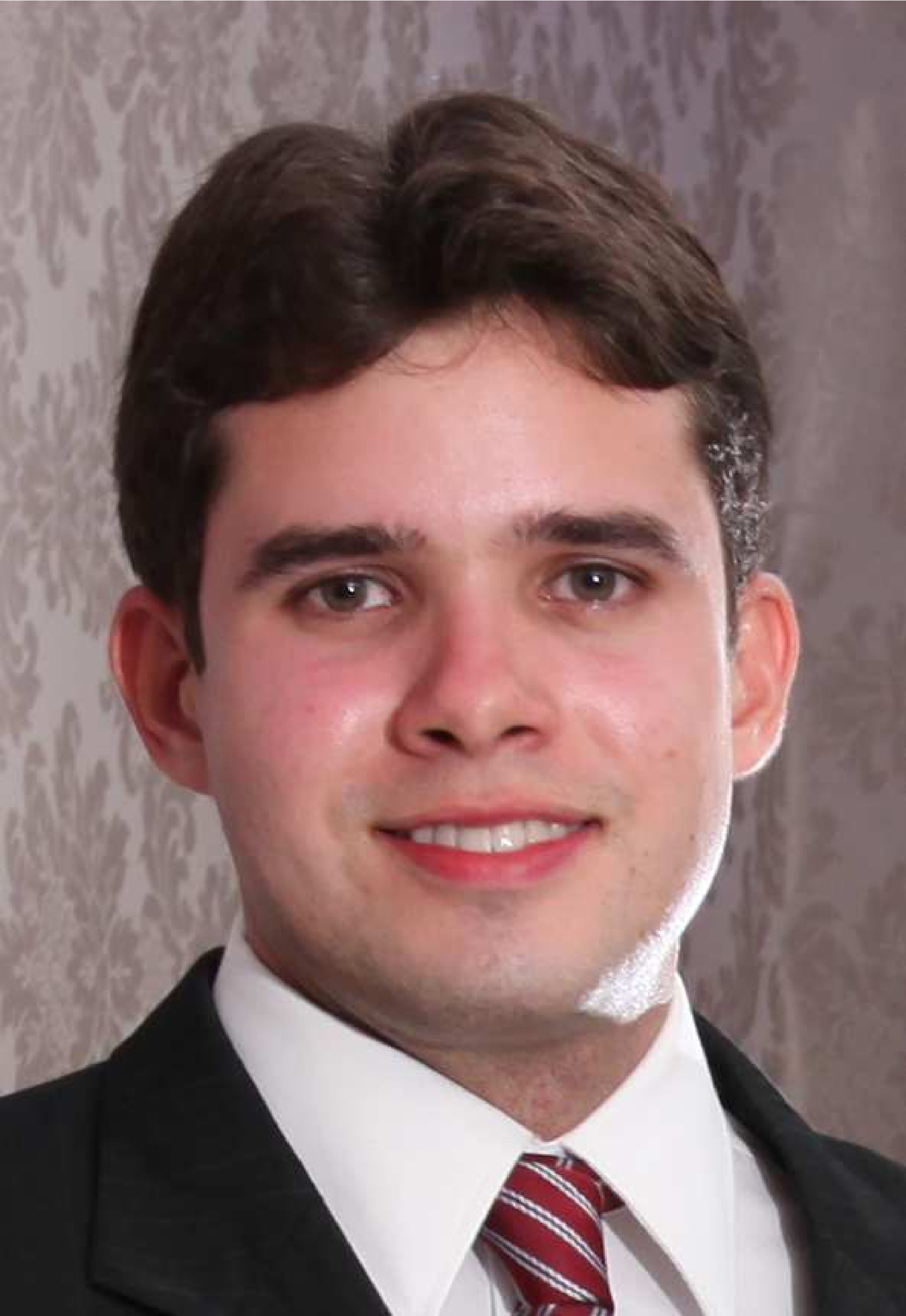}}]{Victor F. Monteiro} received the double B.Sc. degree in General Engineering, from the École Centrale Lyon, France and in Telecommunications Engineering (magna cum laude), from the Federal University of Ceará (UFC), Fortaleza, Brazil, in 2013. In 2015 and 2019, he received the M.Sc. and PhD degrees in Telecommunications Engineering from UFC, respectively. He is currently a Postdoctoral researcher at the Wireless Telecommunications Research Group (GTEL), UFC, where he works in projects in cooperation with Ericsson Research. In 2016, he was an invited researcher at Ericsson Research in Lulea, Sweden, where he worked for 6 months on the topic of LTE-NR Dual Connectivity. Besides, in 2017/2018, he spent one year at Ericsson Research in Stockholm, Sweden, where he worked on the topics of mobility management and channel hardening. From 2010 to 2012, he took part, in France, of the Eiffel Excellence Scholarship Program, established by the French Ministry of Foreign Affairs. His research interests include machine learning, 5G architecture and protocols, 5G measurement and reporting procedures, mobility management and radio resource allocation. 
\end{IEEEbiography}

\begin{IEEEbiography}[{\includegraphics[width=1in,height=1.25in,clip,keepaspectratio]{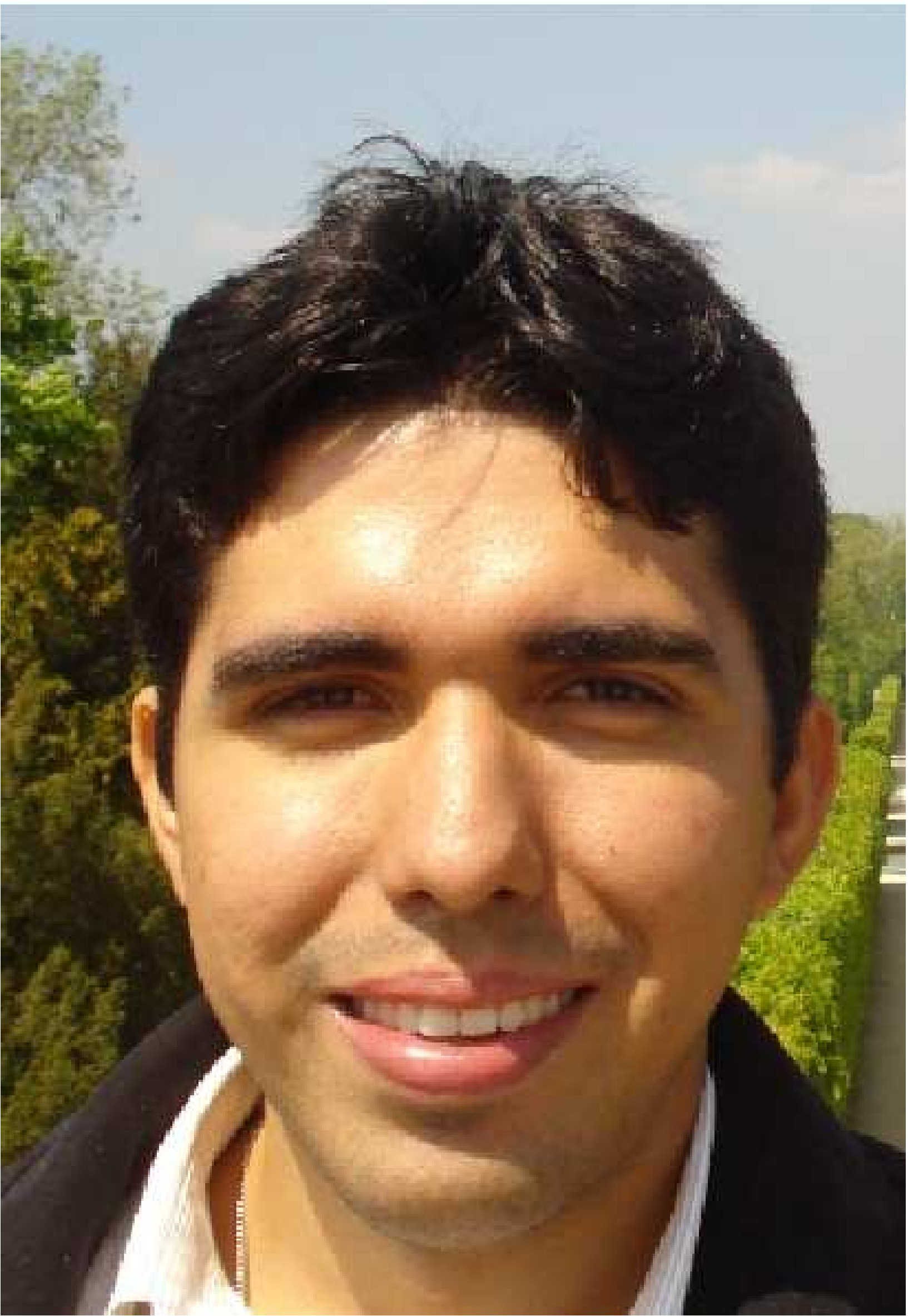}}]{Fco. Rafael M. Lima} received the B.Sc. degree with honors in Electrical Engineering in 2005, and M.Sc. and D.Sc. degrees in Telecommunications Engineering from the Federal University of Ceará, Fortaleza, Brazil, in 2008 and 2012, respectively. In 2008, he has been in an internship at Ericsson Research in Lulea, Sweden, where he studied scheduling algorithms for LTE system. Since 2010, he has been a Professor of Computer Engineering Department of Federal University of Ceará, Sobral, Brazil. Prof. Lima is also a senior member of IEEE and senior researcher at the Wireless Telecom Research Group (GTEL), Fortaleza, Brazil, where he works in projects in cooperation with Ericsson Research. He has published several conference and journal articles as well as patents in the wireless telecommunications field. His research interests include radio resource allocation algorithms for QoS guarantees for 5G and beyond 5G networks in scenarios with multiple services, resources, antennas and users.
\end{IEEEbiography}

\begin{IEEEbiography}[{\includegraphics[width=1in,height=1.25in,clip,keepaspectratio]{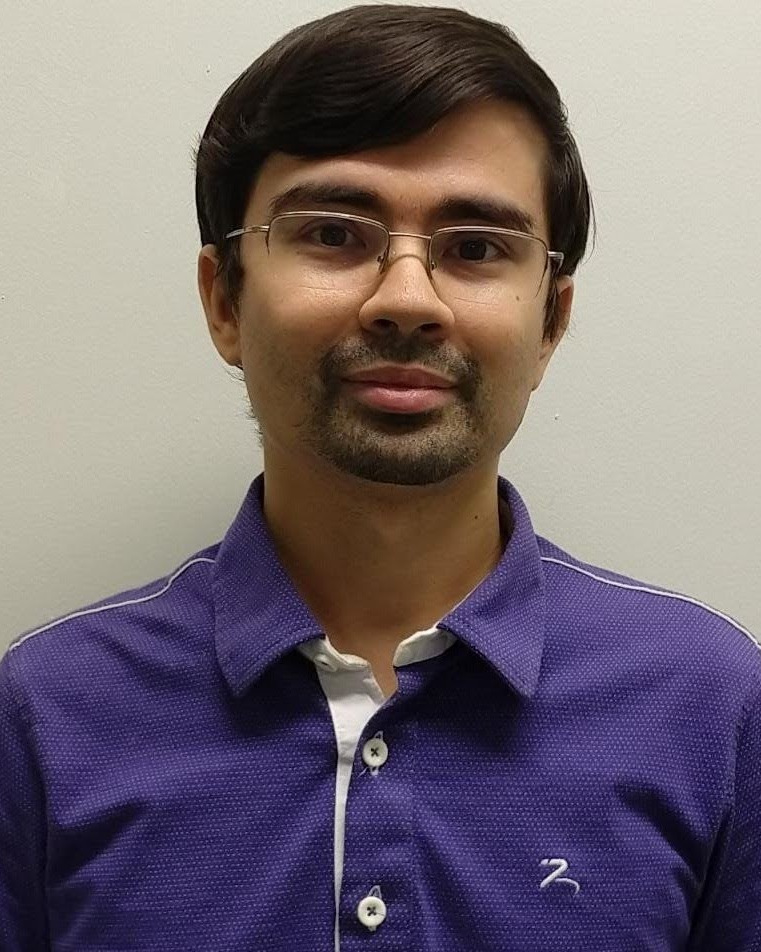}}]{Darlan C. Moreira} received B.Sc. degree in electrical engineering in 2005 from the Federal University of Ceará (UFC), Fortaleza, Brazil, in 2005. In 2007 and 2020 he received the M.S. and Ph.D. degrees, respectively, in Teleinformatics Engineering from the same institution. In 2007, he was an invited researcher at Ericsson Research in Kista, Sweden, where he worked for 3 months. In 2010, he was an invited researcher at Supélec in Gif-sur-Yvette, France, for 6 months. Since 2005 he has been a researcher at Wireless Telecommunications Research Group (GTEL), Brazil, where he has been working in projects within the technical cooperation between GTEL and Ericsson Research. Some topics of his research interests include machine learning, MIMO transceiver design, channel estimation, interference management, integrated access and backhaul, and modeling and simulation of cellular communication.
\end{IEEEbiography}

\begin{IEEEbiography}[{\includegraphics[width=1in,height=1.25in,clip,keepaspectratio]{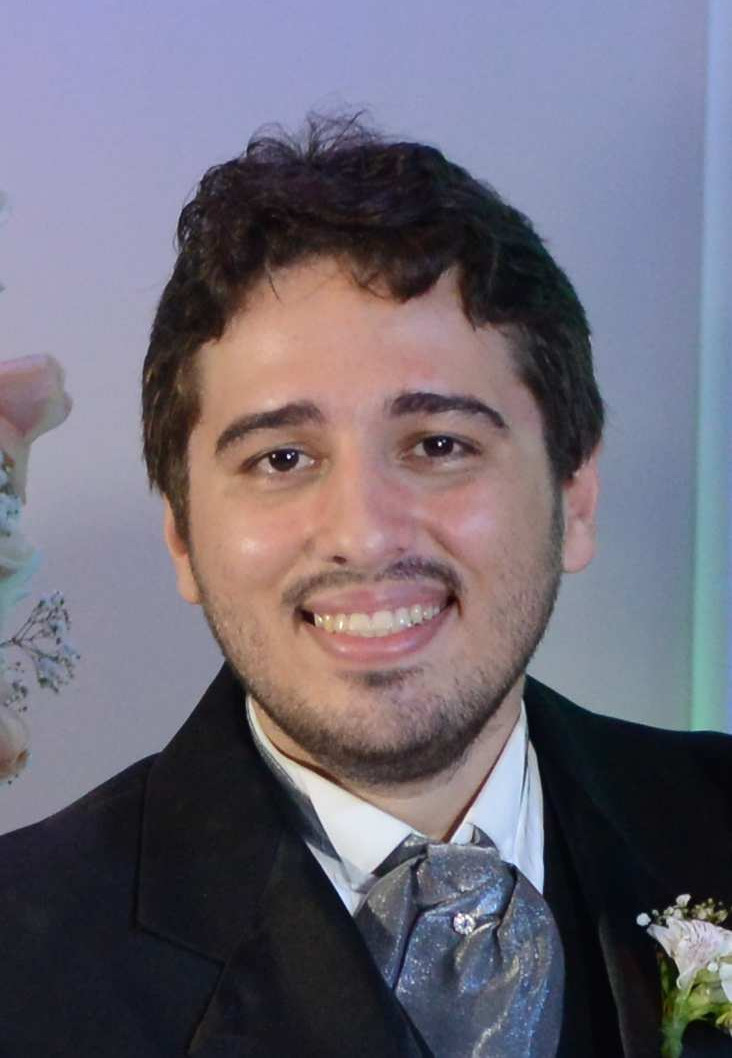}}]{Diego A. Sousa} 
	received the B.Sc. degree in Computer Engineering in University of Cear\'{a} (UFC), Sobral, Brazil, in 2011.
	He received M.Sc. and Ph.D. degree in Teleinformatics Engineering from the UFC, Fortaleza, Brazil, in 2013 and 2018, respectively.
	Since 2013, he has been a researcher at the Wireless Telecom Research Group (GTEL), UFC, participating of projects in a technical and scientific cooperation with Ericsson Research.
	Also, since 2013, he has been a Professor of the Federal Institute of Education, Science, and Technology of Cear\'{a} (IFCE), Paracuru, Brazil.
	His research interests include numerical optimization, 5G networks, coordinated scheduling, radio resource management for QoS/QoE provisioning.
\end{IEEEbiography}

\begin{IEEEbiography}[{\includegraphics[width=1in,height=1.25in,clip,keepaspectratio]{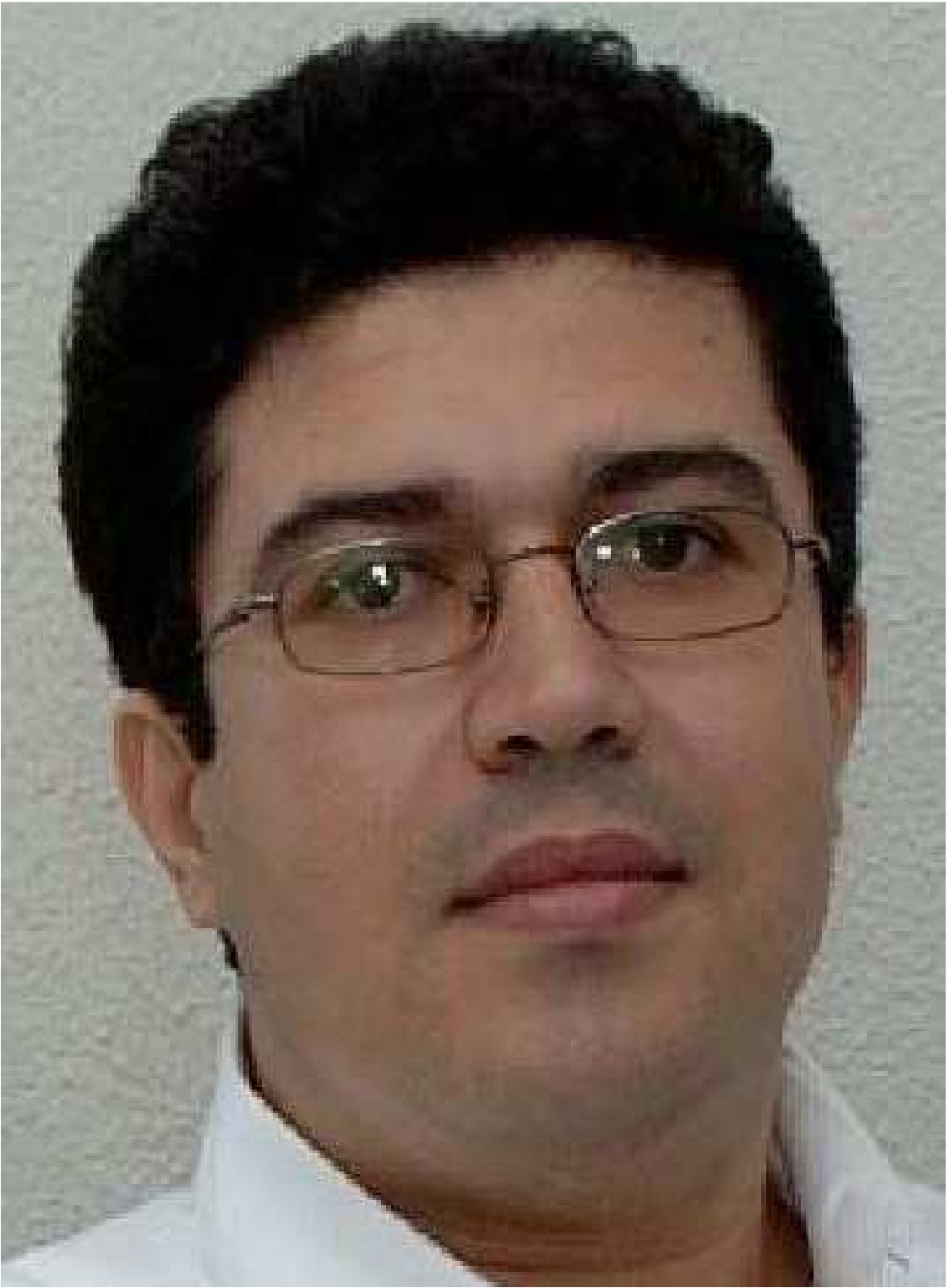}}]{Tarcisio F. Maciel} received his B.Sc. and M.Sc. degrees in Electrical Engineering from the Federal University of Ceará (UFC) in 2002 and 2004, respectively, and his Dr.Ing. degree from Technische Universitat Darmstat (TUD), Germany, in 2008, also in Electrical Engineering. Since 2001, he has actively participated in several projects in a technical and scientific cooperation between the Wireless Telecommunications Research Grupo (GTEL), UFC, and Ericsson Research. From 2005 to 2008, he was a research assistant with the Communications Engineering Laboratory, TUD. Since 2008, he has been a member of the Post-Graduation Program in Teleinformatics Engineering, UFC. In 2009, he was a Professor of computer engineering with UFC-Sobral and sine 2010, he has been a Professor with the Center of Technology, UFC. His research interests include radio resource management, numerical optimization and multi-user/multi-antenna communications. 
\end{IEEEbiography}

\begin{IEEEbiography}[{\includegraphics[width=1in,height=1.25in,clip,keepaspectratio]{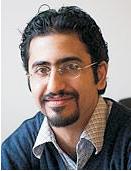}}]{Behrooz Makki} (Senior Member, IEEE) received the Ph.D. degree in communication engineering from Chalmers University of Technology, Gothenburg, Sweden. He was a Postdoctoral Researcher with the Chalmers University of Technology from 2013 to 2017. He is currently working as a Senior Researcher with Ericsson Research, Gothenburg, Sweden. He has co-authored 67 journal papers, 47 conference papers, and 90 patent applications. His current research interests include integrated access and backhaul, green communications, millimeter wave communications, NOMA, and backhauling. Behrooz is a recipient of the VR Research Link Grant, Sweden, in 2014, the Ericsson’s Research Grant, Sweden, in 2013, 2014, and 2015, the ICT SEED Grant, Sweden, in 2017, as well as the Wallenbergs Research Grant, Sweden, in 2018. Also, he is a recipient of the IEEE Best Reviewer Award for IEEE TRANSACTIONS ON WIRELESS COMMUNICATIONS in 2018, and the IEEE Best Editor Award for IEEE WIRELESS COMMUNICATIONS LETTERS in 2020. He is currently working as an Editor for IEEE TRANSACTIONS ON COMMUNICATIONS, IEEE WIRELESS COMMUNICATIONS LETTERS, and IEEE COMMUNICATIONS LETTERS. He was a member of the European Commission projects ``mm-Wave Based Mobile Radio Access Network for 5G Integrated Communications'' and ``ARTIST4'' as well as various national and international research collaborations.
\end{IEEEbiography}

\begin{IEEEbiography}[{\includegraphics[width=1in,height=1.25in,clip,keepaspectratio]{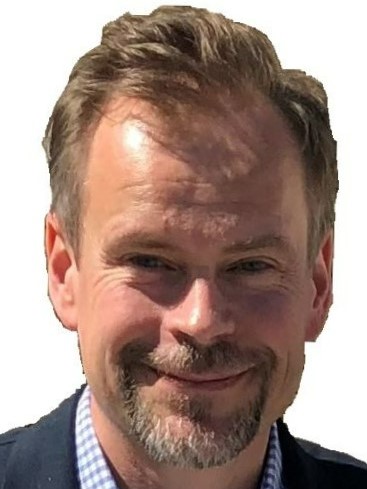}}]{Hans Hannu} is a Master Research Engineer at Ericsson Research since 2011. He holds a M. Sc. in Electrical Engineering / Signal Processing from Luleå University of Technology, 1998, Sweden. Hans works in areas as service realization and performance over cellular networks, including radio access and network transport protocol, with algorithm development, such as data transmission scheduling, and project management. He has co-authored 47 granted patents.
\end{IEEEbiography}

\end{document}